\pdfoutput=1
\documentclass[draft]{agujournal2019}
\usepackage{url}
\usepackage{amsmath}

\draftfalse

\makeatletter
\def\ps@headings{\def\@oddfoot{\centerline{\small --\the\c@page--}}
  \let\@evenfoot\@oddfoot
  \let\@oddhead\@empty
  \let\@evenhead\@empty}
\ps@headings
\makeatother

\begin{document}

\title{Paleoclimate Boundary Conditions as an Out-of-Sample Test for the Forced Response of Ocean Climate Emulators}

\authors{Adam Subel \affil{1}, Laure Zanna \affil{1}}

\affiliation{1}{Courant Institute School of Mathematics, Computing, and Data Science, New York University, New York, USA}

\correspondingauthor{Adam Subel}{adam.subel@nyu.edu}

\begin{keypoints}
\item AI ocean emulators can generalize to out-of-sample orbital forcings, reproducing the upper-ocean, large-scale forced climate response
\item Emulators linearly superimpose their responses to individual boundary forcings and struggle to represent slow internal dynamics
\item Optimizing emulators for skill using mean squared error does not guarantee accurate dynamical responses
\end{keypoints}

\begin{abstract}
AI weather emulators benefit from clear objectives and metrics, which have led to the rapid development of models that outperform traditional benchmarks. In contrast, long-term climate emulators must reliably reproduce forced responses over months to centuries, while relying on training objectives that span a small number of model time steps. We assess autoregressive, full-depth ocean emulators using data from the midHolocene experiment of a numerical climate model to examine their skill in responding to surface forcings from an in-distribution, out-of-sample climate. We demonstrate that these emulators generalize to new orbital forcings, reproducing the spatial structure of the large-scale response as well as changes in seasonal patterns and in the spatial structure of ocean variability, while underestimating their amplitude. Baselines that infer the ocean state directly from the boundary forcings also recover much of the large-scale pattern, but only near the surface, and capture neither the seasonal nor the variability changes, indicating that these require some representation of dynamics. Despite these successes, the emulators fail to reproduce the slow, internally driven evolution of the ocean interior. We then show that the emulators' total forced response is well reconstructed by linearly composing their independent responses to each forcing component. Tracking response across training epochs, we find that convergence on mean state metrics in the training climate does not guarantee that the emulators capture the dynamics necessary for a skillful response. Together, these experiments establish the midHolocene as a controlled, ground-truthed setting for diagnosing forced-response failures before emulators are pushed to out-of-distribution climates.
\end{abstract}

\section*{Plain Language Summary}

AI models for weather forecasting benefit from tackling well-defined challenges, which has enabled the rapid development of models that are competitive against state-of-the-art baselines.  The goals we hope to address with long-term climate emulators are broad and less easily quantified. Building trust and ensuring physical reliability in these rapidly improving tools requires examining how they respond to forcings in settings the emulator has not seen during training. In this work, we use data from a numerical model run under midHolocene conditions (6000 years before the present) to evaluate a full-depth ocean emulator. We use the midHolocene as our test case since the changes across climates primarily reflect shifts in the mean patterns of surface forcings and do not require the emulator to act on values far outside the training data.  We show that the emulator successfully generalizes to these new conditions, capturing large-scale climate patterns, seasonal changes, and ocean variability, but struggles to reproduce interactions occurring in the ocean interior. Comparing the emulator across different stages of training, we show that improvement on common metrics used to evaluate weather and climate emulators, such as mean squared error, does not guarantee better skill in the midHolocene tests.

\section{Introduction}
Rapid growth in machine learning for weather forecasting has driven the development of a wide range of emulator architectures and approaches, which outperform traditional baselines on short to medium range tasks \cite{bi2023accurate,bonev2025fourcastnet,kochkov2024neural,couairon2024archesweather,lang2024aifs,lam2023learning}. Recently, multiple groups have built on this to develop emulators that can run stably for climate timescales, decades to centuries, for the atmosphere \cite{watt2025ace2,chapman2025camulator,guan2025lucie}, sea ice \cite{gregory2026floenet}, and the ocean \cite{dheeshjith2025samudra}. In addition, atmospheric emulators have been coupled to a numerical slab ocean \cite{clark2024ace2},  a sea surface temperature (SST) emulator \cite{cresswell2025deep}, and promising initial progress has shown that coupling an atmosphere emulator to a full-depth ocean emulator can produce realistic rollouts for more than a century \cite{duncan2025samudrace}. Compared to traditional numerical models, which are resource-intensive and often require significant institutional knowledge to use successfully, rolling out these emulators is low-cost and requires minimal specialized expertise. Assuming we faithfully reproduce the underlying system, skillful autoregressive emulators could lower technical barriers and act as a step towards enabling the creation of low-cost surrogate models for scientists. The successful construction of these surrogates would  allow researchers to rapidly test new ideas and explore how perturbations affect the long-term climate state. 

While more recent work looks beyond mean climate statistics, deeper evaluations of the dynamical response to perturbations in emulators are not yet standard practice. Some studies have explored emulators' ability to generate extremes, finding that AI models can reproduce extremes within the training distribution but fail to extrapolate to novel events \cite{sun2025can,sun2025predicting}. Others have performed dynamical tests on atmospheric emulators, confirming physical responses to idealized, localized perturbations \cite{hakim2024dynamical}. Initial work has also estimated the response in the top of the atmosphere radiation imbalance to changes in SSTs using Green's functions computed with emulators, though disagreements with the reference model remain \cite{wu2025applying,van2025reanalysis}.  The most common test case for out-of-sample generalization has been climate warming scenarios \cite{kochkov2024neural,clark2024ace2,dheeshjith2024transfer,chapman2025camulator,guan2025lucie}. While building emulators to support warming runs is an important end goal, these testbeds pose out-of-distribution challenges for AI emulators.

Rather than validating emulators over shorter, transient historical periods or on warming experiments such as 1\% CO$_2$ forcings, which pose challenges even for traditional models  \cite{deser2012communication,milinski2020large}, we look to recent work that suggests using paleoclimate experiments as an additional validation framework for numerical models before moving to future climate \cite{burls2022increasingly,tierney2020past}. With this in mind, we use midHolocene data to explore the emulator's dynamic response in an out-of-sample but in-distribution test case. Within the midHolocene testbed, we conduct a series of tests to diagnose not just what the response is, but how the emulator uses individual boundary forcings to produce the full response. This work is, to our knowledge, the first instance of an autoregressive emulator being trained and evaluated on paleoclimate data; however, our primary goal is to understand the strengths and weaknesses of current emulators in forced-response settings and to highlight the importance of dynamical tests.

We demonstrate the strengths and weaknesses of emulators in capturing dynamical responses through three key results. First, the emulator demonstrates the ability to generalize, appropriately adjusting to new orbital forcing (Section \ref{sec:orbital_forcing}) and capturing changes in the major modes of ocean variability (Section \ref{sec:variability}). Second, we show that the emulator captures regional features of the upper ocean temperature response to midHolocene forcings (Section \ref{sec:Overall_Response}). We further investigate and find independent responses to each boundary forcing, which all linearly superimpose to reproduce the full emulator response (Section \ref{sec:component_forcings_linearity}). Finally, by comparing skill across epochs, we explore the relationship between mean squared error and the emulator's skill in reproducing the true response (Section \ref{sec:epoch_sensitivity}). Across the work, we compare the emulator to two baselines that map the boundary forcing directly to the ocean state, to better isolate where the emulator utilizes internal dynamics to improve the response, and find that it outperforms both in capturing changes in the seasonal cycle, the spatial patterns of temporal variability, and the amplitude and depth of the response (Section \ref{sec:baselines}).

\section{Methods}
\subsection{Data}

This work uses data from the piControl and midHolocene numerical experiments with the Community Earth System Model 2 (CESM2) \cite{danabasoglu2020community} to train and evaluate AI ocean emulators. The primary data source is from the ocean component of these simulations. The ocean component of CESM2 is Parallel Ocean Program 2 (POP2), with a nominal horizontal resolution of $1^\circ$, which corresponds to a uniform $1.25^\circ$ in the zonal direction and a spacing in the meridional direction ranging from $0.27^\circ$ to $0.53^\circ$. POP2 has 60 fixed vertical levels with a maximum depth of 5500 m.

From all datasets, we extract potential temperature, $\theta_O$, and salinity, $S$, as state variables. We additionally extract the boundary surface forcings:  total heat flux coming into the ocean at its surface, $\operatorname{hfds}$, and the two components of surface stress at the interface between the ocean and either the atmosphere or sea ice,  $\tau_u$ (zonal) and $\tau_v$ (meridional). The available CESM2 midHolocene experiment data do not directly provide total surface shear stress at the ocean surface; rather, they provide the stress at the bottom of the atmosphere and at the bottom of the sea ice separately. For this midHolocene run, we first process the stress below the sea ice and the atmospheric stress independently before combining the fields. The total surface stress is calculated by combining the atmospheric and sub-ice stress fields, with local sea ice concentration as the weighting factor. For example, given a sea ice concentration of 0.50, we take an equal average of the two stress components. The midHolocene$-$piControl differences in Figure \ref{fig:Holo_Pi_Comp}, Panels B and C, show that the combined field varies smoothly across the seasonal ice margin, with no small-scale structure imprinted by the ice concentration weighting. This ice-concentration weighting is the same convention CESM2 applies to produce the $\operatorname{hfds}$ field it provides directly for both experiments, so the reconstruction follows CESM2's own definition. We use monthly means for all aforementioned variables. 

We regrid the data to reduce the computational burden during training and to place it on a more regular grid.  This regridding is done in two parts: first vertically and then horizontally. We use a conservative vertical regridding to map our data from the native 60 levels to 19 target levels (2.5, 10, 22.5, 40, 65, 105, 165, 250, 375, 550, 775, 1050, 1400, 1850, 2400, 3100, 4000, 5000, 6000m) as in \citeA{dheeshjith2025samudra}. We note that the vertical bounds of the target grid include a cell below the original vertical grid of CESM2, which we fill with NaNs.  This choice reflects an effort to use data across models in the development of this work. While we do not include data from multiple numerical models in this manuscript, we retain the extra depth channel to avoid retraining a significant number of models and to maintain flexibility for future work. We conducted tests to confirm that performance was not affected, consistent with our expectation that an emulator in this setting ignores the unused layers. Because the network output is masked, the 6000m channels contribute no gradient information during training and do not affect the loss or the learned weights. To ensure we produce realistic values, we created partial grid cells by defining a new cell thickness, $dz^*$, rather than using a fixed cell depth per layer. The steps for regridding a target variable, $\psi(x,y,z)$, from an initial vertical grid at depth $z$, to a coarse vertical grid at depth, $z^*$ are as follows:

\begin{align}
    \delta(x,y,z,z^*) &= \left(\min(z^*_l, z_l) - \max(z^*_u, z_u)\right) H\left(\min(z^*_l, z_l) - \max(z^*_u, z_u)\right), \\
    dz^*(x,y,z^*) &= \sum_z \delta(x,y,z,z^*), \\
    \psi^*(x,y,z^*) &= \frac{\sum_z \delta(x,y,z,z^*)\psi(x,y,z)}{dz^*(x,y,z^*)}.  
\end{align}

\noindent Here, the subscripts $\cdot_u$ and $\cdot_l$ represent the upper and lower boundaries of a vertical cell, respectively, and $H$ is the Heaviside function. We define $\delta(x,y,z,z^*) [m]$ as the fraction of cell, $z$, that overlaps a target coarse cell, $z^*$. Summing over $z$  gives us the total thickness of the target coarse layer, $dz^*(x,y,z^*)$. Using the computed $\delta$ and $dz^*$, we compute the coarsened field, $\psi^*(x,y,z^*)$.  After regridding the data onto the target vertical grid, we regrid it horizontally onto a target Gaussian grid with a nominal horizontal resolution of $1^\circ$. We use the conservative normed method from the xESMF package to regrid and assign values to a coarse cell whenever it overlaps any ocean cells in the native grid \cite{zhuang_jiawei_2023_8356796}. 

\subsection{Training Features and Datasets}
We define three classes of input/output features for our neural networks (see Section \ref{sec:architecture}), which we delineate as (1) state variables, (2) dynamic boundary variables, and (3) computed insolation. For the forward propagator, we define the state variables, $\boldsymbol{\Phi}$, to be the thermodynamic ocean variables potential temperature, $\theta_{O}~ [^{\circ}C]$, and salinity, $S~ [\operatorname{psu}]$. We then define the dynamic boundary conditions, $\boldsymbol{\tau}$, to be surface zonal stress, $\tau_u~[N/m^2]$, meridional stress, $\tau_v~[N/m^2]$, and total heat flux, $\operatorname{hfds}~[W/m^2]$. We also add computed total solar insolation explicitly as an additional input channel,  $\operatorname{I}~[W/m^2]$. We compute insolation using the CLIMLAB Python package \cite{rose2018climlab}. We use the exact orbital parameters from the corresponding CESM2 experiments to ensure that the change in insolation between piControl and midHolocene is accurate. Specifically, for the piControl run, we use an eccentricity of 0.016764, longitude of perihelion of $280.33^\circ$, and obliquity angle of $23.459^\circ$. For the midHolocene run, we use an eccentricity of 0.018682, longitude of perihelion of $180.87^\circ$, and obliquity angle of $24.105^\circ$ \cite{otto2017pmip4}. Our total input vector to the network at time $t$ is $[\boldsymbol{\Phi}_t,\boldsymbol{\tau}_t,\operatorname{I}_t]$ and our output after a time step of one month is $\boldsymbol{\Phi}_{t+\Delta t}$. 

We train emulators independently for each dataset, piControl, and midHolocene, taking the first 3500 samples to be the training data. These 3500 samples correspond to the time window between January year 601 and August year 892 for the piControl experiment and between January year 301 and August year 592 for the midHolocene experiment. We then take the next 200 samples as validation data during training. The available CMIP CESM2 data are stored as monthly means, which we use as the time step for our emulators. For the tests examining climatology or variance changes, we use years 1100 to 1200 for the piControl experiment and years 595-695 for the midHolocene experiment. For the midHolocene we allow for slight overlap between the validation and test sets due to the limited amount of data; however, all responses and comparisons are computed over the final 50 years of the test set and exclude any overlap with the validation data.

\subsection{Architecture}
\label{sec:architecture}
For forward emulation, this work uses the ConvNEXT architecture from Samudra \cite{dheeshjith2025samudra}, a UNet that replaces convolutional blocks with ConvNEXT blocks, following the approach in \citeA{liu2022convnet}. We use average pooling for the downsampling block and bilinear interpolation for the upsampling block. ConvNEXT and UNets, more broadly, are architectures that scale well to large parameter counts and large receptive fields. The computational cost is managed by having each sequential layer operate on a coarser, horizontally aggregated latent representation of the dataset. At its minimum, the horizontal shape of the latent space is reduced to 11 $\times$ 22 from 180 $\times$ 360 at the input resolution. Residual connections sum the inputs to a coarsening block and the outputs from an upsampling block. These connections help preserve details from the finer horizontal resolutions. The ConvNEXT blocks themselves modify a fully convolutional block by replacing the second standard convolutional layer with two pointwise convolution operators, the first of which has a channel width equal to twice the block's final number of output channels. We chose the channel widths for the ConvNEXT blocks to be 250, 350, 450, and 600, which represent slight changes from  \citeA{dheeshjith2025samudra} following more recent development and yield a network with approximately 85 million trainable parameters. Rather than passing in two time steps simultaneously as done for Samudra, we use a single previous state to predict a single future state. For ease of training, we also replaced the batch normalization layers in the ConvNEXT blocks with dynamic tanh layers \cite{zhu2025transformers}, which serve as a normalization and yield comparable performance. 

\subsection{Training Procedure}
To train the emulator, we use a mean squared error (MSE) loss, scaled by estimates of each variable's contribution to the loss. This loss is aggregated over multiple recurrent time steps to better constrain the emulator's behavior during long rollouts. 

For a set of predicted states, $\boldsymbol{\tilde{\Phi}}$, and true states, $\boldsymbol{\Phi}$, we define the loss as a weighted, pointwise MSE summed over rollout step, $t$, channel, $c$, and spatial dimensions, $x$ and $y$, as follows:

\begin{equation}    
    \mathcal{L}(\boldsymbol{\Phi},\boldsymbol{\tilde{\Phi}}) = \sum_{t=1}^T{\sum_{c=1}^{C}W(i,c,t)\sum_{x,y}\lambda_A(x,y)\left(\boldsymbol{\Phi}_t(c,x,y)-\boldsymbol{\tilde{\Phi}}_t(c,x,y)\right)^2}.
    \label{eq: Loss}
\end{equation}

\noindent Here, $\lambda_A(x,y) = \sqrt{A(x,y)/\max_{x,y}{A(x,y)}}$ is a weighting based on the square root of the grid cell area, $A(x,y)$. We take the square root of the grid cell area to prevent weights from vanishing near the poles. $T$ is the number of recurrent steps. We train the emulator using a rollout schedule: 4 recurrent passes, then 8 after the first 15 epochs of training.  $C=38$ is the number of output channels, and $W(i,c,t)$ is a per-channel, per-time-step weight that updates after each batch, $i$, which we call a dynamic weighting. The dynamic weighting aims to balance the gradient contributions from each channel, ensuring that fast-evolving surface states and slowly evolving deep-ocean variables contribute equally to weight updates. A wide range of predefined ad hoc scalings have been used throughout the literature, with choices such as channel standard deviation \cite{watt2025ace2,lam2023learning}, variance of the time step residuals \cite{kochkov2024neural}, or estimates of MSE from a model earlier in development \cite{bi2023accurate}. In our testing, similar approaches either were not strong enough to reduce the imprinting of fast timescales on slower dynamics, an issue in prior works \cite{subel2024building,dheeshjith2024transfer}, or led to poor convergence. Instead, we construct our weighting so that, over the course of training, we upweight the channels and rollout steps with lower MSEs. The idea is that in slowly evolving channels, features are more predictable and the signal we want to extract would negligibly contribute to loss as compared to faster evolving channels. This measure of MSE is imperfect, but it serves our goal to push the model towards learning small signals that otherwise might be overshadowed by noise. The dynamic weights are updated between batches as follows:

\begin{equation}
    W(i,c,t) =  \frac{1}{N_{smooth}} \left((N_{smooth}-1)W(i-1,c,t) + \left(\frac{1}{N_{xy}}\sum_{x,y}\left(\boldsymbol{\Phi}_t(c,x,y)-\boldsymbol{\tilde{\Phi}}_t(c,x,y)\right)^{2}\right)^{-1/2}\right).
\end{equation}

Here, $N_{smooth}$ denotes a smoothing period that prevents the weights from fluctuating excessively; it is set to $N_{smooth} = 100$. The mean is taken over the batch and over the $N_{xy}$ unmasked ocean points, so that each new weight is the reciprocal of the root-mean-square error for that channel and rollout step. Land points are masked and excluded from the mean, squared differences that are identically zero are floored at $10^{-8}$ before inverting, and the resulting weights are clipped so that the largest is at most $500$ times the smallest. We set the weighting value for the initial batch to be $W(0,c,t) = 1$ for all channels. We note that in development, the inclusion of this loss weighting had a larger impact on the bias over the validation dataset and limited impacts on the actual response metrics themselves. 

\subsection{Experimental Design}

\subsubsection{piControl Forcing Experiment}
We select the CESM2 piControl as the source of training data for the base emulator because the fixed levels of carbon dioxide, orbital parameters, and atmospheric aerosols establish a stable climate with stationary statistics. Training on a control climate ensures that the emulator learns the ocean's internal dynamics and natural variability independently of forced trends. Consequently, an emulator trained exclusively on the piControl dataset serves as a stationary baseline for evaluating out-of-sample generalization under altered boundary conditions.

\subsubsection{midHolocene Forcing Experiment}
\label{sec:Holocene_Experiments}
We select the CESM2 midHolocene experiment to provide boundary forcings for out-of-sample testing of emulators trained on CESM2 piControl data. Under the Paleoclimate Modeling Intercomparison Project 4 \cite{kageyama2018pmip4}, the midHolocene experiment primarily explores the effects of changes in Earth's orbit around the Sun. There is a slight reduction in the global CO$_2$ concentration from $284.3~\mathrm{ppm}$ in piControl to $264.4~\mathrm{ppm}$ in the midHolocene; however, the net cooling from this reduction is small, especially compared to the current value of CO$_2$, which is measured at the Mauna Loa Baseline Observatory at around $420~\mathrm{ppm}$ \cite{NOAA_GML_MLO_CO2}.

\begin{figure}[htbp]
    \includegraphics[width=\linewidth]{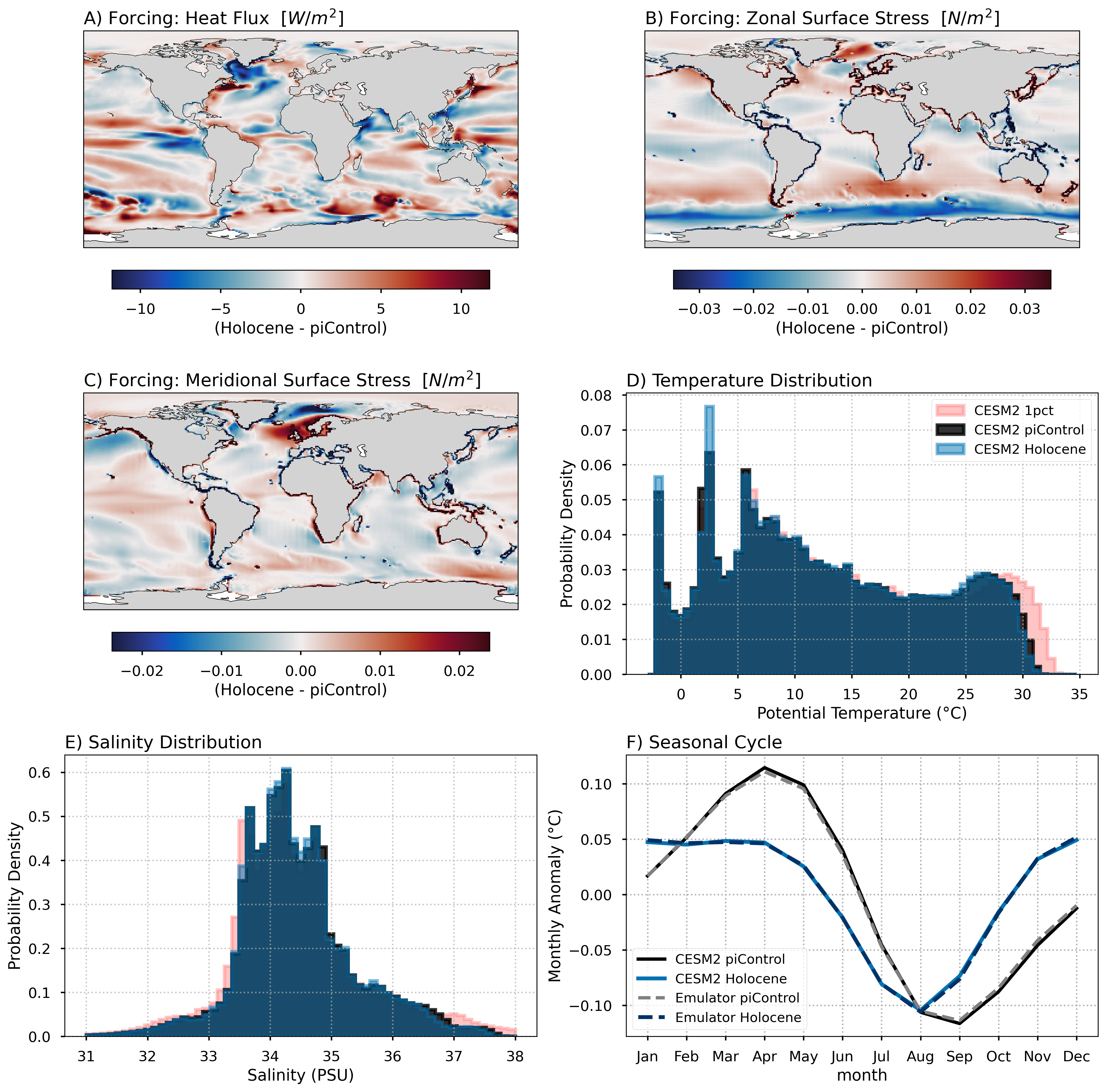}
    \caption{Differences between the piControl and midHolocene CESM2 data. Panels A-C show the difference between midHolocene and piControl in the time-mean boundary forcing components for net heat flux, $\operatorname{hfds}$ (A),  surface zonal stress, $\tau_u$ (B), and surface meridional stress, $\tau_v$ (C), respectively. The spatiotemporal standard deviation of each component in the piControl experiment is $91.17 ~ [W/m^2]$, $0.085 ~[N/m^2]$, and $0.065 ~[N/m^2]$  for $\operatorname{hfds}$, $\tau_u$, and $\tau_v$  respectively. In the midHolocene experiment, the standard deviations are similar,  $93.36 ~ [W/m^2]$, $0.092 ~[N/m^2]$, and $0.0465 ~[N/m^2]$ for $\operatorname{hfds}$, $\tau_u$, and $\tau_v$, respectively. D and E: distribution of temperature and salinity values in the upper 1000m of the ocean for the piControl (black), midHolocene (blue), and 1\% CO$_2$ forcing (red) CESM2 experiments. The 1\% CO$_2$ data are used to provide a reference of the differences in a typical out-of-distribution test case. The data for the 1\% CO$_2$ forcing experiment are taken from years 50-90, which correspond to the period surrounding the doubling. F: seasonal cycle over the upper 200m computed as the average monthly deviation from the annual climatology for CESM2 experiments (solid lines), and for the emulators trained and tested on each dataset, $\mathcal{F}_\mathrm{pi}$ and $\mathcal{F}_\mathrm{mH}$ (dashed lines).  }
    \label{fig:Holo_Pi_Comp}
\end{figure}

The midHolocene experiment exhibits a substantial shift in the surface forcing patterns compared to piControl (Figure \ref{fig:Holo_Pi_Comp}, Panels A-C), but remains within the distribution sampled by piControl for the key model variables, unlike commonly used out-of-distribution warming experiments (Figure \ref{fig:Holo_Pi_Comp}, Panels D-E). We include a reference distribution from a 1\% CO$_2$ warming run using the same CESM2 model, selecting the 40-year period centered on doubling (year 70). While we do find larger differences in the structure of the distributions when comparing more granular regions (Figures S1 and S2), in each region at most $0.01\%$ of values fall outside the piControl range for both potential temperature and salinity. In contrast, the 1\% CO$_2$ warming run has up to $0.7\%$ of values for both variables falling outside that range, and therefore not present in the training data. However, the midHolocene experiment presents a generalization challenge, as the seasonal cycle shows both a decrease in amplitude and a phase shift due to changes in Earth's orbital parameters during the midHolocene (Figure \ref{fig:Holo_Pi_Comp}, panel F). Further differences between the two simulations are discussed in \citeA{otto2020comparison}. 

\subsubsection{Ensemble Generation}
\label{sec:ensemble_gen}

We denote a rollout of the emulator, $\Omega$ as a function of
\begin{itemize}
    \item the neural network emulator used, $\mathcal{F}$, 
    \item the initial conditions provided, $\boldsymbol{\Phi}^{[0]}$,
    \item the boundary forcings applied, $\boldsymbol{\tau}$, and 
    \item the prescribed insolation, $I$.
\end{itemize} 

We compute temperature and salinity responses relative to the training climate when applying forcings coming from a new CESM2 target experiment. To estimate the mean response, we use $N_E = 5$ ensemble members. When forced with monthly climatological boundary conditions, the spread of the zonally averaged response across five ensemble members, measured as the area-weighted mean of the pointwise standard deviation, does not exceed $6\times10^{-6}\,^\circ$C for both $\mathcal{F}_{\mathrm{pi}}$  and $\mathcal{F}_{\mathrm{mH}}$ in any basin at the checkpoints we report. This spread is four orders of magnitude below the maximum spread in the CESM2 response across basins over ten 25-year chunks of each experiment, $4.5\times10^{-2}\,^\circ$C. This motivates applying significance testing only to the CESM2 experiments, and it is why we quote checkpoint spread rather than ensemble spread as the uncertainty on our reported values. We use initial conditions from the final 40 years of a 100-year emulator rollout conducted in the training climate to ensure that our response does not accumulate bias relative to a CESM2 state. 

We denote this 100-year emulation as: 
\begin{equation}       \Omega(\mathcal{F}_{\mathrm{pi}},\boldsymbol{\Phi}_{\mathrm{pi}}^{[0]};\boldsymbol{\tau}_{\mathrm{pi}}, \operatorname{I}_{\mathrm{pi}}), \nonumber
\end{equation} where $\boldsymbol{\Phi}_{\mathrm{pi}}^{[0]}$ and $\boldsymbol{\tau}_{\mathrm{pi}}$ are taken directly from the piControl CESM2 experiment.  When generating these ensembles, we use the boundary forcing from the monthly-mean climatology of the CESM2 experiments. We denote a climatological forcing with an overbar, $\overline{\boldsymbol{\tau}}_{\mathrm{pi}}$. In all cases, the subscript of $\boldsymbol{\tau}$
indicates which CESM2 experiment's climatology is used for the rollout.

 To compute the response of an emulator for a given initial condition, we compare the response between a perturbed ensemble, $\mathcal{E}$, of rollouts, 
 \begin{equation}
     \mathcal{E}(\mathcal{F}_{\mathrm{pi}},\boldsymbol{\Phi}_{\mathcal{F}_{\mathrm{pi}}}^{[0]};\boldsymbol{\tau}_{\mathrm{mH}}, \operatorname{I}_{\mathrm{mH}})=[{\Omega}^1(\mathcal{F}_{\mathrm{pi}},\boldsymbol{\Phi}_{\mathcal{F}_{\mathrm{pi}}}^{[0]};\boldsymbol{\tau}_{\mathrm{mH}}, \operatorname{I}_{\mathrm{mH}}),\allowbreak ..., {\Omega}^{N_E}(\mathcal{F}_{\mathrm{pi}},\boldsymbol{\Phi}_{\mathcal{F}_{\mathrm{pi}}}^{[0]};\boldsymbol{\tau}_{\mathrm{mH}}, \operatorname{I}_{\mathrm{mH}})], \nonumber 
 \end{equation}  
 and the corresponding unperturbed ensemble of rollouts, 
 \begin{equation}
     \mathcal{E}(\mathcal{F}_{\mathrm{pi}},\boldsymbol{\Phi}_{\mathcal{F}_{\mathrm{pi}}}^{[0]};\boldsymbol{\tau}_{\mathrm{pi}}, \operatorname{I}_{\mathrm{pi}})=[{\Omega}^1(\mathcal{F}_{\mathrm{pi}},\boldsymbol{\Phi}_{\mathcal{F}_{\mathrm{pi}}}^{[0]};\boldsymbol{\tau}_{\mathrm{pi}}, \operatorname{I}_{\mathrm{pi}}),\allowbreak..., {\Omega}^{N_E}(\mathcal{F}_{\mathrm{pi}},\boldsymbol{\Phi}_{\mathcal{F}_{\mathrm{pi}}}^{[0]};\boldsymbol{\tau}_{\mathrm{pi}}, \operatorname{I}_{\mathrm{pi}})], \nonumber
 \end{equation}
 using the same set of initial conditions.  We then average the response, $R(t,x,y,z)$, over all ensemble members as:
\begin{align}
    R(t,x,y,z) &=  \mathcal{E}(\mathcal{F}_{\mathrm{pi}},\boldsymbol{\Phi}_{\mathcal{F}_{\mathrm{pi}}}^{[0]};\boldsymbol{\tau}_{\mathrm{mH}}, \operatorname{I}_{\mathrm{mH}}) -\mathcal{E}(\mathcal{F}_{\mathrm{pi}},\boldsymbol{\Phi}_{\mathcal{F}_{\mathrm{pi}}}^{[0]};\boldsymbol{\tau}_{\mathrm{pi}}, \operatorname{I}_{\mathrm{pi}})\\ 
    &=\frac{1}{N_E}\sum_{n=1}^{N_E} \left({\Omega}^n\left(\mathcal{F}_{\mathrm{pi}},\boldsymbol{\Phi}_{\mathcal{F}_{\mathrm{pi}}}^{[0]};\boldsymbol{\tau}_{\mathrm{mH}}, \operatorname{I}_{\mathrm{mH}}\right)-{\Omega}^n\left(\mathcal{F}_{\mathrm{pi}},\boldsymbol{\Phi}_{\mathcal{F}_{\mathrm{pi}}}^{[0]};\boldsymbol{\tau}_{\mathrm{pi}}, \operatorname{I}_{\mathrm{pi}}\right)\right). \nonumber
\end{align}

\subsubsection{Baselines}
\label{sec:baselines}

Because the boundary forcings, $\operatorname{hfds}$, $\tau_u$, and $\tau_v$, are prescribed from the parent CESM2 experiment as a function of both the surface atmosphere and ocean state,  part of the response can in principle be inferred directly from the forcing without any representation of internal dynamics. To estimate this, we compare the emulator against two baselines that map the boundary forcings directly to the ocean state. The first is a local linear regression operator (LRO), which predicts the state at each grid column from the surface forcing at that column, with no horizontal communication, no internal dynamics, and no memory. The second is a forcing-only network, which keeps the emulator's architecture, nonlinearity, and global receptive field, but takes only the boundary forcings as input, again with no state and no memory. Both are fit on the same piControl data used to train $\mathcal{F}_{\mathrm{pi}}$ and applied to the midHolocene forcings, and both are described in more detail in Section S2. Skill that the LRO reproduces reflects a direct local imprint of the forcing; skill that the forcing-only network reproduces but the LRO does not requires a nonlinear or non-local mapping from the forcing, but no internal dynamics; and skill that neither reproduces requires the evolving ocean state. We do not include any null baselines such as persistence or climatology as all results are a difference between climates, for which each baseline corresponds to a response of zero everywhere by construction.

\section{Evaluating Out-of-Sample Generalization}
We show that the AI emulators in this work, trained on either the piControl or midHolocene experiment, can reproduce changes to climatology, variability, and mean states when driven by out-of-sample boundary forcings. As described in Section \ref{sec:Holocene_Experiments}, rolling out a piControl emulator in a midHolocene climate requires both generalizing to new orbital forcings and capturing large-scale adjustments in the spatial patterns of the ocean structure on century timescales. These changes represent the cumulative response to shifts in surface forcing patterns.

We follow the procedure outlined in Section \ref{sec:ensemble_gen} to perturb the different forcing components after training an emulator on the piControl dataset alone. Capturing the long-term trends in deep-ocean processes remains an unsolved challenge, and we do not propose a solution in this work \cite{dheeshjith2025samudra}. As such, we focus on the upper ocean response above 1000m. In the following sections we report metrics from a single representative checkpoint per emulator. We use an early checkpoint for the piControl emulators, taken before the late-training skill degradation, and a later checkpoint for the midHolocene emulator, which shows no such degradation. The full epoch and seed dependence is given in Section \ref{sec:epoch_sensitivity}. There, we show that these metrics are most robust over the regions of highest skill and more variable elsewhere; we therefore restrict our conclusions to emphasizing broad, consistent contrasts between the two emulators. Comparing faster upper-ocean processes allows us to estimate the system's response by using the difference in ocean states between the CESM2 midHolocene and piControl experiments, as we expect the system to have equilibrated on the century timescales we investigate.

\subsection{Reproducing a New Seasonality}
\label{sec:orbital_forcing}
We first examine the emulator's ability to adjust to the new orbital forcing of the midHolocene simulation, the primary generalization challenge between the two experiments in this work. Information on the seasonality of the midHolocene experiment is contained in the dynamic boundary forcings, but is most explicitly encoded in the solar insolation boundary channel (Figure S3). For this test, we compare two rollouts: one using initial conditions and boundary forcings from the piControl climate, $\Omega(\mathcal{F},\boldsymbol{\Phi}_{\mathrm{pi}}^{[0]};\boldsymbol{\tau}_{\mathrm{pi}},\operatorname{I}_{\mathrm{pi}})$, and the other from midHolocene climate, $\Omega(\mathcal{F},\boldsymbol{\Phi}_{\mathrm{mH}}^{[0]};\boldsymbol{\tau}_{\mathrm{mH}},\operatorname{I}_{\mathrm{mH}})$. We generate this set of rollouts both for the emulators trained on piControl, $\mathcal{F}_{\mathrm{pi}}$, and midHolocene, $\mathcal{F}_{\mathrm{mH}}$ data.

\begin{figure}
    \centering
    \includegraphics[width=.75\linewidth]{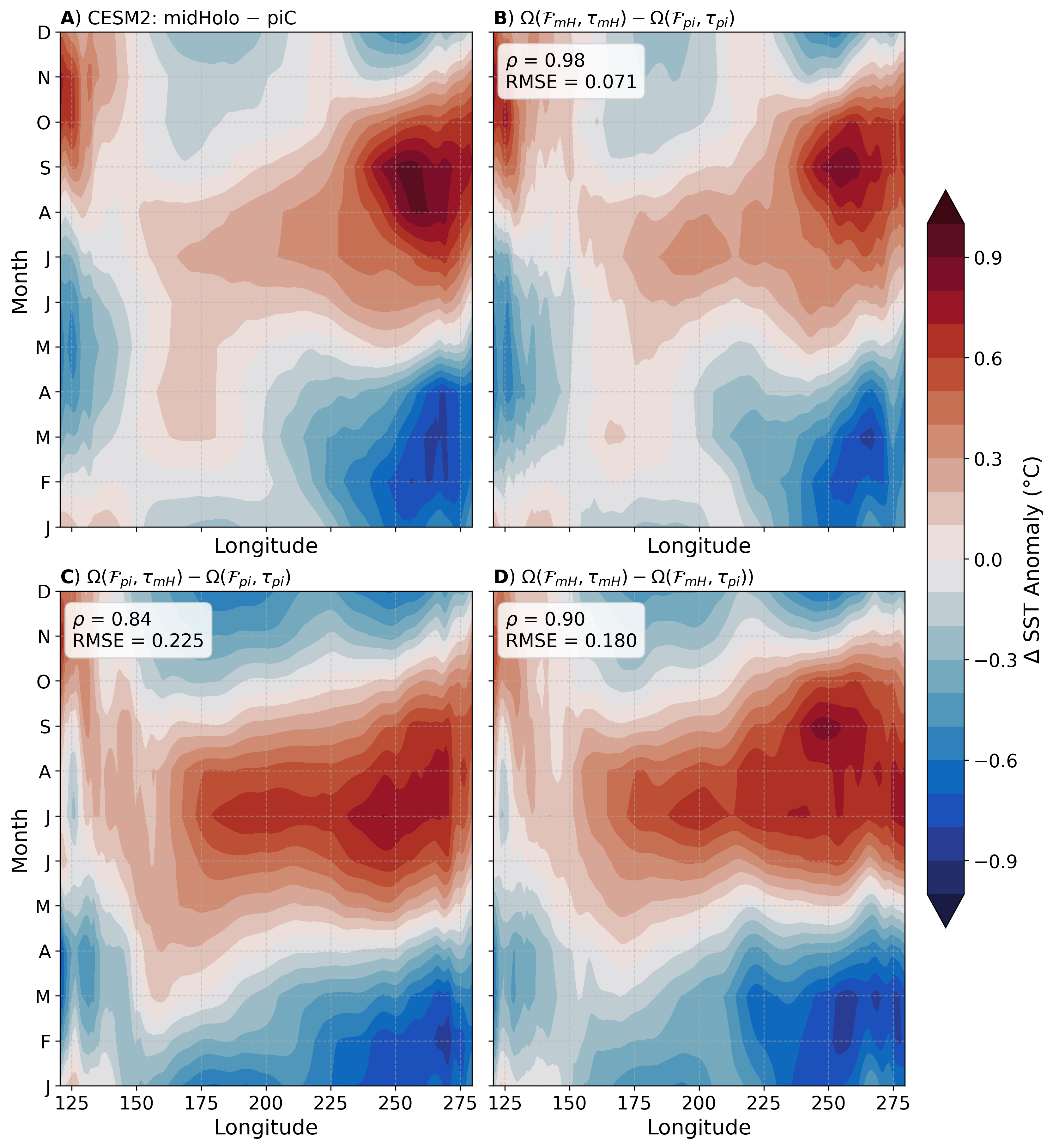}
    \caption{Comparison of the tropical Pacific seasonal SST anomaly changes between the midHolocene and piControl.  We define SST as the uppermost layer (0-5m) and compute anomalies as the difference between the monthly and annual climatologies. We then compute the meridional average of these anomalies between $1.5^\circ \mathrm{S}-1.5^\circ \mathrm{N}$. A: difference between the midHolocene and piControl CESM2 numerical experiments.  For B-D, we use the title to indicate the networks and rollout parameters, using abbreviated notation defined in Section \ref{sec:ensemble_gen}. For each rollout, we use initial conditions and insolation from the same experiment listed for the boundary forcings (e.g., where we state $\boldsymbol{\tau}_{\mathrm{pi}}$, we also use $\boldsymbol{\Phi}_{\mathrm{pi}}^{[0]}$ and $\operatorname{I}_{\mathrm{pi}}$).  B: difference between $\mathcal{F}_{\mathrm{mH}}$ rolled out for the midHolocene climate and $\mathcal{F}_{\mathrm{pi}}$ rolled out for the piControl climate. C: difference between $\mathcal{F}_{\mathrm{pi}}$ rolled out for the midHolocene and piControl climate. D: same difference as C, but for $\mathcal{F}_{\mathrm{mH}}$. We evaluate the climatology over the final 50 years of each 100-year rollout. For each panel, we compute the correlation, $\rho$, and RMSE with respect to the CESM2 difference in A.} 
    \label{fig:Pacific_SST}
\end{figure}

We first focus on the Tropical Pacific, following \citeA{otto2020comparison}, who identified significant differences in the CESM2 climatologies between the piControl and midHolocene experiments. In the Tropical Pacific, changes in the potential temperature climatology relative to piControl are of similar amplitude to the seasonal cycle itself. These anomalies align with the difference in insolation and represent both a shift in phase and a reduction in total magnitude, as shown globally in Figure \ref{fig:Holo_Pi_Comp} panel F. In Figure \ref{fig:Pacific_SST}, we reproduce the results shown by \citeA{otto2020comparison} in their Figure S4, and compare the monthly mean anomalies of sea surface temperature (SST) relative to the annual mean. We take the upper ocean layer (0-5m) as SST and average over the same latitudinal extent: $1.5^\circ \mathrm{S} - 1.5^\circ \mathrm{N}$. To compute monthly anomalies, we use the last 50 years of a 100-year rollout for the emulators and the last 300 years for each numerical experiment.

Both emulators ($\mathcal{F}_{\mathrm{pi}}$ and $\mathcal{F}_{\mathrm{mH}}$) capture the change in seasonality in the tropical Pacific SSTs between the piControl and midHolocene climates (Figure \ref{fig:Pacific_SST} A) as shown by comparing the rollouts of each emulator under forcings from the two climates (Figure \ref{fig:Pacific_SST} C-D). Under the midHolocene forcings, the two emulators produce warmer anomalies in July, August, and September (JAS) and cooler anomalies in December, January, and February (DJF) than under the piControl forcings. The changes in anomaly strength are in line with insolation changes between the midHolocene and piControl climates. This is reflected in the correlations to the true change of 0.84 for $\mathcal{F}_{\mathrm{pi}}$ and 0.90 for $\mathcal{F}_{\mathrm{mH}}$. The emulators produce smoother responses than the ground truth: the largest amplitude increases are spread over a wider latitude band than in the CESM2 response, which shows the greatest changes east of 200$^\circ$E. We rule out training-climate bias as a source of error in the responses, showing that the difference between $\Omega(\mathcal{F}_\mathrm{mH},\boldsymbol{\Phi}_{\mathrm{mH}}^{[0]};\boldsymbol{\tau}_{\mathrm{mH}},\operatorname{I}_{\mathrm{mH}})$ and $\Omega(\mathcal{F}_{\mathrm{pi}},\boldsymbol{\Phi}_{\mathrm{pi}}^{[0]};\boldsymbol{\tau}_{\mathrm{pi}},\operatorname{I}_{\mathrm{pi}})$  strongly agrees with the truth, with a correlation of 0.98 and a low RMSE of $0.07^\circ \mathrm{C}$ with respect to the CESM2 experiments (Figure \ref{fig:Pacific_SST}, panel B).  

\begin{figure}
    \centering
    \includegraphics[width=\linewidth]{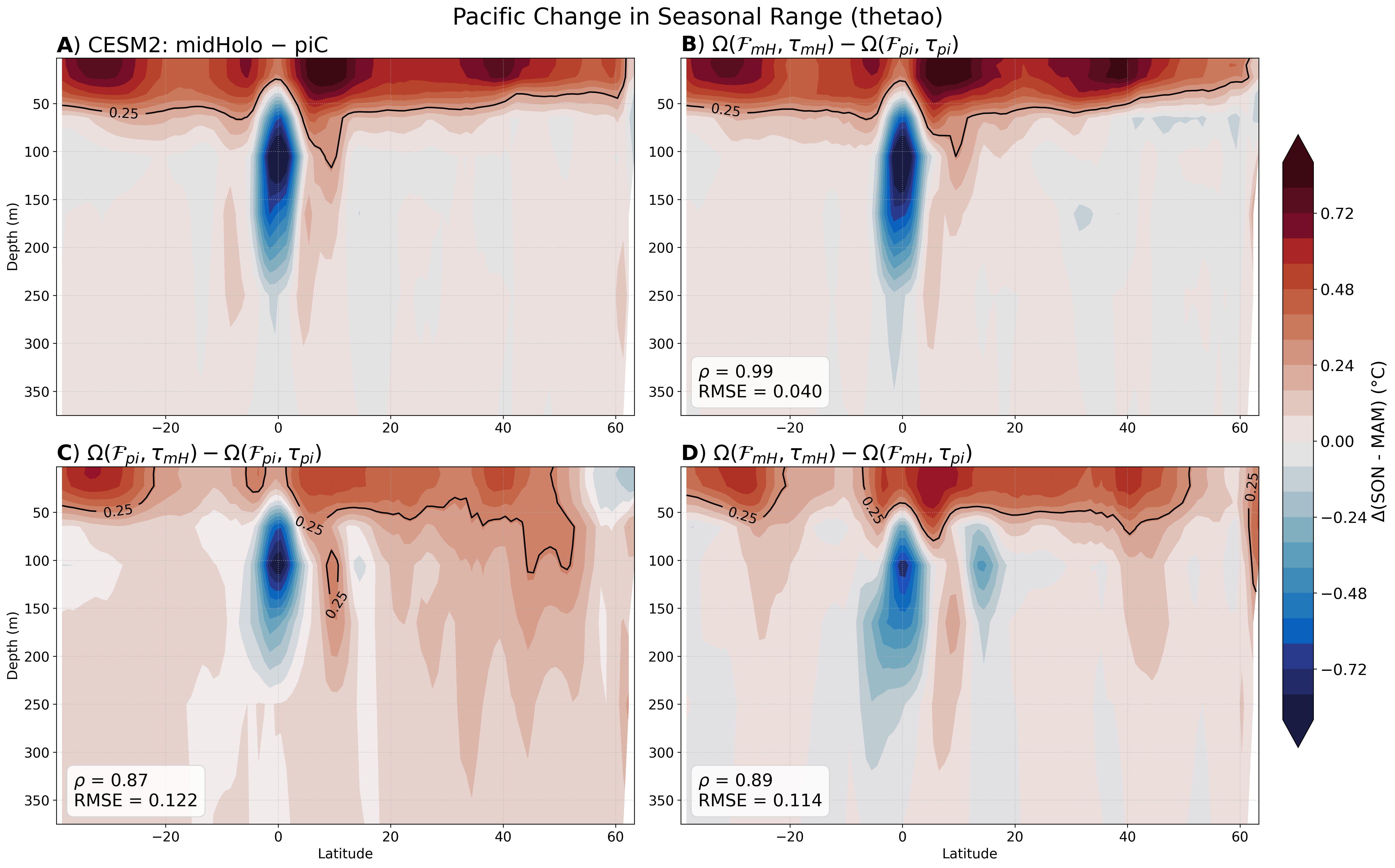}
    \caption{Depth profiles detailing the change in the seasonal range of potential temperature (SON minus MAM) between the midHolocene and piControl climates (Equation \ref{eq:pacific_sst_anoms}) in the Pacific Ocean. A: change in range between the two CESM2 experiments. For B-D: we use the title to indicate the networks and rollout parameters using  abbreviated forms of the  notation defined in Section \ref{sec:ensemble_gen}. For each rollout, we use initial conditions and insolation from the same experiment listed for the boundary forcings (e.g., where we state $\boldsymbol{\tau}_{\mathrm{pi}}$, we also use $\boldsymbol{\Phi}_{\mathrm{pi}}^{[0]}$ and $\operatorname{I}_{\mathrm{pi}}$).   B: difference between the climatology of $\mathcal{F}_{\mathrm{mH}}$ rolled out for the midHolocene climate and  $\mathcal{F}_{\mathrm{pi}}$ rolled out for the piControl climate. C: difference between $\mathcal{F}_{\mathrm{pi}}$ rolled out for the midHolocene and the piControl climate. D:  same difference  for $\mathcal{F}_{\mathrm{mH}}$. We evaluate the climatology over the final 50 years of each 100-year rollout. For each panel, we compute the correlation, $\rho$, and RMSE with respect to the CESM2 difference in A.}
    \label{fig:seasonal_cycle}
\end{figure}

To extend the analysis below the surface, we compare the change in the seasonal range of potential temperature over the upper 500m of the Pacific Ocean between boreal spring (MAM) and fall (SON) in Figure \ref{fig:seasonal_cycle}. Skill in this metric ensures that the response in Figure \ref{fig:Pacific_SST} comes from changes across the upper ocean rather than a superficial linking of the surface forcings and surface state. Using the seasonal climatology of potential temperature over the last 50 years of a 100-year rollout, we compute the change in the range of the seasonal anomalies as follows: 
\begin{multline}
    \Delta\left(\mathrm{SON} - \mathrm{MAM} \right) = \left(\Omega(\cdots;\boldsymbol{\tau}_{\mathrm{mH}},\operatorname{I}_{\mathrm{mH}})_{\theta_O} ^{\mathrm{SON}} - \Omega(\cdots;\boldsymbol{\tau}_{\mathrm{mH}},\operatorname{I}_{\mathrm{mH}})_{\theta_O}^{\mathrm{MAM}}\right) - \\\left(\Omega(\cdots;\boldsymbol{\tau}_{\mathrm{pi}},\operatorname{I}_{\mathrm{pi}})_{\theta_O} ^{\mathrm{SON}} - \Omega(\cdots;\boldsymbol{\tau}_{\mathrm{pi}},\operatorname{I}_{\mathrm{pi}})_{\theta_O}^{\mathrm{MAM}}\right).
    \label{eq:pacific_sst_anoms}
\end{multline}
Here we denote the season over which we compute the climatology with a superscript and we use the rollout notation to highlight the differences we compare. We include a reference that uses data from the respective CESM2 experiments (Figure  \ref{fig:seasonal_cycle}, panel A) and compare the same difference between these experiments using a set of rollouts from each emulator (Figure  \ref{fig:seasonal_cycle}, Panels C-D).

As above, both emulators capture the broad vertical structure of the change in the seasonal cycle when generalizing to a climate with unseen orbital parameters, with correlations of $0.87$ and $0.89$ for $\mathcal{F}_{\mathrm{pi}}$ and $\mathcal{F}_{\mathrm{mH}}$ respectively (Figure \ref{fig:seasonal_cycle}, Panels C and D). To quantify how well the emulator reproduces the magnitude of the response, we use the amplitude ratio, the area-weighted ratio of the standard deviations of the emulator and true response ($\sigma_{\text{emu}}/\sigma_{\text{truth}}$). Here $\sigma_{\text{emu}}$ is the spatial standard deviation of the emulator response and $\sigma_{\text{truth}}$ is the standard deviation of the true response. This ratio is one when amplitudes match and values below one indicate an underestimation of the response magnitude. Because it reflects the full spatial variance, the ratio can fall below one even where peak values are well reproduced. While the pattern is well reproduced, both emulators underrepresent the magnitude of the change and smooth out subsurface features, recovering amplitude ratios of $0.66$ and $0.72$. This magnitude deficit is specific to the cross-climate generalization rather than a baseline bias: on their own training climates the emulators reproduce the seasonal range almost exactly, with a correlation of $0.99$ and an amplitude ratio of $1.03$ for the difference shown in panel B. To help with comparing panels, we include a contour where $\Delta\left(\mathrm{SON} - \mathrm{MAM} \right)=0.25^\circ \mathrm{C}$. The depth of the 0.25$^\circ$C amplitude increase is similar across all panels over most latitudes, but in Panels C and D the contour outcrops at the surface and extends too deep at high latitudes away from the equator. Further, we find that this seasonal response is not purely driven by boundary forcings as both the LRO and forcing only baseline fail to capture this response with low correlations of 0.13 and 0.27 respectively (Figure S4). 

Returning to the difference between $\Omega(\mathcal{F}_{\mathrm{mH}}, \Phi^{[0]}_{\mathrm{mH}}; \tau_{\mathrm{mH}}, \operatorname{I}_{\mathrm{mH}})$ and $\Omega(\mathcal{F}_{\mathrm{pi}}, \Phi^{[0]}_{\mathrm{pi}}; \tau_{\mathrm{pi}}, \operatorname{I}_{\mathrm{pi}})$ (Figure \ref{fig:seasonal_cycle}, panel B), we do not see the diffuse features or reduced amplitude found in the generalization tests (panels C and D), but again find some spurious weak cooling in the northern latitudes. This additional test, as above, verifies the emulators' ability to produce unbiased rollouts on their training climates. Both the surface and subsurface measures of the seasonal cycle show that the emulators can capture the broad structure and phase changes to seasonality over the upper ocean when driven by forcings from a climate with out-of-sample orbital parameters.  We include results for the Atlantic in Figure S5, which show similar strengths and limitations to those in the Pacific.

\subsection{Capturing Changes to Variability}
\label{sec:variability}
\begin{figure}
    \centering
    \includegraphics[width=1\linewidth]{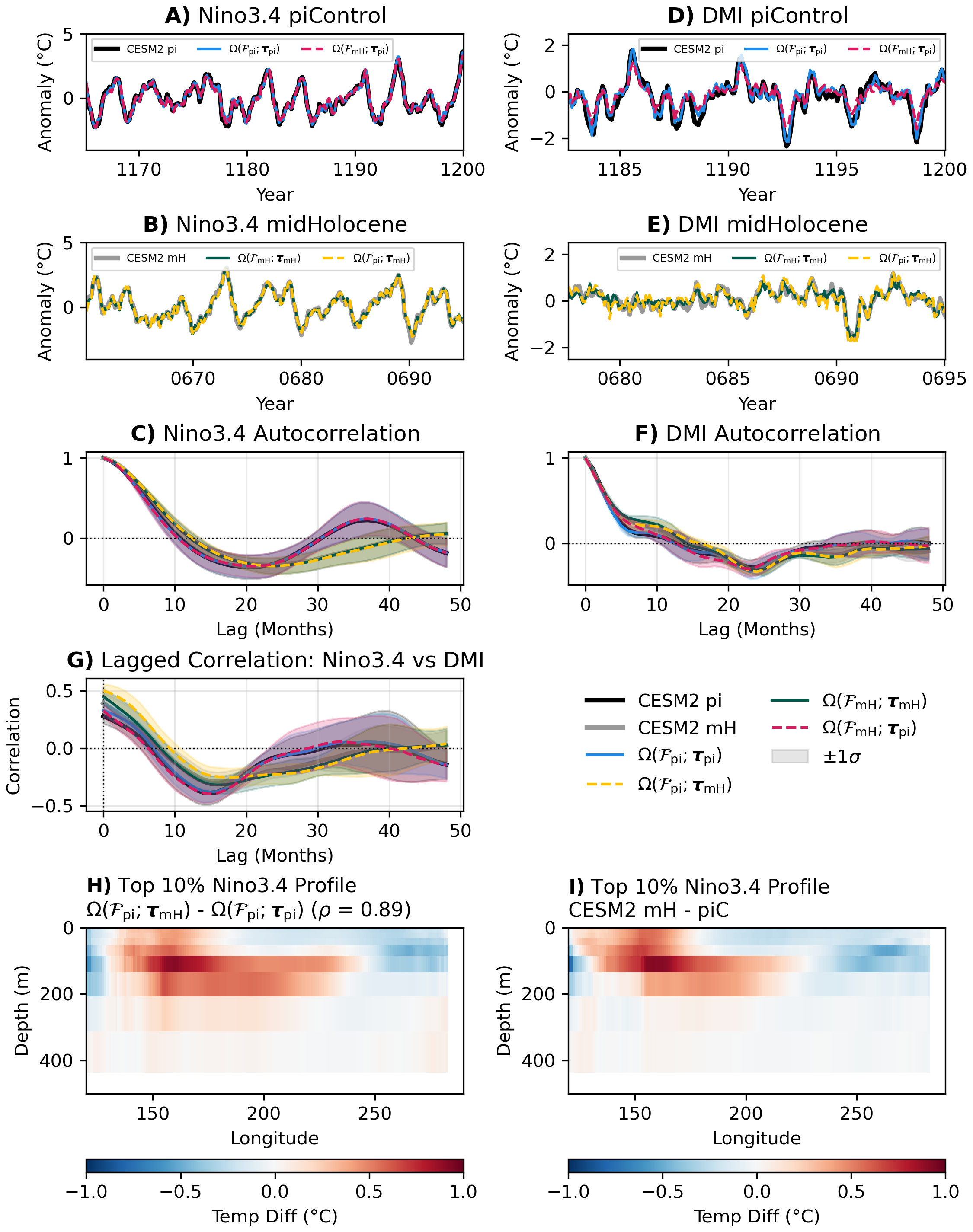}
    \caption{Comparison of dominant modes of ocean variability across the piControl and midHolocene experiments. A-C: the timeseries of Ni\~{n}o3.4 in the piControl climate (A), the midHolocene climate (B), and the lagged autocorrelation (C). D-F: similar results for the DMI, a measure of the Indian Ocean Dipole. G: the lagged correlation between Ni\~{n}o3.4 and the DMI. H and I: upper ocean structural changes between the two experiments, highlighting the difference in anomaly patterns between the midHolocene and piControl experiments for $\mathcal{F}_{\mathrm{pi}}$ and the CESM2 data in H and I, respectively. We include the spatial correlation between H and I in the title of H.}
    \label{fig:Variability}
\end{figure}

Having shown that the emulator can capture direct changes in seasonality, we examine its ability to capture changes in the system's variability. To do this, we first examine the temporal autocorrelation of the El Ni\~{n}o Southern Oscillation (ENSO) and the Indian Ocean Dipole (IOD)  as well as the cross-correlation between the two (Figure \ref{fig:Variability}). ENSO represents the largest mode of climate variability, and capturing its behavior is key for many climate feedbacks \cite{mcphaden2006enso}. We find that in the midHolocene experiment, this mode becomes less regular with a broader peak in the temporal power spectrum (Figure S8), while the IOD does not show significant changes. We include the IOD as a second major signal of variability, because it governs conditions in the Indian Basin and represents an independent set of dynamics from those governing ENSO \cite{saji1999dipole}.  

To quantify these climate modes, we use Ni\~{n}o3.4 for ENSO and the Dipole Mode Index (DMI) for the IOD. We chose these indices for their fidelity in representing their respective modes, but also because they are known to decouple most of the spurious correlation between modes \cite{saji1999dipole}. Ni\~{n}o3.4 is given as the SST anomaly relative to the monthly climatology averaged over the region between $170-120^\circ\mathrm {W}$ and $5^\circ \mathrm{S}-5^ \circ \mathrm{N}$.  The DMI is computed as the difference between the area-averaged SST anomalies in the western ($50-70^\circ \mathrm{E}$ and $10^\circ \mathrm{S}-10^\circ \mathrm{N}$)  and eastern ($90-110^\circ \mathrm{E}$ and $10-0^\circ \mathrm{S}$) Indian Ocean. 

\begin{figure}
    \centering
    \includegraphics[width=\linewidth]{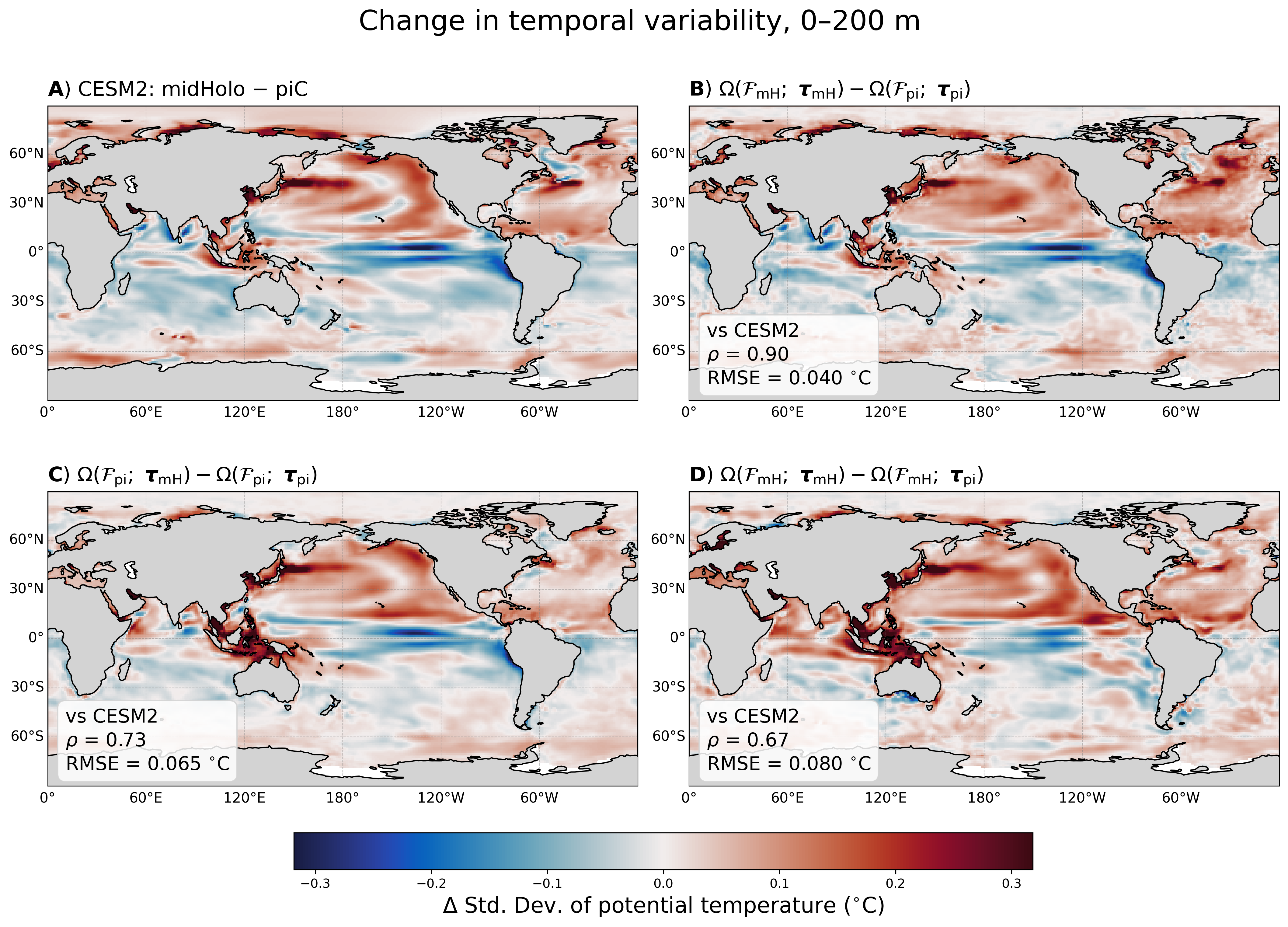}
    \caption{Comparison of the change in the potential temperature variability over the upper 200m. A: true difference between the two CESM2 experiments. For B-D, we use the title to indicate the networks and rollout parameters, using abbreviated notation defined in Section \ref{sec:ensemble_gen}. For each rollout, we use initial conditions and insolation from the same experiment listed for the boundary forcings (e.g., where we state $\boldsymbol{\tau}_{\mathrm{pi}}$, we also use $\boldsymbol{\Phi}_{\mathrm{pi}}^{[0]}$ and $\operatorname{I}_{\mathrm{pi}}$).  B: difference between variability when $\mathcal{F}_{\mathrm{mH}}$ is rolled out for the midHolocene climate and $\mathcal{F}_{\mathrm{pi}}$ is rolled out for the piControl climate. C: difference for $\mathcal{F}_{\mathrm{pi}}$ rolled out for the midHolocene and piControl climate. D: the same difference  for $\mathcal{F}_{\mathrm{mH}}$. We evaluate the temporal variability over the final 50 years of each 100-year rollout. For each panel, we compute the correlation, $\rho$, and RMSE with respect to the CESM2 difference in A.}
    \label{fig:spatial_variability}
\end{figure}

Figure \ref{fig:Variability} shows the timeseries of Ni\~{n}o3.4 (panels A and B) and the DMI (panels D and E) from CESM2, $\mathcal{F}_{\mathrm{pi}}$, and $\mathcal{F}_{\mathrm{mH}}$. Panels A and D show rollouts,  $ \Omega(\mathcal{F}_\mathrm{\cdot},\boldsymbol{\Phi}_{\mathrm{pi}}^{[0]};\boldsymbol{\tau}_{\mathrm{pi}},\operatorname{I}_{\mathrm{pi}})$,  and Panels B and E show rollouts,  $\Omega(\mathcal{F}_\mathrm{\cdot},\boldsymbol{\Phi}_{\mathrm{mH}}^{[0]};\boldsymbol{\tau}_{\mathrm{mH}},\operatorname{I}_{\mathrm{mH}})$.  We use these rollouts rather than the climatological-forcing ensembles of Section \ref{sec:ensemble_gen} because climatological boundary forcings damp out variability relative to the CESM2 references. Both the Ni\~{n}o3.4 and DMI timeseries computed from the emulator rollouts strongly agree with the timeseries from CESM2, with correlations all above 0.98 for Ni\~{n}o3.4 and above 0.91 for the DMI. While the Ni\~{n}o3.4 index is well constrained locally by the surface forcings and is reproduced at 0.90 by a local linear operator fit to piControl (Section S2), the same operator cannot capture the DMI under midHolocene forcing, with a correlation of 0.23, or the subsurface structure of strong El  Ni\~{n}o events, with a correlation of $-0.03$. A forcing-only network, which retains the emulator's architecture but no internal dynamics (Section S2), reproduces both, so agreement on these indices is a check that the emulator remains well represented out of sample rather than evidence of learned dynamics. We look to the spatial maps in Figure \ref{fig:spatial_variability} discussed below to further separate the emulator and the baselines. Given the agreement between the emulators and CESM2 for the timeseries of the two indices, both $\mathcal{F}_{\mathrm{pi}}$ and $\mathcal{F}_{\mathrm{mH}}$ capture the autocorrelation of each index in both the in- and out-of-sample climates (Figure \ref{fig:Variability}, Panels C and F). The emulators additionally capture the lagged correlation between indices, though the spread between subsets of the data remains large (Figure \ref{fig:Variability}, panel G). 

We next examine the imprint of ENSO over the upper 500m of the Tropical Pacific, below the surface at which the boundary forcings are applied. For Panels H and I of Figure \ref{fig:Variability}, we take the ocean anomalies for the 10\%  of states with the largest Ni\~{n}o3.4 index and average them to get a mean profile of Ni\~{n}o events. We then compute the difference in these profiles between the midHolocene and piControl experiments: $\Omega(\mathcal{F}_\mathrm{pi},\boldsymbol{\Phi}_{\mathrm{mH}}^{[0]};\boldsymbol{\tau}_{\mathrm{mH}},\operatorname{I}_{\mathrm{mH}})-\Omega(\mathcal{F}_\mathrm{pi},\boldsymbol{\Phi}_{\mathrm{pi}}^{[0]};\boldsymbol{\tau}_{\mathrm{pi}},\operatorname{I}_{\mathrm{pi}})$. For brevity, we include only results from CESM2 and  $\mathcal{F}_\mathrm{pi}$ in Figure \ref{fig:Variability} but the same plot for $\mathcal{F}_\mathrm{mH}$, along with subsurface profiles for the IOD region, is included in Figure S10 and the same comparison against both baselines in Figure S11. As we expect to see a strong zonal structure of the anomaly, we compute the meridional average between $5^\circ \mathrm{S}-5^\circ \mathrm{N}$. We find that the emulator skill reflects an accurate subsurface ENSO structure, showing a correlation of 0.89 with the CESM2 difference across the largest outlier events, though the forcing-only network reaches 0.88 for the same difference. The difference is not perfect, with $\mathcal{F}_{\mathrm{pi}}$ exhibiting a slight eastward bias as compared to CESM2, with the weighted center of positive differences shifting from $161^\circ \mathrm{E}$ to $172^\circ \mathrm{E}$. This spatial shift is accompanied by a minor reduction in the amplitude of negative differences over the eastern Pacific ($230-280^\circ \mathrm{E}$), which decreases from $-0.21^\circ \mathrm{C}$ in CESM2 to $-0.19^\circ \mathrm{C}$ in $\mathcal{F}_{\mathrm{pi}}$. 

To assess decadal variability, we examined the emulator's ability to reproduce the Atlantic Multidecadal Oscillation (AMO, which corresponds to the difference between the North Atlantic SST anomaly and the global temperature anomaly), Figure S12. The emulators are both able to capture the same timescales and overall structure of the index, with correlations to the true timeseries above 0.78 for all emulator rollouts; however, the magnitude of the anomalies is uniformly reduced as compared to CESM2. The weaker representation of this Atlantic mode is consistent with findings in Sections \ref{sec:Overall_Response} and \ref{sec:epoch_sensitivity}, showing that behavior in the North Atlantic remains more epoch- and model-dependent than features in the tropical Pacific, which represent more dominant signals for the emulator to learn.

We also examine maps of the spatial structure of variability changes over the upper 200m in Figure \ref{fig:spatial_variability}, to go beyond the bulk indices and better understand local changes. We measure this as the difference between the standard deviation of potential temperature from rollouts in a midHolocene and piControl climate. In the tropical Pacific and Indian Ocean, where we find the ENSO and the IOD, the spatial patterns of variability change for $\mathcal{F}_{\mathrm{pi}}$ (Figure \ref{fig:spatial_variability}, panel C) and $\mathcal{F}_{\mathrm{mH}}$ (Figure \ref{fig:spatial_variability}, panel D) reflect changes seen in the CESM2 experiments (Figure \ref{fig:spatial_variability}, panel A). Both $\mathcal{F}_{\mathrm{pi}}$ and $\mathcal{F}_{\mathrm{mH}}$ capture the broad increase in Northern Hemisphere and decrease in Southern Hemisphere variability, with global correlations of 0.73 and 0.67 for $\mathcal{F}_{\mathrm{pi}}$ and $\mathcal{F}_{\mathrm{mH}}$. Neither the LRO nor the forcing-only network capture this change with low correlations of 0.15 and 0.29 respectively (Figure S14). The correlation is strongest for both emulators in the Pacific, at 0.81 and 0.71, and weakest in the Indian Ocean, at 0.56 and 0.41 for $\mathcal{F}_{\mathrm{pi}}$ and $\mathcal{F}_{\mathrm{mH}}$ respectively. We find the largest differences between emulators occur in the Southern Hemisphere, where the correlation falls from 0.72 for $\mathcal{F}_{\mathrm{pi}}$ to 0.47 for $\mathcal{F}_{\mathrm{mH}}$ in the Southern Ocean and from 0.76 to 0.25 in the South Atlantic. Some of this disagreement may come from the random seed rather than from differences in the data, so we do not claim this pattern is purely a function of the different training data. Further, we find that while some skill drops off over 200-1000m, both emulators still capture the large-scale variability changes here with correlations of 0.45 and 0.46 for $\mathcal{F}_{\mathrm{pi}}$ and $\mathcal{F}_{\mathrm{mH}}$ respectively (Figure S15), while both baselines degrade further (Figure S16). In the deeper portion of the upper ocean, biases of the emulators on their own climate may limit the response skill as the correlation between the difference of the two emulators evaluated on their own climate (Panel B) drops to 0.60 as compared to the value of 0.90 over the upper 200m. The ability to reproduce variability changes demonstrates that the emulators capture key dynamical features across climates, while also highlighting outstanding issues at high latitudes and noisier patterns of change across the globe. 

\subsection{Forced Response between piControl and midHolocene Climates}
\label{sec:Overall_Response}
\begin{figure}[htbp]
    \makebox[\textwidth][c]{\includegraphics[width=7.5in]{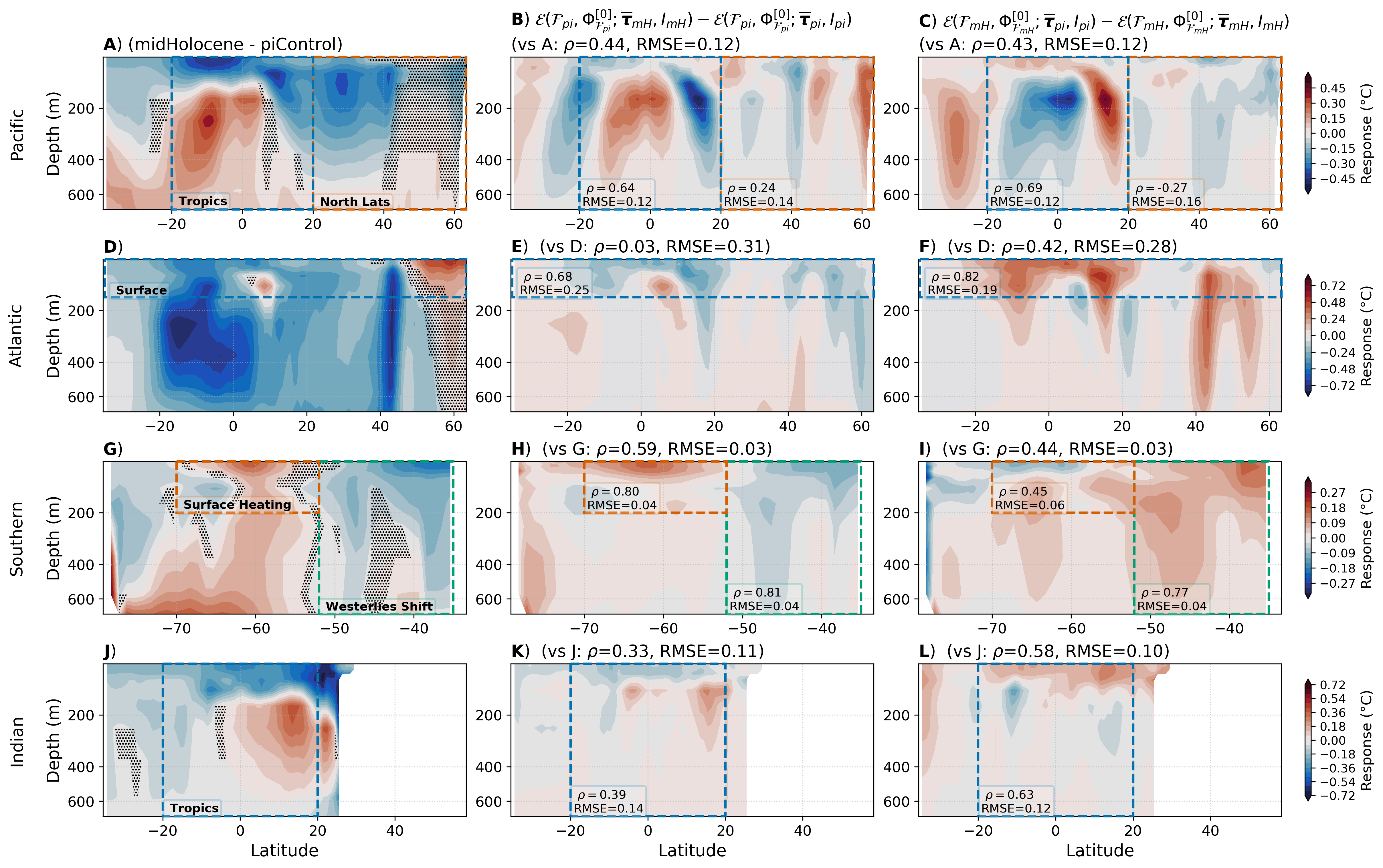}}
    \caption{Comparison of the zonally averaged potential temperature response in each ocean basin when perturbing an emulator using climatological forcings taken from midHolocene/piControl CESM2 data. A, D, G, and J: difference between the CESM2 midHolocene and piControl experiments as an estimate of the true response. B, E, H, and K: response of the emulator trained on piControl data when perturbed with midHolocene boundary forcings, $R(\mathcal{F}_{\mathrm{pi}})$. C, F, I, and L: response of the emulator trained on midHolocene data when perturbed with piControl boundary forcings, $R(\mathcal{F}_{\mathrm{mH}})$. We show responses in the Pacific (A-C), Atlantic (D-F), Southern (G-I), and Indian Ocean (J-L). For each emulator response, we include the correlation, $\rho$,  and RMSE with respect to the true change in the respective basin, A,D,G, and J. Note that for $R(\mathcal{F}_{\mathrm{mH}})$, we flip the sign of the true response before computing the correlation and RMSE. We additionally highlight specific regions in each basin, and again include the correlation and RMSE over each region with respect to the change between CESM2 experiments. We stipple the CESM2 panels to mark regions where the reference response is not significant relative to internal variability (two-tailed Student's $t$-test at the 95\% level, with internal variability estimated from ten 25-year chunks of each experiment).}
    \label{fig:Full_Responses_All_Basins}
\end{figure}

\begin{figure}[htbp]
    \makebox[\textwidth][c]{\includegraphics[width=7.5in]{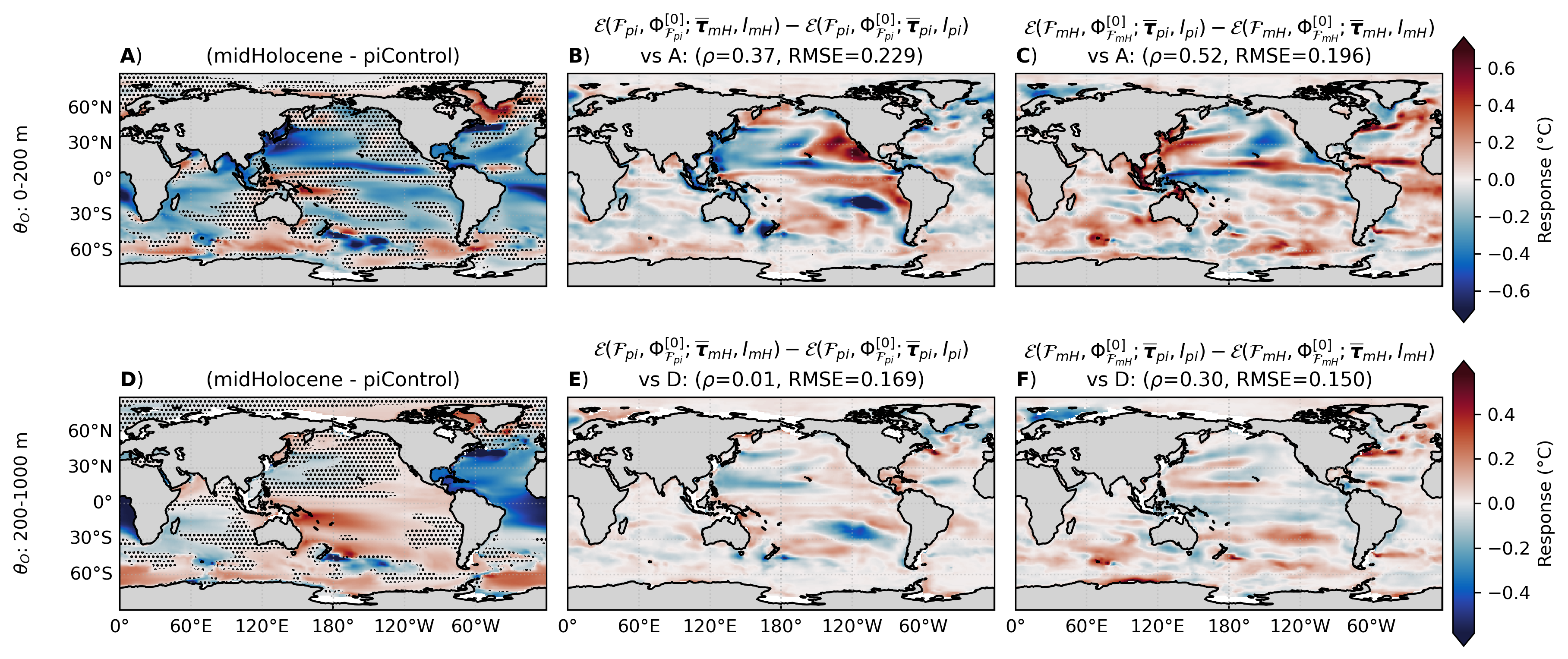}}
    \caption{Comparison of the depth-averaged potential temperature response over the upper 200m (A-C) and 200-1000m (D-F) when perturbing an emulator with climatological forcings from the midHolocene/piControl CESM2 experiments. A and D: difference between the CESM2 midHolocene and piControl experiments as an estimate of the true response. B and E: responses of the emulator trained on piControl data when perturbed with midHolocene boundary forcings, $R(\mathcal{F}_{\mathrm{pi}})$. C and F: responses of the emulator trained on midHolocene data when perturbed with piControl boundary forcings, $R(\mathcal{F}_{\mathrm{mH}})$.  For each emulator response, we include the correlation, $\rho$, and RMSE with respect to the difference between CESM2 experiments in A and D. Note that for $R(\mathcal{F}_{\mathrm{mH}})$, we flip the sign of the true response before computing the correlation and RMSE. We stipple the CESM2 panels to mark regions where the reference response is not significant relative to internal variability (two-tailed Student's $t$-test at the 95\% level, with internal variability estimated from ten 25-year chunks of each experiment).}
    \label{fig:Full_Responses_Map}
\end{figure}

The ocean emulator reproduces coherent regional responses to new boundary forcings, shown in the zonally averaged depth profiles in Figure \ref{fig:Full_Responses_All_Basins} and the map of the upper ocean responses in Figure \ref{fig:Full_Responses_Map}. For these experiments, we compare the final 25-year response between an ensemble of rollouts from the base piControl climate and an ensemble using boundary forcings from the midHolocene, each run for 150 years, $ R(\mathcal{F}_{\mathrm{pi}}) = \mathcal{E}(\mathcal{F}_{\mathrm{pi}},\boldsymbol{\Phi}_{\mathcal{F}_\mathrm{pi}}^{[0]};\overline{\boldsymbol{\tau}}_{\mathrm{mH}},\operatorname{I}_{\mathrm{mH}}) - \mathcal{E}(\mathcal{F}_{\mathrm{pi}},\boldsymbol{\Phi}_{\mathcal{F}_\mathrm{pi}}^{[0]};\overline{\boldsymbol{\tau}}_{\mathrm{pi}},\operatorname{I}_{\mathrm{pi}})$. For the emulator trained on midHolocene data, $\mathcal{F}_{\mathrm{mH}}$, we run the same experiment using initial conditions from a midHolocene rollout, $\boldsymbol{\Phi}_{\mathcal{F}_\mathrm{mH}}^{[0]}$ and piControl boundary forcings: $  R(\mathcal{F}_{\mathrm{mH}}) =\mathcal{E}(\mathcal{F}_{\mathrm{mH}},\boldsymbol{\Phi}_{\mathcal{F}_\mathrm{mH}}^{[0]};\overline{\boldsymbol{\tau}}_{\mathrm{pi}},\operatorname{I}_{\mathrm{pi}}) - \mathcal{E}(\mathcal{F}_{\mathrm{mH}},\boldsymbol{\Phi}_{\mathcal{F}_\mathrm{mH}}^{[0]};\overline{\boldsymbol{\tau}}_{\mathrm{mH}},\operatorname{I}_{\mathrm{mH}})$. Between the two climates, the largest changes are driven either directly by insolation or by its latitudinal redistribution \cite{otto2020comparison}. In CESM2, these changes manifest as a northward shift in westerlies in the Southern Ocean, which leads to a slowdown of transport through the Drake Passage and a slight increase in transport closer to the Pacific subtropical gyre. In the Atlantic, changes in insolation strengthen the westerlies over the subtropical gyres and modestly increase Atlantic Meridional Overturning Circulation (AMOC) strength. While we do not directly represent circulation changes in our emulators for this task, changes in mixing and transport appear in the resulting responses of thermodynamic variables. To keep descriptions clear and consistent when describing the sign of the response, we describe changes from the perspective of $R(\mathcal{F}_{\mathrm{pi}})$. The two emulators are perturbed relative to opposite reference climates, and so we define, for convenience, a sign convention, $R(\mathcal{F}_{\mathrm{pi}}) \leftrightarrow -R(\mathcal{F}_{\mathrm{mH}})$. While we cannot verify the exact reversibility or linearity of the applied forcing without conducting additional numerical experiments, we expect much of the large-scale upper ocean response to be mirrored. As such, comparisons between $R(\mathcal{F}_{\mathrm{pi}})$ and $-R(\mathcal{F}_{\mathrm{mH}})$ are made under this convention and are not designed to establish the validity of this assumption for the numerical model, but are the best comparisons available without access to additional numerical experiments. Values in parentheses beside each metric give its range across training checkpoints prior to the late-training decline described in Section \ref{sec:epoch_sensitivity}. We exclude one checkpoint for $\mathcal{F}_{\mathrm{mH}}$, which becomes unstable in the Northern Atlantic leading to nonphysical response magnitudes.

Both emulators capture the correct sign of the response over much of the Pacific, primarily at lower latitudes near the tropics (Figure \ref{fig:Full_Responses_All_Basins}, Panels A-C). Between $20^\circ \mathrm{S}-20^\circ \mathrm{N}$ and across the upper 1000m, both $ R(\mathcal{F}_{\mathrm{pi}}) $ and $ R(\mathcal{F}_{\mathrm{mH}}) $ show strong correlations to the  CESM2 response, with values of 0.64 (0.64-0.83) and 0.69 (0.64-0.79), respectively, both above the respective basin-wide correlations of 0.44 (0.44-0.65) and 0.43 (0.38-0.67). We again look at the amplitude ratios to quantify the strength of the response. Both emulators reproduce roughly two-thirds of the true response amplitude over the Pacific Ocean, with amplitude ratios of $0.62$ ($0.39$-$0.63$) for $R(\mathcal{F}_{\mathrm{pi}})$ and $0.65$ ($0.35$-$0.82$) for $R(\mathcal{F}_{\mathrm{mH}})$. The damping is smaller in the tropics, where the amplitude ratios rise to $0.73$ for both emulators ($0.49$-$0.73$ and $0.43$-$0.85$) and the emulator peak responses fall even closer to that of CESM2. Against a true range of $-0.43$ to $0.38\,^\circ$C, $R(\mathcal{F}_{\mathrm{pi}})$ spans $-0.51$ to $0.32\,^\circ$C and $R(\mathcal{F}_{\mathrm{mH}})$ spans $-0.47$ to $0.47\,^\circ$C, so the peak magnitudes are reproduced more faithfully than the amplitude ratio alone would suggest. In the depth profile, the response is composed of a near-surface cooling that extends deeper into the ocean around $15^\circ \mathrm{N}$ and a subsurface warming that is strongest south of the equator. The strong mean response aligns with the emulator's ability to reproduce changes in climate anomalies over that region (Figure \ref{fig:seasonal_cycle}) and the vertical structure of ENSO anomalies (Figure \ref{fig:Variability}, panel H).  The response skill drops off at higher northern latitudes, as, while the emulators capture the broad cooling, both emulators erroneously introduce warming just north of $20^\circ \mathrm{N}$. This misplaced warming leads, in part, to low correlation values of 0.24 and -0.27 over the region between $20-65^\circ \mathrm{N}$ and the upper 1000m for $ R(\mathcal{F}_{\mathrm{pi}}) $ and $ R(\mathcal{F}_{\mathrm{mH}}) $, respectively.  We hypothesize that the incorrect warming feature is tied to an outstanding failure to capture long-term, basin-wide cooling. As seen in the true response, this cooling should have compensated for the local heating seen in both emulator responses. The anomalous warming coincides with a positive heat flux change near the Kuroshio current (Figure \ref{fig:Holo_Pi_Comp},  panel A), which helps explain why the emulators produce a signal of the opposite sign. Additionally, we find the learned response to be mostly independent of which experiment is used to train the emulator, with a correlation of 0.76 between $ R(\mathcal{F}_{\mathrm{pi}}) $ and $ -R(\mathcal{F}_{\mathrm{mH}}) $ over the entire Pacific and an even higher correlation of 0.88 in the tropics. This corresponds to both emulators producing similarly correct tropical features and struggling at higher latitudes. In the supplement, we show the same results for salinity, finding a slightly weaker level of agreement in the Pacific (Figure S17). 

In the Atlantic, emulators struggle to capture the extensive basin-wide cooling as well as the depth and magnitude of the strong cold plume at around $45^\circ \mathrm{N}$ (Figure \ref{fig:Full_Responses_All_Basins}, Panels D-F). The response at high latitudes shows the strongest inter-epoch variance, discussed further in Section \ref{sec:epoch_sensitivity}. Across the upper 150m, the emulators capture the broad cooling with correlations to the truth of 0.68 (0.65-0.74) and 0.82 (0.65-0.86) for $ R(\mathcal{F}_{\mathrm{pi}}) $ and $ R(\mathcal{F}_{\mathrm{mH}}) $, respectively. Given the large spread across checkpoints, we do not consider the difference between the two emulators to be significant. Even in this region the magnitude is somewhat reduced, with amplitude ratios of $0.60$ ($0.50$-$0.80$) and $0.80$ ($0.36$-$0.84$) for $R(\mathcal{F}_{\mathrm{pi}})$ and $R(\mathcal{F}_{\mathrm{mH}})$, respectively. We do find that the learned structures in this region are again similar across the two emulators, with a correlation between $ R(\mathcal{F}_{\mathrm{pi}}) $ and $ -R(\mathcal{F}_{\mathrm{mH}}) $ of 0.88. Additionally, the below-surface warming at around $10^\circ \mathrm{N}$ is correctly located, but extends too deep into the ocean. These subsurface errors are likely due to the emulators again failing to accumulate background cooling, which limits the depth of the warming in the true response and is a large source of error in this basin. These errors in the response are also consistent with the emulator's dampened representation of AMO variability (Section \ref{sec:variability}), pointing to a broader deficit in the representation of internal Atlantic dynamics. 

In the Southern Ocean, the emulator captures wind-driven cooling down to 1000m between $52-35^\circ \mathrm{S}$ and heat-flux driven warming around $60^\circ \mathrm{S}$. The limitations again stem from a failure to accumulate changes in the deeper ocean, as seen in the disconnect between surface warming and the lower depths. We explore additional details of the Southern Ocean response in Section \ref{sec:component_forcings}. In the Indian Ocean, the emulator broadly reproduces the response, but the features exhibit little local structure and the magnitude is the most strongly reduced of any basin, with a basin-wide amplitude ratio of 0.33 ($0.25$-$0.42$). Here the surface peak is also strongly reduced, with the maximum emulator cooling of 0.29$^\circ$C well below that of CESM2 at 0.69$^\circ$C. We note that across all basins, the weaker amplitude and lack of accumulated trends is not an artifact of a shorter rollout length compared against an equilibrated model difference. We find that both the pattern correlation and the amplitude ratio converge after roughly the first 75 years in every basin (Table S1).

Figure \ref{fig:Full_Responses_Map} shows the depth-averaged response over the upper 200m (panels A-C) and 200-1000m (panels D-F). As discussed above, we find that $ R(\mathcal{F}_{\mathrm{pi}}) $ and $ R(\mathcal{F}_{\mathrm{mH}}) $ correlate with the true response more strongly near the surface (0.37 for piControl and 0.52 for midHolocene) and struggle to capture any structure in the deeper ocean (0.01 for piControl and 0.30 for midHolocene). As with the depth profiles, the emulator captures some of the sign and structure of the response, albeit to a lesser degree, but fails to account for the broad cooling over the globe. In both emulators, the warming extending from the Kuroshio connects to a small warming signal in the western Pacific, which lacks balancing from accumulated cooling in the emulator response. As shown in prior studies, the accumulation of small trends over long rollouts remains an unresolved challenge with ocean emulators  \cite{dheeshjith2025samudra}. 

Having mapped the response, we next ask what drives the remaining errors. One potential source of these disagreements is the omission of freshwater flux ($\operatorname{wfo}$) as a boundary forcing, a design choice adopted from previous work \cite{dheeshjith2025samudra}. To investigate whether this missing driver is responsible, we provide additional results in Section S1 from an emulator trained with $\operatorname{wfo}$. Comparing the same zonally averaged potential temperature response to that in Figure \ref{fig:Full_Responses_All_Basins}, the emulator trained with wfo performs modestly better in the Pacific, with the correlation improving from 0.44 to 0.65 and the RMSE dropping from 0.12 to 0.10. This improvement is not consistent across basins. Training with wfo also raises the correlation in the Atlantic from 0.03 to 0.22 and in the Indian basin from 0.33 to 0.50, but lowers it in the Southern Ocean from 0.59 to 0.46, including over the high-latitude Westerlies Shift and Surface Heating regions. More importantly, in every basin the skill is insensitive to the midHolocene freshwater signal itself. Holding wfo at the piControl climatology during the perturbation rather than applying the full midHolocene wfo changes the correlation and RMSE negligibly (Table S2). Taken together, adding the freshwater boundary condition neither reliably improves the high-latitude response nor responds to the midHolocene freshwater forcing, indicating that the remaining discrepancies are not explained by the missing surface boundary condition.

To contextualize these correlations, we consider how much of the emulator's response can be explained by a direct mapping from the surface forcing to the ocean state, rather than by learned dynamics. Comparing the emulator to the baselines from Section \ref{sec:baselines} separates the local, linear response to the forcing, the learned response to the forcing using the same architecture as the emulator, and the response driven by a representation of internal dynamics. In the tropical Pacific, the emulator outperforms the LRO with correlations of $0.64$ and $0.36$ respectively. The emulator similarly outperforms the LRO in the Atlantic surface region, with correlations of $0.68$ and $0.43$. In both regions the forcing-only network performs comparably to the emulator, with correlations of $0.77$ and $0.64$, indicating that the response is not a local imprint of the forcing but is still recoverable from the forcing without internal dynamics. By contrast, much of the Southern Ocean response is captured by the LRO, with similar correlations of $0.59$ for the emulator and $0.61$ for the LRO. This suggests that much of the response here in the upper ocean reflects a largely local, linear adjustment to the surface forcing rather than learned internal dynamics. While the emulator outperforms the baselines in some regions, the pattern correlations are similar in most basins; however, the baselines' skill is concentrated near the surface, and it is in the amplitude and the depth of the response that the emulator separates from them. The emulator recovers a larger fraction of the true response amplitude in both the Pacific and the Southern Ocean, with amplitude ratios of $0.62$ and $0.46$ as compared to $0.41$ and $0.27$ for the LRO and $0.38$ and $0.25$ for the forcing-only network. The contrast is sharpest over the Surface Heating region of the Southern Ocean, where the LRO matches the emulator's correlation, $0.82$ against $0.80$, while recovering less than a third of the response amplitude, $0.29$ against $0.87$. We quantify the depth of the response using the centroid depth, $z_c$. We form the response amplitude at each level, $\sigma(z)$, as the latitude-weighted root mean square of the zonally averaged response, and take its first moment with depth,
\begin{equation}
z_c \;=\; \frac{\sum_z \sigma(z)\,z\,\Delta z}{\sum_z \sigma(z)\,\Delta z}.
\label{eq:centroid}
\end{equation}
This measures where the bulk of the response is located rather than just the overall amplitude or pattern. In the Southern Ocean, the CESM2 response has a centroid depth of $442$m, which the emulator places at $339$m while the LRO and the forcing-only network place it at $259$m and $215$m. In the Pacific the same ordering holds, at $329$m for CESM2 as compared to $258$m, $196$m, and $190$m. This ordering holds in every basin and region we examine (Figure S21). Over the Westerlies Shift region of the Southern Ocean, where the LRO matches the emulator's correlation at $0.81$, its response sits $95$m too shallow while the emulator's centroid falls within $1$m of CESM2. We probe the linearity of the emulator more directly in the following sections, decomposing the forced response into its individual forcing contributions and quantifying how the emulator linearly superimposes them.

\section{Component Responses and Epoch Sensitivity}
To go beyond simply presenting the emulators' skill, we investigate how boundary forcing components drive changes in the emulator's ocean state and explore the sensitivity of the results across epochs and emulators. Addressing the former, we decompose the response by forcing the emulator with individual components of the dynamical boundary conditions.  In Section \ref{sec:component_forcings}, we find that the emulator is able to separate out the wind and heat flux responses.

To address the latter, in Section \ref{sec:epoch_sensitivity}, we explore the sensitivity of response metrics across epochs. Through this study, we show that commonly used MSE or bias-based measures of skill do not reliably predict the emulator's response skill. Across checkpoints, response skill varies substantially and  is not predicted by test RMSE. We further find that ensemble spread varies across epochs in a manner uncorrelated with response skill and inconsistent across seeds. This suggests that training and validating models for these use cases may require a broader range of evaluations than are typically employed prior to final testing. Despite sensitivity to epoch, this investigation reinforces the conclusions in Sections \ref{sec:orbital_forcing} and \ref{sec:Overall_Response}, as the results are qualitatively consistent across epochs, especially in regions where we claim the emulator has higher skill.

\subsection{Contributions from Forcing Components}
\label{sec:component_forcings}
We compare the response of the emulator trained on the full set of dynamic boundary forcings to that of an emulator trained on an individual component to better understand how emulators use boundary forcing information. For this section, we denote the emulator trained on all forcings as $\mathcal{F}_{\mathrm{pi}}^{\mathrm{All}}=\mathcal{F}_{\mathrm{pi}}$ and an emulator trained on a single dynamic component as $\mathcal{F}_{\mathrm{pi}}^{\cdot}$, where $\cdot$ corresponds to a single boundary channel. For example, the network trained only on heat flux will be denoted as $\mathcal{F}_\mathrm{pi}^{\operatorname{hfds}}$. For this example, we perturb only the heat flux and examine the response for the full emulator: 
\begin{equation}   R(\mathcal{F}_{\mathrm{pi}}^{\mathrm{All}};\overline{\operatorname{hfds}}_{\mathrm{mH}})= \allowbreak \mathcal{E}(\mathcal{F}_{\mathrm{pi}}^{\mathrm{All}},\boldsymbol{\Phi}_{\mathcal{F}_\mathrm{pi}}^{[0]};\allowbreak [\overline{\tau_u}]_{\mathrm{pi}},\allowbreak [\overline{\tau_v}]_{\mathrm{pi}},\allowbreak \overline{\operatorname{hfds}}_\mathrm{mH},\operatorname{I}_{\mathrm{pi}}) - \mathcal{E}(\mathcal{F}_{\mathrm{pi}}^{\mathrm{All}},\boldsymbol{\Phi}_{\mathcal{F}_\mathrm{pi}}^{[0]};[\overline{\tau_u}]_{\mathrm{pi}},[\overline{\tau_v}]_{\mathrm{pi}},\overline{\operatorname{hfds}}_\mathrm{pi},\operatorname{I}_{\mathrm{pi}}). \nonumber
\end{equation}
For another example, we again perturb heat flux but examine the  response for the single-component emulator: 
\begin{equation}    R(\mathcal{F}_\mathrm{pi}^{\operatorname{hfds}};\overline{\operatorname{hfds}}_{\mathrm{mH}})=\mathcal{E}(\mathcal{F}_\mathrm{pi}^{\operatorname{hfds}},\boldsymbol{\Phi}_{\mathcal{F}_\mathrm{pi}}^{[0]};\overline{\operatorname{hfds}}_\mathrm{mH},\operatorname{I}_{\mathrm{pi}}) - \mathcal{E}(\mathcal{F}_\mathrm{pi}^{\operatorname{hfds}},\boldsymbol{\Phi}_{\mathcal{F}_\mathrm{pi}}^{[0]};\overline{\operatorname{hfds}}_\mathrm{pi},\operatorname{I}_{\mathrm{pi}}). \nonumber
\end{equation}
In this section, the response of the full emulator when perturbed with all boundary forcings is written explicitly as $R(\mathcal{F}_{\mathrm{pi}}^{\mathrm{All}};\overline{\boldsymbol{\tau}}_\mathrm{mH},\operatorname{I}_\mathrm{mH})$.  Note, we write the components of surface stress in brackets to preserve subscript notation for the climate.

\begin{figure}[htbp]
    \makebox[\textwidth][c]{\includegraphics[width=7.5in]{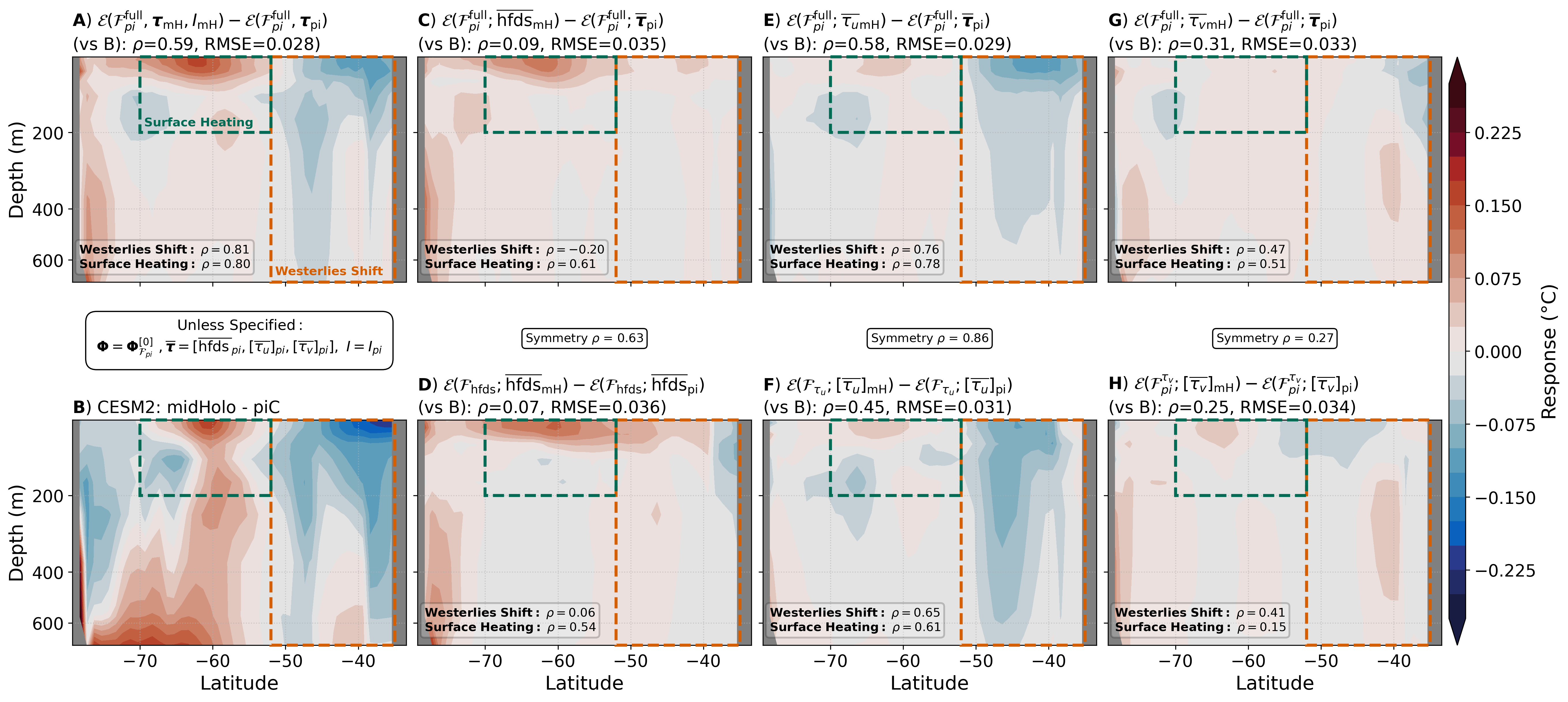}}
    \caption{Comparison of the zonally averaged potential temperature response in the  Southern Ocean when perturbing piControl trained emulators with boundary forcing components of climatological values from the midHolocene CESM2 data. The title of each panel follows the notation in Section \ref{sec:ensemble_gen}; however, for brevity, unless specified, the rollouts are initialized from a piControl rollout, $\boldsymbol{\Phi}_{\mathrm{pi}}^{[0]}$, each forcing component comes from piControl climatology, $\overline{\boldsymbol{\tau}}=[\overline{\operatorname{hfds}}_{\mathrm{pi}},[\overline{\tau_u}]_{\mathrm{pi}},[\overline{\tau_v}]_{\mathrm{pi}}]$, and the insolation is computed with piControl orbital parameters, $\operatorname{I}_{\mathrm{pi}}$. A: total response of $\mathcal{F}_{\mathrm{pi}}^{\mathrm{All}}$ when all boundary forcing components are taken from the midHolocene climatology: $R(\mathcal{F}_{\mathrm{pi}}^{\mathrm{All}};\overline{\boldsymbol{\tau}}_{\mathrm{mH}},\operatorname{I}_{\mathrm{mH}})$. B: difference between the CESM2 midHolocene and piControl runs to provide a reference. C, E, and G: response of the $\mathcal{F}_{\mathrm{pi}}^{\mathrm{All}}$ emulator when perturbing individual components of the boundary forcing, (e.g. $R(\mathcal{F}_{\mathrm{pi}}^{\mathrm{All}};\overline{\operatorname{hfds}}_{\mathrm{mH}})$).  D, F, and H: response of an emulator trained on a single forcing component to the corresponding boundary component (e.g. $R(\mathcal{F}_{\mathrm{pi}}^{\operatorname{hfds}};\overline{\operatorname{hfds}}_{\mathrm{mH}})$). We show the response for $\overline{\operatorname{hfds}}$ (C and D), $\overline{\tau_u}$ (E and F), and $\overline{\tau_v}$ (G and H). For each response, we show the correlation, $\rho$, and the RMSE relative to the ground-truth response, B. In the middle of the plot, we include the correlation between the component emulator and $\mathcal{F}_{\mathrm{pi}}^{\mathrm{All}}$ forced by the respective boundary component (e.g. $\rho(R(\mathcal{F}_{\mathrm{pi}}^{\mathrm{All}};\overline{\operatorname{hfds}}_{\mathrm{mH}}),R(\mathcal{F}_{\mathrm{pi}}^{\operatorname{hfds}};\overline{\operatorname{hfds}}_{\mathrm{mH}}))$. Within each emulator response panel, we additionally include the correlation for the two highlighted regions, again with respect to B.} 
    \label{fig:component_responses}
\end{figure}

In the physical ocean system, each dynamic forcing component drives different physical pathways, which often lead to independent response structures. For example, an increase in zonal surface stress may directly drive Sverdrup transport or induce mixing with deeper waters. For heat flux, strong cooling may cause water to sink, whereas added heat may remain near the surface, without transport by winds or internal circulation to move the warmed water to greater depths. We also find, when conducting the experiments described above, that the full emulator, $\mathcal{F}_{\mathrm{pi}}^{\mathrm{All}}$, and single component emulators, $\mathcal{F}_{\mathrm{pi}}^{\cdot}$,  also generate independent responses for each forcing component;  however, the magnitude is often larger for the single component networks. We use the Southern Ocean to highlight this in Figure \ref{fig:component_responses} but include plots for all basins in Figures S24-S26. As discussed in Section \ref{sec:Overall_Response}, the emulators reproduce key features of the Southern Ocean response. Here, we focus on the upper ocean warming near $60^\circ\mathrm{S}$ driven by heat flux and the cold plume between $52^\circ-35^\circ \mathrm{S}$ that arises from the northward shift of the westerlies.

When perturbing only $\operatorname{hfds}$ as described earlier in the section, both $R(\mathcal{F}_{\mathrm{pi}}^{\mathrm{All}};\overline{\operatorname{hfds}}_\mathrm{mH})$ and $R(\mathcal{F}_\mathrm{pi}^{\operatorname{hfds}};\overline{\operatorname{hfds}}_\mathrm{mH})$ show near-surface warming across most latitude bands (Figure \ref{fig:component_responses} Panels C-D). This signal is strongest at $60^\circ \mathrm{S}$ and matches the corresponding signal in the total response and true estimates (Figure \ref{fig:component_responses}
 Panels A and B).  We quantify this response by computing the correlation and the maximum response value over the upper $ 200 \, \mathrm {m} $ and between $ 70^\circ$ and $52^\circ\, \mathrm{S}$, thereby isolating the warming signal.  Here, $R(\mathcal{F}_{\mathrm{pi}}^{\mathrm{All}};\overline{\boldsymbol{\tau}}_\mathrm{mH},\operatorname{I}_\mathrm{mH})$,  $R(\mathcal{F}_{\mathrm{pi}}^{\mathrm{All}};\overline{\operatorname{hfds}}_\mathrm{mH})$, and $R(\mathcal{F}_\mathrm{pi}^{\operatorname{hfds}};\overline{\operatorname{hfds}}_\mathrm{mH})$ all show a strong correlation to the difference between CESM2 experiments, with correlation values of 0.80, 0.61, and 0.54, respectively. While other component responses may show similar correlations, we find response magnitudes that are far weaker than the truth, with maximum values all below $0.05^\circ \mathrm{C}$ compared to the true maximum of $0.16^\circ \mathrm{C}$. In contrast, $R(\mathcal{F}_{\mathrm{pi}}^{\mathrm{All}};\overline{\boldsymbol{\tau}}_\mathrm{mH},\operatorname{I}_\mathrm{mH})$,  $R(\mathcal{F}_{\mathrm{pi}}^{\mathrm{All}};\overline{\operatorname{hfds}}_\mathrm{mH})$, and $R(\mathcal{F}_\mathrm{pi}^{\operatorname{hfds}};\overline{\operatorname{hfds}}_\mathrm{mH})$ all reproduce more accurate values of $0.16^\circ \mathrm{C}$, $0.12^\circ \mathrm{C}$, and $0.13^\circ \mathrm{C}$, respectively.

The largest signal in the Southern Ocean response comes from the northward shift in westerlies seen in  $\tau_u$. This introduces near-surface cooling, which we see reflected in the full response of $R(\mathcal{F}_{\mathrm{pi}}^{\mathrm{All}};\overline{\boldsymbol{\tau}}_\mathrm{mH},\operatorname{I}_\mathrm{mH})$  (Figure \ref{fig:component_responses} panel A) as well as the component responses $R(\mathcal{F}_{\mathrm{pi}}^{\mathrm{All}};[\overline{\tau_u}]_\mathrm{mH})$ and $R(\mathcal{F}_\mathrm{pi}^{\tau_u};[\overline{\tau_u}]_\mathrm{mH})$ (Figure \ref{fig:component_responses} Panels E and F, respectively). For the zonal surface stress responses, we focus on the region above $1000m$ and between $52-35^\circ \mathrm{S}$. All responses to a perturbed zonal surface stress show a relatively high correlation of 0.81, 0.76, and 0.65 for $R(\mathcal{F}_{\mathrm{pi}}^{\mathrm{All}};\overline{\boldsymbol{\tau}}_\mathrm{mH},\operatorname{I}_\mathrm{mH})$, $R(\mathcal{F}_{\mathrm{pi}}^{\mathrm{All}};[\overline{\tau_u}]_\mathrm{mH})$, and $R(\mathcal{F}_\mathrm{pi}^{\tau_u};[\overline{\tau_u}]_\mathrm{mH})$, respectively. Both heat flux responses, $R(\mathcal{F}_{\mathrm{pi}}^{\mathrm{All}};\overline{\operatorname{hfds}}_\mathrm{mH})$ and $R(\mathcal{F}_\mathrm{pi}^{\operatorname{hfds}};\overline{\operatorname{hfds}}_\mathrm{mH})$, show a poor correlation with the truth, showing that the emulator is unable to fill information gaps using other forcing components. While $R(\mathcal{F}_\mathrm{pi}^{\tau_v};[\overline{\tau_v}]_\mathrm{mH})$ (Figure \ref{fig:component_responses} panel H) shows a higher correlation with the truth, the strength of the response, $-0.049^\circ \mathrm{C}$, is too weak compared to the value from CESM2, $-0.21^\circ \mathrm{C}$. In the Southern Ocean the full and single-component $\tau_v$ responses also agree less closely than for the other forcings ($\rho = 0.27$), consistent with $\tau_v$ being a weaker and less spatially coherent driver in this basin whose response is less consistent across emulators.

Looking at other basins (Figures S24-S26), we see a similar pattern reemerge. We find that the component responses of $\mathcal{F}_{\mathrm{pi}}^{\mathrm{All}}$ correlate with those of the individually trained emulators in active forcing regions, mirroring the results in the Southern Ocean. However, in the Pacific basin, the single-component emulators are better able to learn the relationship between forcing components and show stronger responses in the tropics as compared to component responses of $\mathcal{F}_{\mathrm{pi}}^{\mathrm{All}}$. In the other basins shown in the supplement, responses to  $\operatorname{hfds}$ contribute to the total response in a similar manner to $\tau_v$ in the Southern Ocean. Although small in magnitude, the component response to heat flux still captures important features not present in the response to either component of surface stress (Table S4). In the Atlantic, we expect the most nonlinear responses at high latitudes and find this reflected in the disagreement between emulators, with distinct response patterns at latitudes above $40^\circ \mathrm{N}$. The high latitudes are also where the true midHolocene response is most strongly mediated by sea-ice feedbacks \cite{otto2020comparison}, processes that are nonlinear in the true system and that our emulator does not represent. This offers a physical reason to suspect that an additive emulator would miss high-latitude structure, though, absent single-forcing numerical runs, we cannot confirm that the true response there is non-additive.

\subsection{Additivity of the Emulator's Component Responses}
\label{sec:component_forcings_linearity}
We further examine how the emulator combines the response contributions from individual boundary components. We note that this probes additivity of the emulator's response, not the additivity of the underlying CESM2 experiments, which would require single-forcing CESM2 experiments that are not available. We find that across all basins both the amplitude and spatial patterns of the total response, $R \left(\mathcal{F}_{\mathrm{pi}}^{\mathrm{All}},\boldsymbol{\tau}_{\mathrm{mH}},\operatorname{I}_{\mathrm{mH}}\right)$, are well captured by the superposition of component responses. We measure this by evaluating the residual between the full emulator response and the linear superpositions of component forcings, $R\left(\mathcal{F}_{\mathrm{pi}}^{\mathrm{All}},\boldsymbol{\tau}_{\mathrm{mH}},\operatorname{I}_{\mathrm{mH}}\right)-\sum_{\tau\in[\boldsymbol{\tau}_{\mathrm{mH}},\operatorname{I}_{\mathrm{mH}}]}R \left(\mathcal{F}_{\mathrm{pi}}^{\mathrm{All}},\tau_{\mathrm{mH}}\right)$.  This residual provides a measure of the emulator's departure from additivity; a residual small relative to the total response indicates that the emulator combines forcing components close to linearly. 

\begin{figure}
    \centering
    \includegraphics[width=.9\linewidth]{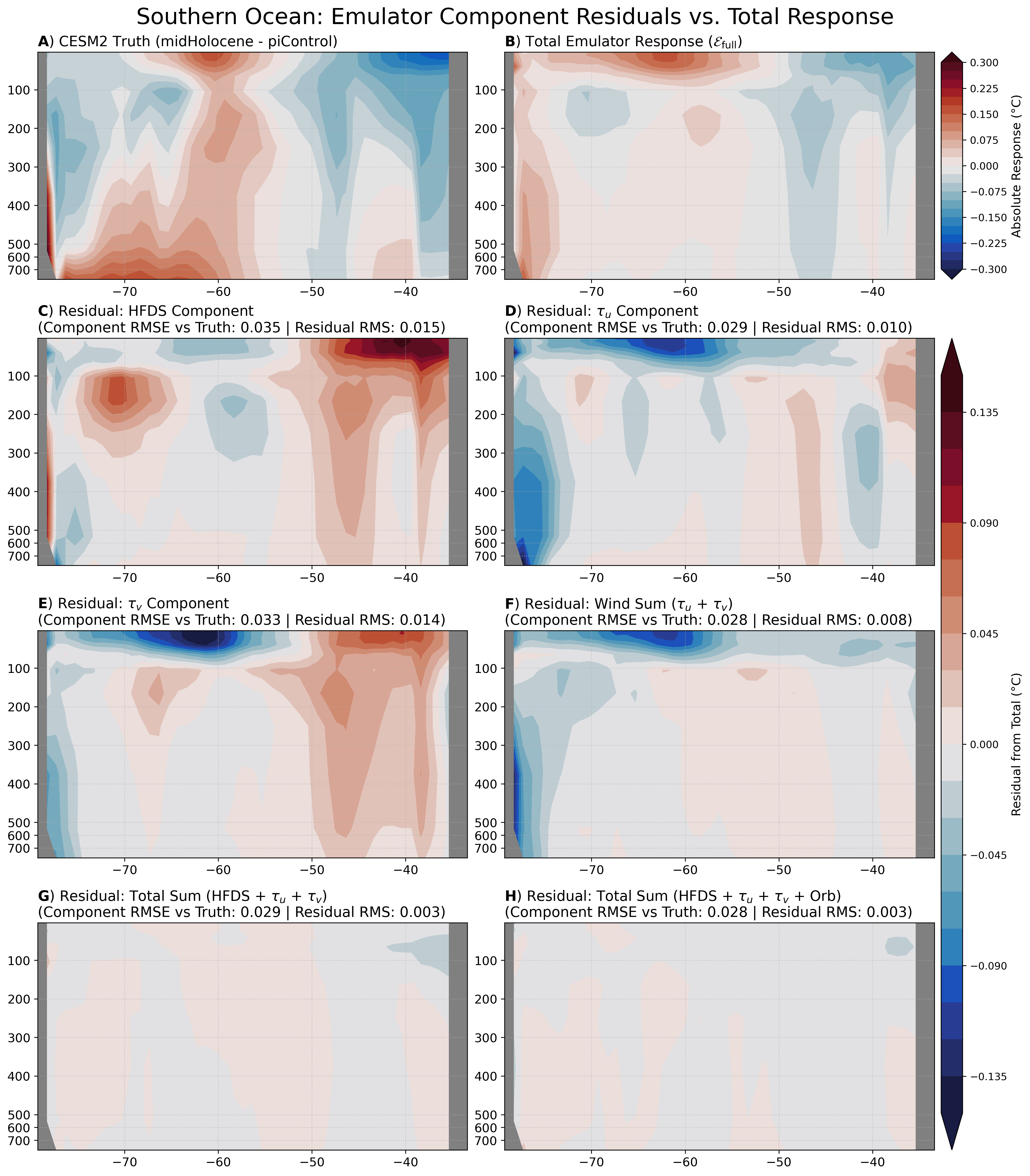}
    \caption{Comparison of the total Southern Ocean potential temperature response to midHolocene boundary forcings against the linear superposition of individual component responses. A: Difference between the CESM2 midHolocene and piControl numerical experiments for reference. B: Total potential temperature response of the full emulator ($\mathcal{F}_{\mathrm{pi}}^{\mathrm{All}}$) to all midHolocene boundary forcings, denoted as $R(\mathcal{F}_{\mathrm{pi}}^{\mathrm{All}};\overline{\tau}_{\mathrm{mH}},\operatorname{I}_{\mathrm{mH}})$. C-E: Residuals computed between a component response of the emulator to the total response. The component responses are heat flux (C: $R(\mathcal{F}_{\mathrm{pi}}^{\mathrm{All}};\overline{hfds}_{\mathrm{mH}})$), zonal surface stress (D: $R(\mathcal{F}_{\mathrm{pi}}^{\mathrm{All}};[\overline{\tau_{u}}]_{\mathrm{mH}})$), and meridional surface stress (E: $R(\mathcal{F}_{\mathrm{pi}}^{\mathrm{All}};[\overline{\tau_{v}}]_{\mathrm{mH}})$). F-H: Residuals computed between linear superpositions of component responses of the emulator to the total response. The superpositions are both surface stress components (F), all dynamic boundary forcings consisting of zonal surface stress, meridional surface stress, and heat flux (G), and all boundary forcings including solar insolation (H).}
    \label{fig:residuals_Southern}
\end{figure}

Focusing on the Southern Ocean basin, in Figure \ref{fig:residuals_Southern} we show the residual of individual boundary component responses with respect to the total zonally averaged potential temperature response. All of the standalone component responses, Panels C-E, are missing details of the total response, particularly in magnitude, with residual amplitudes close to that of the total response. As we begin to add the component responses together, this disagreement reduces, both in amplitude and spatial pattern. Starting with the response to zonal surface stress, which corresponds to the total response most strongly with a correlation of 0.81 and RMSE of 0.010, we find that adding in the meridional surface stress response brings the sum closer to the full response with a correlation of 0.89 and RMSE of 0.008. Adding the heat flux response accounts for the remaining residual, raising the correlation to 0.99 and lowering the RMSE to 0.003. The response to orbital changes contributes negligibly to the reconstruction of the annual-mean response in this basin, leaving the correlation and RMSE unchanged at 0.99 and 0.003. In the other basins (Table S4), the contributions from the orbital forcing are small, but still larger than the negligible contribution in the Southern Ocean. Although it contributes weakly to the equilibrium response, the orbital forcing is necessary for capturing the seasonal cycle changes in Figure \ref{fig:seasonal_cycle} with the correlation in Panel C dropping from 0.87 to 0.52 for $\mathcal{F}_{\mathrm{pi}}$ (Figure S3). That the linearly superimposed responses capture not just the structure but the amplitude of the overall emulator response suggests that, for perturbations of this magnitude, the emulator combines forcing responses approximately linearly, consistent with the forcing pathways interacting only weakly through its internal state.

We further explore this in the Pacific and Atlantic oceans, with the RMSE and correlation values in Table S4 and residual maps for the other basins in Figures S27-S28. We find that the linear sum of responses shows a similar monotonic relationship as we see in the Southern Ocean. In the Pacific, the correlation and RMSE improve from 0.79 and 0.053 to 0.99 and 0.013 when adding the remaining component responses to the zonal surface stress response. In the Atlantic, we see a similar improvement from 0.82 and 0.035 to 0.99 and 0.008. These results reinforce that the differences between the individual component responses in Section \ref{sec:component_forcings} are meaningfully independent and do not overlap in magnitude or structure with respect to the total response.

\subsection{Response Sensitivity to Epoch}
\label{sec:epoch_sensitivity}
Having decomposed the relationship between component forcings and the overall response, we shift our focus to the relationship between what we optimize (MSE, and with it in-distribution skill) and what we evaluate (the forced response). MSE has been used as a well-behaved, easy-to-optimize loss function and is the first choice as groups tackle new challenges in weather and climate \cite{subel2024building,dheeshjith2025samudra,bi2023accurate,bire2025ocean,chapman2025camulator,chattopadhyay2024oceannet,watt2025ace2,clark2024ace2,lam2023learning,duncan2025samudrace}. Despite this, an emulator that perfectly optimizes for MSE over multiple recurrent passes is not guaranteed to find the best model for answering science questions. In forecasting, this has led to the rise of stochastic methods, which optimize scoring rules that converge to an underlying distribution rather than the distribution's mean as does MSE \cite{gneiting2007strictly,subich2025fixing}. These methods fall into two categories: 1)  ensembles methods that optimize a proper score such as the continuous ranked probability score \cite{kochkov2024neural,lang2024aifs,bonev2025fourcastnet}; and 2) diffusion-based models \cite{price2025probabilistic,couairon2024archesweather}.

\begin{figure}
    \centering
    \includegraphics[width=\linewidth]{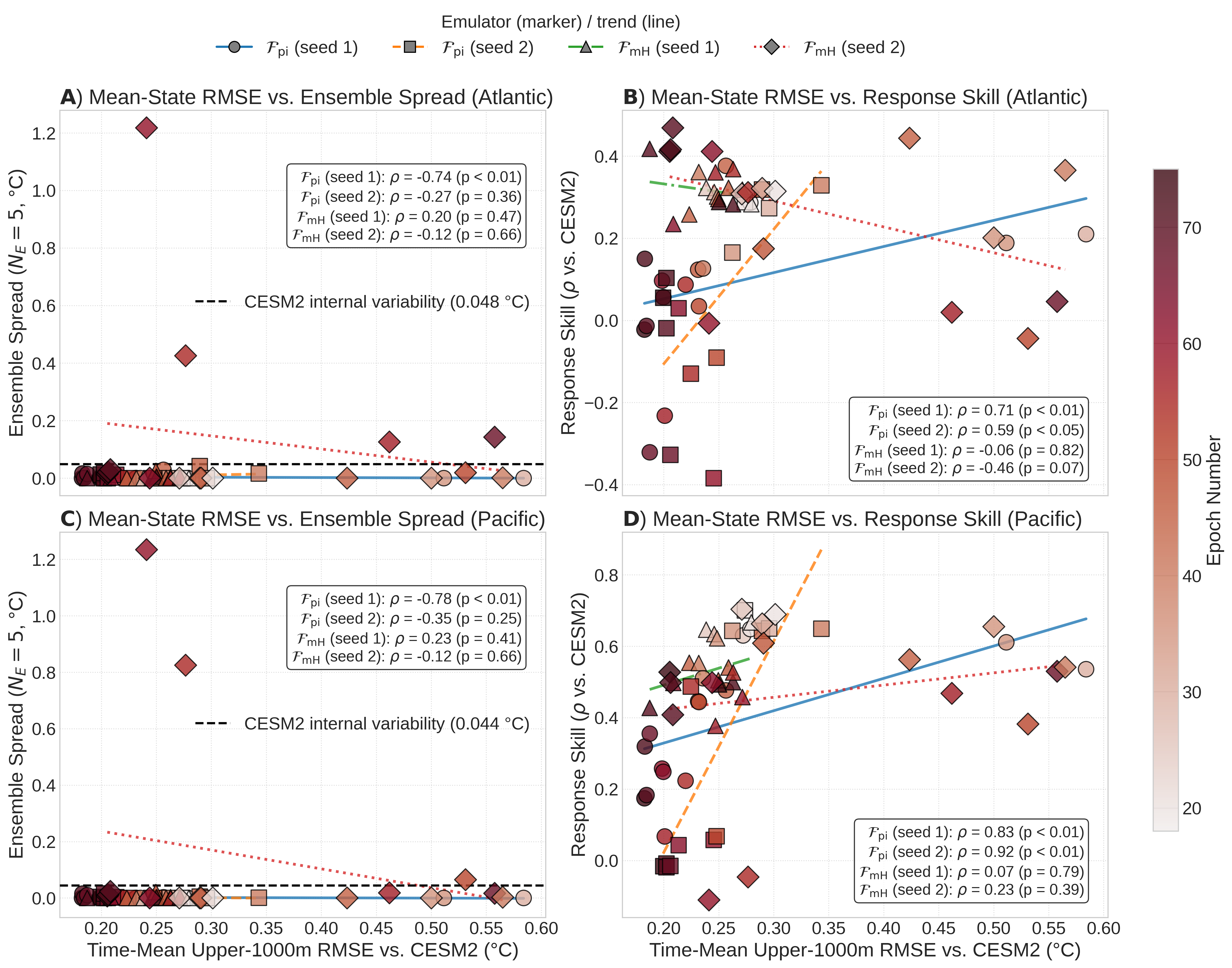}
    \caption{Epoch sensitivity of out-of-sample response skill and spread as a function of epoch. In all panels, response metrics are plotted against the RMSE of the upper 1000m depth-averaged mean state in the referenced basin. The mean states are computed over a 100 year rollout of each emulator using held out data from the training experiment (e.g. $\Omega(\mathcal{F}_{\mathrm{pi}},\boldsymbol{\Phi}_{\mathrm{pi}}^{[0]};\boldsymbol{\tau}_{\mathrm{pi}},\operatorname{I}_{\mathrm{pi}})$) and RMSE is evaluated against the same 100 year period from CESM2 data. We compute response metrics for the Atlantic (A and B) and the Pacific (C and D) separately. For ensemble spread (A and C), we use the spatial mean of the pointwise standard deviation across five initial conditions for each zonally averaged response. We compare the spreads to an estimate from the two CESM2 experiments (dashed line). For this estimate, we use the standard deviation of the response over ten 25-year chunks of the CESM2 datasets. Skill (B and D) is measured as the correlation between the zonally averaged response from the emulator and the zonally averaged difference between CESM2 experiments, as done in Section \ref{sec:Overall_Response}.  Results are plotted for two random seeds of $\mathcal{F}_{\mathrm{pi}}$ (circles and triangles) and $\mathcal{F}_{\mathrm{mH}}$ (squares and diamonds).  Given that checkpoints from a single training run are not independent samples, we use the rank correlations and $\rho$ values shown in each panel as descriptive measures of the checkpoints
and not a true significance test.}
    \label{fig:epoch_comparison}
\end{figure}

The relationship between MSE metrics and an emulator's response skill is even less direct. To better understand this relationship for the experiments in this paper, we reproduced the results using checkpoints from multiple epochs and, for each emulator, two independent random seeds (Figure \ref{fig:epoch_comparison}). Across the checkpoints we examine, test RMSE spans a relatively narrow, largely converged range, with the RMSE of most checkpoints falling between $0.18$ and $0.36\,^\circ$C, yet response skill varies substantially within that range. Across the four emulators, the rank correlation between test RMSE and response skill is inconsistent in both magnitude and sign, ranging from approximately $+0.9$ to $+0.1$ in the Pacific and from $+0.7$ to $-0.5$ in the Atlantic. We therefore do not interpret the sign or magnitude of this relationship too strongly. The robust conclusion is that checkpoints that appear equivalent or improved by test RMSE are not equivalent in response skill, confirming that MSE remains an unreliable metric for predicting response skill. We additionally find that ensemble spread varies across epochs in a manner that does not track response skill, and that the relationship between spread and RMSE is itself seed-dependent, being anti-correlated for one piControl seed but not the other. This inconsistency reinforces that standard validation metrics do not reliably predict response skill. This points to a need to incorporate the methods mentioned above, or novel developments, into the pipeline for climate emulators. These conclusions align with results from emulating reduced-order models of climate data in \citeA{falasca2025neural}. They also suggest that the common practice of selecting the checkpoint with the lowest validation loss or test RMSE may not be well-suited to these problems \cite{lam2023learning,watt2025ace2,chapman2025camulator}.

We find that across the basins examined in Section \ref{sec:Overall_Response}, response skill converges early in training, plateauing over roughly the first fifty epochs, with the second piControl seed degrading after epoch 45. Over the Pacific basin the plateau correlations span 0.44-0.67 and 0.64-0.70 for the two piControl seeds and 0.19-0.67 and 0.54-0.70 for the two midHolocene seeds. Over the Atlantic basin they span 0.03-0.38, 0.17-0.33, 0.24-0.36, and 0.19-0.44, respectively. After these epochs, skill declines for both piControl seeds, with mean correlations falling to 0.21 and 0.12 in the Pacific basin and to $-0.02$ and $-0.08$ in the Atlantic basin, leaving essentially no basin-wide skill in the Atlantic. This decline is reflected across each subregion in Figure \ref{fig:Full_Responses_All_Basins}. Over these same later epochs the midHolocene emulators degrade less, and their weakest checkpoints recover at later epochs rather than remaining degraded. Due to the strong late-training degradation, which we identify through the out-of-sample response itself, we report results for the piControl emulators from the early, pre-degradation regime instead of the lowest-loss checkpoint; checkpoints within this regime themselves span a wide range of response skill, so this choice characterizes the emulator's strengths and weaknesses rather than the degraded behavior that emerges late in training. We further note that no checkpoint performs best across all regions for each emulator, and that we have not selected for the best performance in any region. Given the range of performance across epochs and seeds as well as the identification of a region of collapsed skill using the test data, we discuss trends and broadly consistent behaviors rather than claiming that highly local results from any particular checkpoint are indicative of a repeatable success or failure.

\section{Conclusion}

Many studies have focused on advancing state-of-the-art weather emulators and developing robust methodological tests for such emulators \cite{bi2023accurate,kochkov2024neural,price2025probabilistic,lang2024aifs,bonev2025fourcastnet,sun2025can,sun2025predicting,zhang2025numerical}; however, work on ocean and climate applications remains underdeveloped. Fortunately, many of the advances in weather emulation have allowed for rapid improvement in climate atmosphere emulation \cite{watt2025ace2,chapman2025camulator,guan2025lucie}, short-term ocean forecasting \cite{huang2025fuxi,cui2025forecasting}, regional ocean modeling \cite{chattopadhyay2024oceannet},  decadal to multi-century ocean emulation \cite{guo2025data,dheeshjith2025samudra}, and coupled climate emulation \cite{duncan2025samudrace}. 
Weather and climate emulators both learn a discrete propagator for geophysical fluid dynamics, yet their validation requirements differ fundamentally. Weather forecasting permits real-time verification against observations, whereas climate emulation must contend with regimes beyond the reach of any numerical baseline. In the absence of ground truth for such conditions, validation cannot rely solely on standard trajectory metrics; it must instead demonstrate that the emulator preserves physically consistent dynamics outside the training domain.

In this work, we assess an AI autoregressive ocean emulator through targeted dynamical tests, using CESM2 piControl and midHolocene simulations to probe out-of-sample generalization within the training distribution. Verifying that an emulator can capture the subsurface ocean response to varying external forcings is a key step toward positioning such emulators as viable sandboxes for ``what if'' experiments, in the tradition of numerical climate modeling \cite{braconnot2012evaluation}. Here we deliberately choose an in-distribution, out-of-sample starting point to surface the outstanding challenges that must be addressed before moving toward fully out-of-distribution settings.

Future warming scenarios risk large distributional shifts in temperature and salinity (Figure \ref{fig:Holo_Pi_Comp}, Panels D-E), but the midHolocene experiment perturbs the climate system primarily through orbital parameters, producing shifts in atmospheric patterns with comparatively modest mean-state changes. The emulators capture changes in the seasonal cycle of potential temperature, which involve both a shift in phase and a structured change in amplitude. Neither baseline recovers these changes from the boundary forcings alone. We find that emulators also reproduce measures of variability in the new climate, not just the mean response. The emulators reproduce modes of variability such as ENSO,  the Indian Ocean Dipole, and the AMO with a reduced amplitude, although much of the change to modes of variability can be captured through the forcing alone.  We find that the emulators capture changes to the spatial structure of temporal variability over the upper ocean, which are not directly encoded in the changes to surface forcings. Perturbation experiments further confirm the ability of emulators to recover complex upper-1000m anomalies in potential temperature under unseen midHolocene forcing applied to a piControl background state for regions such as the tropical Pacific, Southern Ocean, or the near surface.

The emulators' skill in out-of-sample perturbation experiments is neither uniform nor without significant limitations. While the emulators better capture the amplitude and depth of the response as compared to the two baselines, performance is strongest where the baselines suggest that ocean response is governed by direct atmospheric forcing: in the tropics and near the surface the emulators reproduce changes in seasonality, variability, and time-mean state under midHolocene boundary conditions (Figures \ref{fig:seasonal_cycle}, \ref{fig:Variability}, \ref{fig:Full_Responses_All_Basins}, and \ref{fig:Full_Responses_Map}). Skill degrades substantially at northern high latitudes, where the ocean response is shaped by nonlinear dynamics, including AMOC changes \cite{simpson2023cesm2} and the slow accumulation of heat from small insolation imbalances, rather than by direct atmospheric drivers. This degradation points to a clear limitation: the emulators fail to represent the slow, self-driven evolution of the ocean interior, suppressing the deep heat accumulation seen in the true response. It also points to a possible failure to capture nonlinear feedbacks between the ocean state and surface forcing. 

To diagnose the emulator's out-of-sample behavior, we decompose the total ocean response into contributions from individual boundary-forcing components. The emulator treats each forcing channel largely independently. In the Southern Ocean (Section \ref{sec:component_forcings}), heat fluxes drive near-surface warming while zonal surface stresses drive deeper mixing, without significant artificial cross-contamination between pathways. This independence holds most strongly within a single emulator, where summing its component responses reproduces the full response with a correlation near 0.99 in every basin we examine. We do find, however, that partially forced responses show stronger amplitudes than the full emulator's component responses and reduced stability over long rollouts. Adding freshwater flux improves skill in most basins but degrades the Southern Ocean, so we do not claim new boundary forcings will strictly benefit response skill. As shown in Section \ref{sec:component_forcings_linearity}, the emulator reconstructs the full ocean response by linearly superimposing component responses, suggesting that the forcing pathways interact only weakly through the internal state. This additivity is consistent with the emulator's failure to represent phenomena such as deep ocean heat accumulation and the internally driven circulation changes in the North Atlantic, although we cannot confirm that the true response to the midHolocene changes would itself be nonlinear without single-forcing numerical experiments, such as midHolocene runs perturbed in $\boldsymbol{\tau}$ or $\operatorname{hfds}$ alone, which are not part of the PMIP4 protocol.

These results point to weak feedback of the ocean state on itself, at least at the monthly time steps considered here. This is consistent with the asymmetric roles of the two systems: the atmosphere responds rapidly to slowly evolving boundary conditions, while the ocean integrates atmospheric changes over long timescales, acting as the climate system's memory. By fixing the atmospheric trajectory during training and eliminating uncertainty, the emulator may be conditioned to follow atmospheric forcing rather than internal ocean dynamics, suppressing deep heat accumulation and collapsing the true ensemble spread. As shown in Section \ref{sec:epoch_sensitivity}, minimizing mean squared error, even over extended rollouts, does not by itself guarantee that the emulator will recover these dynamics, nor does a lower test RMSE reliably predict improved response skill.

Although long-term ocean emulation remains sparsely explored, adjacent emulation efforts offer several promising directions. Introducing stochasticity (Section \ref{sec:epoch_sensitivity}) would enable the emulator to represent uncertainty and more faithfully reflect the chaotic nature of the climate system. Coupled emulators, trained jointly across the atmosphere, ocean, and sea ice, may better capture cross-component interactions \cite{duncan2025samudrace}, while physically constrained architectures offer another avenue for improving dynamical fidelity \cite{sha2025improving,watt2025ace2}. While groups have begun to evaluate against dynamical benchmarks for the atmosphere \cite{watt2025ace2,clark2024ace2,hakim2024dynamical}, the ability of emulators to capture nonlinear ocean feedbacks remains an open question essential for climate application. The out-of-sample generalization demonstrated here supports the continued use of ocean emulators for perturbation experiments targeting the upper ocean's directly forced response; extending this reliability to the slow, subsurface interior will require training strategies explicitly designed for those dynamics.

\section*{Open Research Section}

The unprocessed midHolocene CMIP6 data used in this work come from the NCAR Glade filestore and are also available from the Earth System Grid Federation \cite{danabasoglu2019midholocene}. The unprocessed CESM2 piControl \cite{danabasoglu2019picontrol} and CESM2 1\% CO$_2$ \cite{danabasoglu2019onepctco2} data were accessed through the pangeo CMIP6 cloud archive (\url{https://cmip6.storage.googleapis.com/pangeo-cmip6.csv}). The code for training and evaluating the emulators, together with preprocessing scripts for the raw datasets and the model weights for the emulators shown in the main text and several additional epoch checkpoints, is archived on Zenodo \cite{subel2026code} with code at \url{https://github.com/adam-subel/Probing-the-Dynamical-Response-of-Ocean-Climate-Emulators}.

\section*{Conflict of Interest declaration}
The authors declare there are no conflicts of interest for this manuscript.

\acknowledgments
The authors thank Fabrizio Falasca, Alistair Adcroft, and Mitchell Bushuk for helpful discussions, as well as Rory Basinski-Ferris and Andrew Brettin for comments on the manuscript. This material is based upon work supported by the National Science Foundation Graduate Research Fellowship under Grant No. DGE-2234660 and NSF Grant No. RISE-2530958. This research received support through Schmidt Sciences.

\bibliography{references}

\clearpage

\setcounter{section}{0}
\setcounter{equation}{0}
\setcounter{figure}{0}
\setcounter{table}{0}
\renewcommand{\thesection}{S\arabic{section}}
\renewcommand{\theequation}{S\arabic{equation}}
\renewcommand{\thefigure}{S\arabic{figure}}
\renewcommand{\thetable}{S\arabic{table}}

\section*{Supporting Information}

\noindent\textbf{Contents of this file}

\begin{enumerate}
\item Text S1 (Freshwater Flux)
\item  Text S2 (Baselines)
\item  Figures S1 to S28
\item  Tables S1 to S4
\end{enumerate}

\section{Freshwater Flux}
To incorporate freshwater flux as an additional boundary condition, we construct the total freshwater forcing by combining contributions from the atmosphere, land, and sea ice: precipitation ($\operatorname{pr}$), evaporation ($\operatorname{evspsbl}$), sea ice freshwater flux ($\operatorname{siflfwbot}$), and brine drainage ($\operatorname{siflfwdrain}$). 

While these components are directly available for the CESM2 midHolocene experiment, $\operatorname{siflfwbot}$ is absent from the standard piControl outputs. We reconstruct this field using available sea ice mass budget variables as the sum of basal and lateral sea ice melt minus basal and frazil ice growth:

\begin{equation}
    \operatorname{siflfwbot} = \operatorname{sidmassmeltbot} + \operatorname{sidmasslat} - \operatorname{sidmassgrowthbot} - \operatorname{sidmassgrowthwat}.
\end{equation}

Each sub-component is processed independently on its native grid. Following the horizontal regridding procedure described in Section 2.1 of the main text, we map these fields onto the target $1^\circ$ Gaussian grid. We then produce the total freshwater flux, $\operatorname{wfo}$, by combining the terms as follows:
\begin{equation}
    \operatorname{wfo} = \operatorname{pr} - \operatorname{evspsbl} + \operatorname{siconc}\left(\operatorname{siflfwbot} + \operatorname{siflfwdrain}\right).
\end{equation}
\noindent Here $\operatorname{siconc}$ is the sea ice concentration and is used to weight the fluxes coming from under the sea ice. We then train the emulator following the same procedures in the main text sections 2.3 and 2.4 and with the total input vector to the network of $[\boldsymbol{\Phi}_t,\boldsymbol{\tau}_t,\operatorname{wfo}_t,I_t]$.

\section{Baselines}
\label{appendix:lro}

To contextualize the emulator's forced response and separate the internal dynamics from a learned relationship to the forcings, we construct two baselines that predict the state directly from the boundary forcings. The first, a linear regression operator (LRO), is local and linear, predicting the ocean state from the surface forcing at the same grid column, with no horizontal communication, no internal dynamics, and no memory. The second, a forcing-only network, keeps the emulator's architecture, nonlinearity, and global receptive field, removing the ability to represent internal dynamics, and maps the boundary forcings at a given time to the ocean state at that same time. Both are fit on the same 3500 months of piControl data used to train $\mathcal{F}_{\mathrm{pi}}$, and both are applied to the midHolocene forcings.

\subsection{Linear Regression Operator}

Let $S(x,y,z,t)$ denote a state variable ($\theta_O$ or $S$) and $F_p(x,y,t)$ the surface forcings, both sampled monthly over the fit window, where $p$ indexes the individual forcing fields ($\operatorname{hfds}$, $\tau_u$, $\tau_v$, and the solar insolation $\operatorname{I}$). To indicate the monthly climatology of a field we use the notation $\overline{S}^{\,m}$. We denote the time mean climatology as $\overline{S}$ and use $\langle S \rangle^{\,n}$ to indicate annual means. The model assumes that, at each horizontal location and depth, the state anomaly is a linear combination of the local surface forcing anomalies,

\begin{equation}
S'(x,y,z,t) \;\approx\; \sum_{p=1}^{P} \beta_p(x,y,z)\, F'_p(x,y,t) ,
\end{equation}

with no contribution from neighboring locations.

Each diagnostic is defined against a different reference, so we fit three forms of the LRO by taking anomalies about a correspondingly different mean:

\begin{equation}
S'(t) \;=\;
\begin{cases}
\langle S \rangle^{\,n(t)} - \overline{S}, & \text{annual samples,}\\[4pt]
S(t) - \overline{S}, & \text{monthly samples, seasonal cycle retained,}\\[4pt]
S(t) - \overline{S}^{\,m(t)}, & \text{monthly samples, seasonal cycle removed.}
\end{cases}
\label{eq:lro_anomalies}
\end{equation}

Here $n(t)$ and $m(t)$ are the calendar year and calendar month of sample $t$, and we construct $F'_p$ in the same manner as $S'$. The first operator serves the time-mean response, the second the change in the seasonal range, and the third the variability diagnostics. The three differ only in this choice; all other aspects of the fit are identical.

The choice of anomaly also determines which forcings carry information. We exclude insolation from the two LRO cases that fit annual means and monthly climatological anomalies as this field would be zero in both cases. We use an LRO fit to the deviation against the time mean when examining any signal that contains the seasonal cycle. We note that $\beta$ is fit once for each LRO and the seasonal structure of the predicted response comes only from the seasonal structure of the forcing change.

We solve for $\beta$ independently at every grid column and depth level. To regularize consistently across variables and depths, we standardize both the predictors and the target by their temporal standard deviations, $\sigma_{F_p}(x,y)$ and $\sigma_S(x,y,z)$, before fitting. The standardized coefficients $\tilde{\beta}$ minimize a ridge (Tikhonov) objective over the $T$ training samples,

\begin{equation}
\tilde{\beta}(x,y,z) \;=\; \arg\min_{b}\;
\sum_{t=1}^{T} \left( \frac{S'}{\sigma_S}
- \sum_{p} b_p \frac{F'_p}{\sigma_{F_p}} \right)^{\!2}
+ \lambda T \sum_{p} b_p^{2},
\end{equation}

with $\lambda = 0.1$. We scale the penalty by $T$ so that the shrinkage does not depend on the sampling rate and the annual and monthly operators are regularized identically. The physical-unit coefficients are recovered by rescaling,

\begin{equation}
\beta_p(x,y,z) \;=\; \tilde{\beta}_p(x,y,z)\,
\frac{\sigma_S(x,y,z)}{\sigma_{F_p}(x,y)} .
\end{equation}

Land points are filled with zeros prior to fitting so the solver remains well posed; they are masked out of all subsequent diagnostics.

We apply the fitted operator in two ways. For the time-mean and seasonal-range diagnostics, we apply it to the change in the forcing climatology. Let $\Delta F_p(x,y)$ be the difference between the midHolocene and piControl climatologies of forcing $p$, giving the predicted ocean response

\begin{equation}
R(x,y,z) \;=\; \sum_{p=1}^{P} \beta_p(x,y,z)\,\Delta F_p(x,y),
\end{equation}

which we evaluate monthly and then either average over the year or difference between SON and MAM to match the diagnostic. For the variability diagnostics we instead apply the operator to the month-by-month forcing anomalies of a single climate, centered by the same case of Equation \ref{eq:lro_anomalies} that we used to fit it. This yields a timeseries of ocean states from which we compute the Ni\~{n}o3.4 and DMI indices and the subsurface composite. We evaluate this application over windows that overlap the corresponding emulator rollouts, so that we score the two over the same months.

\subsection{Forcing-Only Network}

To isolate the role of internal dynamics, we train the forcing-only network with the same architecture, boundary forcings, output variables, and training data as $\mathcal{F}_{\mathrm{pi}}$. We remove the ocean state from the inputs, so the network maps the boundary forcings alone to the ocean state at the same time,

\begin{equation}
\boldsymbol{\Phi}(t) \;\approx\; \mathcal{B}\!\left(\boldsymbol{\tau}(t), \operatorname{I}(t)\right),
\end{equation}

across all depth levels of both state variables. Because the network takes no state as input, we apply it directly rather than rolling it out, and we take its response to the midHolocene forcings as the difference between its predictions under the two sets of boundary forcings.

Retaining the emulator's architecture is what makes this baseline informative. The network has the same nonlinearity, receptive field, and capacity as $\mathcal{F}_{\mathrm{pi}}$, so any diagnostic it reproduces is one we can recover from the instantaneous forcing without reference to the ocean state. The difference between the two therefore isolates the contribution of internal dynamics in the emulator. 

\begin{table}[h]
\caption{Convergence of the response metrics with rollout length. Pattern correlation ($\rho$) and amplitude ratio ($\sigma/\sigma_{\mathrm{truth}}$) of the zonally averaged, upper-1000\,m potential-temperature response against the CESM2 midHolocene $-$ piControl difference, evaluated over successive 25-year windows of a 150-year rollout. }
\label{tab:rollout_length}
\centering
\begin{tabular}{llcccccc}
\hline
& & \multicolumn{6}{c}{rollout window (years)}\\
\cline{3-8}
Basin & Response & 0--25 & 25--50 & 50--75 & 75--100 & 100--125 & 125--150\\
\hline
\multicolumn{8}{l}{\textit{pattern correlation} $\rho$}\\
Pacific  & $R(\mathcal{F}_{\mathrm{pi}})$ & 0.51 & 0.44 & 0.44 & 0.44 & 0.44 & 0.44\\
         & $R(\mathcal{F}_{\mathrm{mH}})$ & 0.51 & 0.48 & 0.43 & 0.42 & 0.42 & 0.42\\
Atlantic & $R(\mathcal{F}_{\mathrm{pi}})$ & 0.07 & 0.03 & 0.03 & 0.03 & 0.03 & 0.03\\
         & $R(\mathcal{F}_{\mathrm{mH}})$ & 0.18 & 0.27 & 0.40 & 0.41 & 0.41 & 0.41\\
Southern & $R(\mathcal{F}_{\mathrm{pi}})$ & 0.57 & 0.59 & 0.59 & 0.59 & 0.59 & 0.59\\
         & $R(\mathcal{F}_{\mathrm{mH}})$ & 0.42 & 0.38 & 0.43 & 0.44 & 0.44 & 0.44\\
Indian   & $R(\mathcal{F}_{\mathrm{pi}})$ & 0.34 & 0.32 & 0.32 & 0.32 & 0.32 & 0.32\\
         & $R(\mathcal{F}_{\mathrm{mH}})$ & 0.50 & 0.55 & 0.58 & 0.58 & 0.58 & 0.58\\
\hline
\multicolumn{8}{l}{\textit{amplitude ratio} $\sigma/\sigma_{\mathrm{truth}}$}\\
Pacific  & $R(\mathcal{F}_{\mathrm{pi}})$ & 0.54 & 0.62 & 0.62 & 0.62 & 0.62 & 0.62\\
         & $R(\mathcal{F}_{\mathrm{mH}})$ & 0.55 & 0.60 & 0.64 & 0.65 & 0.65 & 0.65\\
Atlantic & $R(\mathcal{F}_{\mathrm{pi}})$ & 0.33 & 0.35 & 0.35 & 0.35 & 0.35 & 0.35\\
         & $R(\mathcal{F}_{\mathrm{mH}})$ & 0.52 & 0.51 & 0.50 & 0.50 & 0.50 & 0.50\\
Southern & $R(\mathcal{F}_{\mathrm{pi}})$ & 0.44 & 0.47 & 0.47 & 0.47 & 0.47 & 0.47\\
         & $R(\mathcal{F}_{\mathrm{mH}})$ & 0.67 & 0.69 & 0.63 & 0.61 & 0.61 & 0.61\\
Indian   & $R(\mathcal{F}_{\mathrm{pi}})$ & 0.31 & 0.32 & 0.32 & 0.32 & 0.32 & 0.32\\
         & $R(\mathcal{F}_{\mathrm{mH}})$ & 0.43 & 0.42 & 0.39 & 0.39 & 0.39 & 0.39\\
\hline
\end{tabular}
\end{table}

\begin{table}[h]
\caption{Pattern correlation ($\rho$) and RMSE [$^\circ$C] of the zonally averaged, upper-1000\,m potential-temperature response against the CESM2 midHolocene$-$piControl difference. The baseline column is the piControl emulator of the main text ($F_{pi}$), which uses no freshwater forcing. The remaining columns are a single piControl-trained emulator that adds freshwater flux as a boundary condition ($F^{FW}_{pi}$), evaluated under the midHolocene perturbation two ways: with the freshwater flux held fixed at the piControl climatology, and with the freshwater flux perturbed to its full midHolocene values. All other boundary forcings are perturbed to midHolocene in both cases. Values are listed for each basin considered in Figure \ref{fig:Full_Responses_All_Basins_wfo}. }
\label{tab:wfo}
\centering
\begin{tabular}{l cc cc cc}
\hline
 & \multicolumn{2}{c}{Baseline} & \multicolumn{4}{c}{wfo emulator ($F^{FW}_{pi}$)} \\
\cline{2-3}\cline{4-7}
 & \multicolumn{2}{c}{(no wfo)} & \multicolumn{2}{c}{freshwater fixed} & \multicolumn{2}{c}{freshwater perturbed} \\
Basin / region & $\rho$ & RMSE & $\rho$ & RMSE & $\rho$ & RMSE \\
\hline
Pacific                            & 0.44 & 0.12 & 0.65 & 0.10 & 0.63 & 0.10 \\
\quad Tropics (Pacific)            & 0.64 & 0.13 & 0.81 & 0.10 & 0.80 & 0.10 \\
\quad North lats (Pacific)         & 0.24 & 0.14 & 0.51 & 0.13 & 0.44 & 0.13 \\
Atlantic                           & 0.03 & 0.31 & 0.22 & 0.30 & 0.24 & 0.30 \\
\quad Surface (Atlantic)           & 0.68 & 0.25 & 0.70 & 0.25 & 0.67 & 0.26 \\
Southern                           & 0.59 & 0.03 & 0.46 & 0.03 & 0.46 & 0.03 \\
\quad Westerlies Shift (Southern)  & 0.81 & 0.04 & 0.69 & 0.05 & 0.69 & 0.05 \\
\quad Surface Heating (Southern)   & 0.80 & 0.04 & 0.72 & 0.04 & 0.71 & 0.04 \\
Indian                             & 0.33 & 0.11 & 0.50 & 0.10 & 0.49 & 0.10 \\
\quad Tropics (Indian)             & 0.39 & 0.14 & 0.61 & 0.12 & 0.60 & 0.13 \\
\hline
\end{tabular}
\end{table}

\begin{table}[h]
\caption{Pattern correlation ($\rho$), amplitude ratio ($\sigma/\sigma_{\mathrm{truth}}$), RMSE
[$^\circ$C] and centroid depth ($z_c$) [m] of the zonally averaged, upper-1000\,m
potential-temperature response against the CESM2 midHolocene$-$piControl difference, for the
piControl emulator ($\mathcal{F}_{\mathrm{pi}}$), the local linear regression operator (LRO) and the
forcing-only network (FO). All three are fit or trained on piControl and driven with the
midHolocene forcing change, and the values are those reported in Figure S21. The centroid depth is
the first moment with depth of the response amplitude, so a value shallower than the CESM2 column
indicates a response confined closer to the surface; it is omitted for the two regions that do not
extend below 200\,m.}
\label{tab:baseline_metrics}
\centering
\resizebox{\textwidth}{!}{
\begin{tabular}{lcccccccccccccc}
\hline
 & \multicolumn{3}{c}{$\rho$} & \multicolumn{3}{c}{$\sigma/\sigma_{\mathrm{truth}}$} & \multicolumn{3}{c}{RMSE} & \multicolumn{4}{c}{$z_c$}\\
\cline{2-4}\cline{5-7}\cline{8-10}\cline{11-14}
Basin / region & $\mathcal{F}_{\mathrm{pi}}$ & LRO & FO & $\mathcal{F}_{\mathrm{pi}}$ & LRO & FO & $\mathcal{F}_{\mathrm{pi}}$ & LRO & FO & CESM2 & $\mathcal{F}_{\mathrm{pi}}$ & LRO & FO\\
\hline
Pacific & 0.44 & 0.32 & 0.50 & 0.62 & 0.41 & 0.38 & 0.13 & 0.14 & 0.13 & 329 & 258 & 196 & 190\\
\quad Tropics (Pacific) & 0.64 & 0.36 & 0.77 & 0.73 & 0.40 & 0.4 & 0.12 & 0.15 & 0.11 & 303 & 240 & 222 & 176\\
\quad North lats (Pacific) & 0.24 & 0.28 & $-$0.41 & 0.40 & 0.5 & 0.2 & 0.14 & 0.14 & 0.16 & 282 & 261 & 160 & 212\\
Atlantic & 0.03 & 0.13 & 0.18 & 0.40 & 0.28 & 0.32 & 0.36 & 0.35 & 0.35 & 419 & 332 & 249 & 229\\
\quad Surface (Atlantic) & 0.68 & 0.43 & 0.64 & 0.60 & 0.5 & 0.5 & 0.24 & 0.29 & 0.25 & -- & -- & -- & --\\
Southern & 0.59 & 0.61 & 0.44 & 0.46 & 0.27 & 0.25 & 0.05 & 0.05 & 0.06 & 442 & 339 & 259 & 215\\
\quad Westerlies Shift (Southern) & 0.81 & 0.81 & 0.65 & 0.54 & 0.42 & 0.36 & 0.03 & 0.04 & 0.05 & 311 & 311 & 216 & 208\\
\quad Surface Heating (Southern) & 0.80 & 0.82 & 0.50 & 0.87 & 0.29 & 0.48 & 0.03 & 0.04 & 0.05 & -- & -- & -- & --\\
Indian & 0.33 & 0.15 & 0.58 & 0.33 & 0.93 & 0.38 & 0.14 & 0.18 & 0.13 & 280 & 273 & 149 & 215\\
\quad Tropics (Indian) & 0.39 & 0.04 & 0.63 & 0.36 & 0.35 & 0.37 & 0.13 & 0.15 & 0.12 & 240 & 229 & 176 & 197\\
\hline
\end{tabular}
}
\end{table}

\begin{table}[h]
\centering
\caption{Correlations and RMSE of reconstructed individual and combined forcing responses against the Full emulator and CESM2 Truth.}
\label{tab:linearity}
\resizebox{\textwidth}{!}{
\begin{tabular}{lcccccccccccc}
\hline
& \multicolumn{12}{c}{piControl$\rightarrow$midHolocene} \\
& \multicolumn{4}{c}{\textbf{Pacific Ocean}} & \multicolumn{4}{c}{\textbf{Atlantic Ocean}} & \multicolumn{4}{c}{\textbf{Southern Ocean}} \\
 & \multicolumn{2}{c}{vs. Full}& \multicolumn{2}{c}{ vs. Truth}& \multicolumn{2}{c}{vs. Full}& \multicolumn{2}{c}{ vs. Truth}& \multicolumn{2}{c}{vs. Full}& \multicolumn{2}{c}{ vs. Truth}\\
\hline
Forcing Combination& $\rho$& RMSE& $\rho$& RMSE& $\rho$ & RMSE & $\rho$ & RMSE & $\rho$& RMSE & $\rho$ & RMSE \\
\hline
Full & - & - & 0.44 & 0.12& - & - & 0.03 & 0.31& - & - & 0.59 & 0.03\\
$\operatorname{hfds}$& 0.55 & 0.08& 0.47 & 0.12& 0.37 & 0.06& 0.31 & 0.30& 0.45 & 0.02& 0.09 & 0.04\\
$\tau_u$ & 0.79 & 0.05& 0.35 & 0.12& 0.82 & 0.04& 0.01 & 0.31& 0.81 & 0.01& 0.58 & 0.03\\
$\tau_v$ & 0.69 & 0.06& 0.21 & 0.13& 0.54 & 0.05& -0.16 & 0.31& 0.58 & 0.01& 0.31 & 0.03\\
$\mathcal{I}$ & 0.41 & 0.08& 0.41 & 0.13& 0.34 & 0.060& 0.00& 0.31& 0.11 & 0.02& 0.15 & 0.03\\
$\tau_u + \tau_v$ & 0.95 & 0.03& 0.36 & 0.12& 0.92 & 0.02& -0.08 & 0.32& 0.89 & 0.01& 0.59 & 0.03\\
hfds + $\tau_u + \tau_v$ & 0.98 & 0.02& 0.42 & 0.12& 0.99 & 0.01& 0.01 & 0.31& 0.99 & 0.003 & 0.57 & 0.03\\
hfds + $\tau_u + \tau_v + \mathcal{I}$ & 0.99 & 0.01& 0.45 & 0.11& 0.99 & 0.01& 0.01 & 0.31& 0.99 & 0.003 & 0.58 & 0.03\\
\hline
\end{tabular}
}
\end{table}

\begin{figure}
    \centering
    \includegraphics[width=\linewidth]{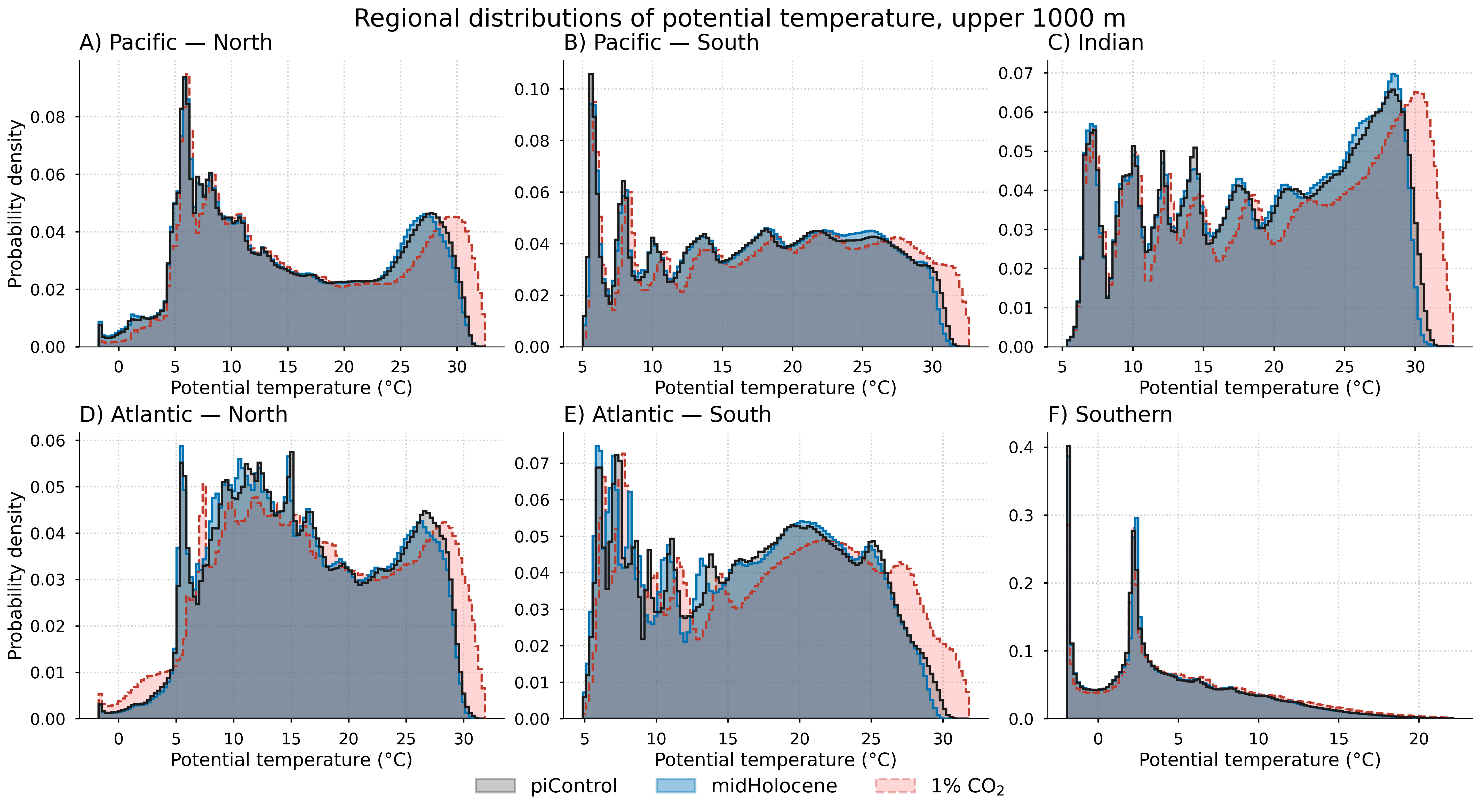}
    \caption{Distributions of potential temperature over the upper 1000\,m for the piControl (grey), midHolocene (blue), and 1\% CO$_2$ (red, dashed) CESM2 experiments, separated by region: the North (A) and South (B) Pacific, the Indian Ocean (C), the North (D) and South (E) Atlantic, and the Southern Ocean (F). The Pacific and Atlantic are split at the equator. We use the same periods as in Figure 1, panels D and E. Each panel sets its own axis limits.}
    \label{fig:regional_pdfs}
\end{figure}

\begin{figure}
    \centering
    \includegraphics[width=\linewidth]{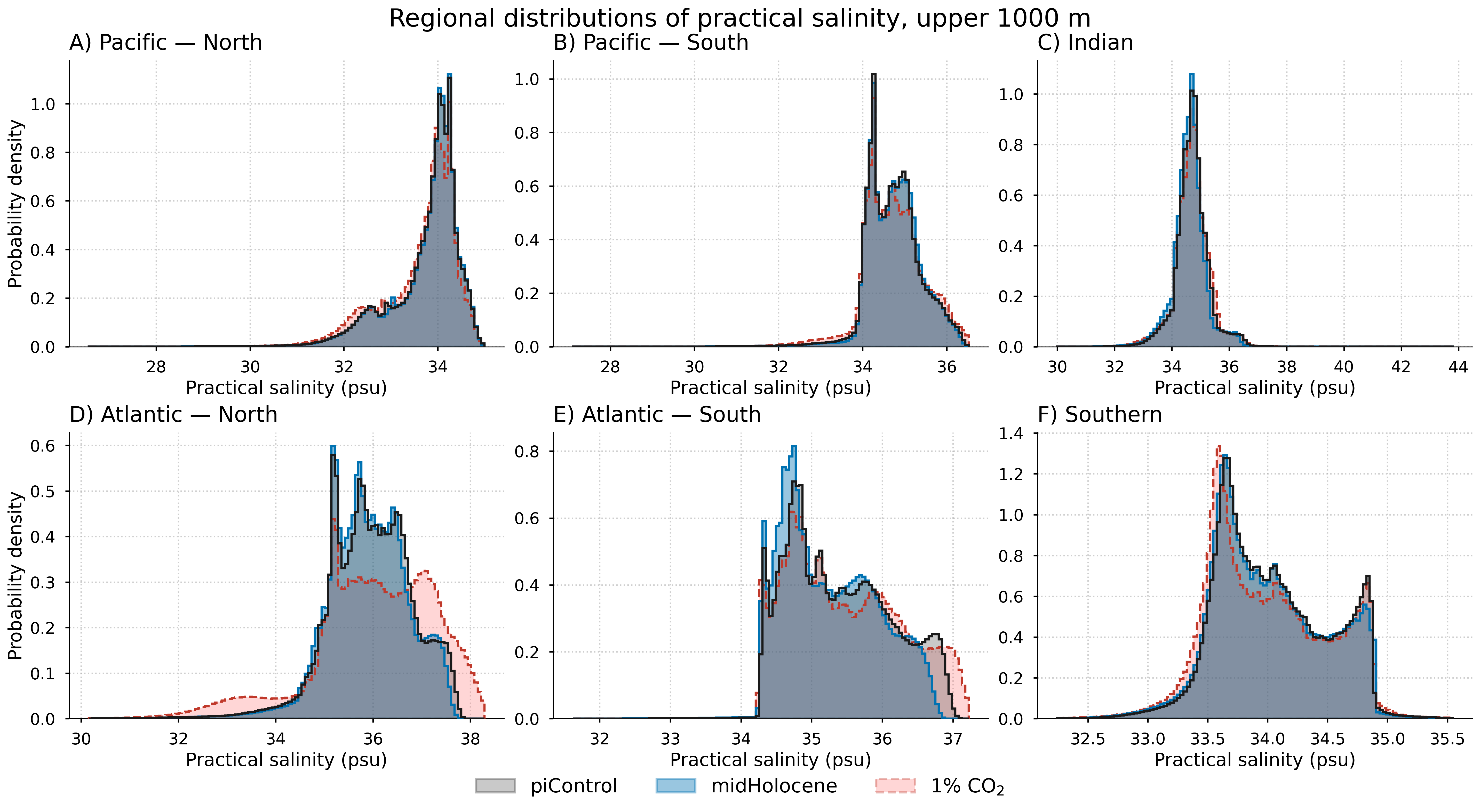}
    \caption{Distributions of salinity over the upper 1000\,m for the piControl (grey), midHolocene (blue), and 1\% CO$_2$ (red, dashed) CESM2 experiments, separated by region: the North (A) and South (B) Pacific, the Indian Ocean (C), the North (D) and South (E) Atlantic, and the Southern Ocean (F). The Pacific and Atlantic are split at the equator. We use the same periods as in Figure 1, panels D and E. Each panel sets its own axis limits.}
    \label{fig:regional_pdfs_so}
\end{figure}

\begin{figure}
    \centering
    \includegraphics[width=\linewidth]{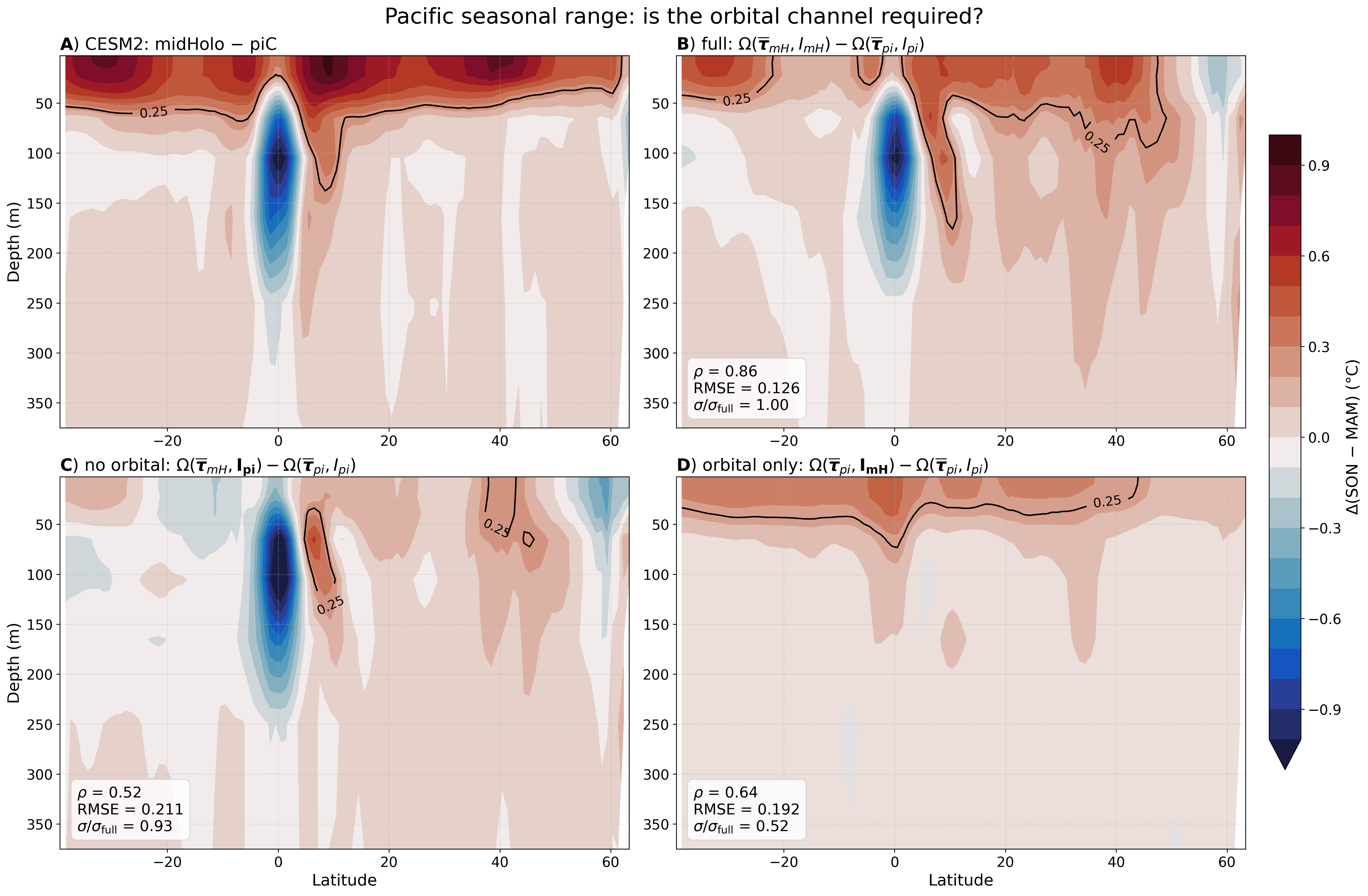}
    \caption{Depth profiles detailing the change in the seasonal range of potential temperature (SON minus MAM) between the midHolocene and piControl climates (Equation 7) in the Pacific Ocean, separating the contribution of the orbital forcing from that of the dynamic surface forcings. A: change in range between the two CESM2 experiments. For B-D we use the title to indicate the rollout parameters using abbreviated forms of the notation defined in Section 2.5.3, where $\overline{\boldsymbol{\tau}}$ denotes the dynamic surface boundary forcings and $I$ the insolation. In contrast to the other seasonal cycle figures, here we set the insolation independently of the boundary forcings. All three panels use $\mathcal{F}_{\mathrm{pi}}$, are initialized from $\boldsymbol{\Phi}_{\mathrm{mH}}^{[0]}$, and are differenced against a common control rollout, $\Omega(\mathcal{F}_{\mathrm{pi}},\boldsymbol{\Phi}_{\mathrm{pi}}^{[0]};\overline{\boldsymbol{\tau}}_{\mathrm{pi}},I_{\mathrm{pi}})$, in the same sense as A. B: both the dynamic forcings and the insolation set to midHolocene values. C: midHolocene dynamic forcings with the insolation held at piControl. D: midHolocene insolation with the dynamic forcings held at piControl. Because the dynamic forcings are taken from the CESM2 midHolocene simulation, they carry the orbital signal indirectly, and C is therefore not free of orbital influence. Black contours mark the $0.25~^\circ$C level. We evaluate the climatology over the final 50 years of each 100-year rollout. For each panel, we compute the correlation, $\rho$, and RMSE with respect to the CESM2 difference in A, and the amplitude ratio $\sigma/\sigma_{\mathrm{full}}$ with respect to the full response in B.}
    \label{fig:seasonal_cycle_ablation}
\end{figure}

\begin{figure}
    \centering
    \includegraphics[width=\linewidth]{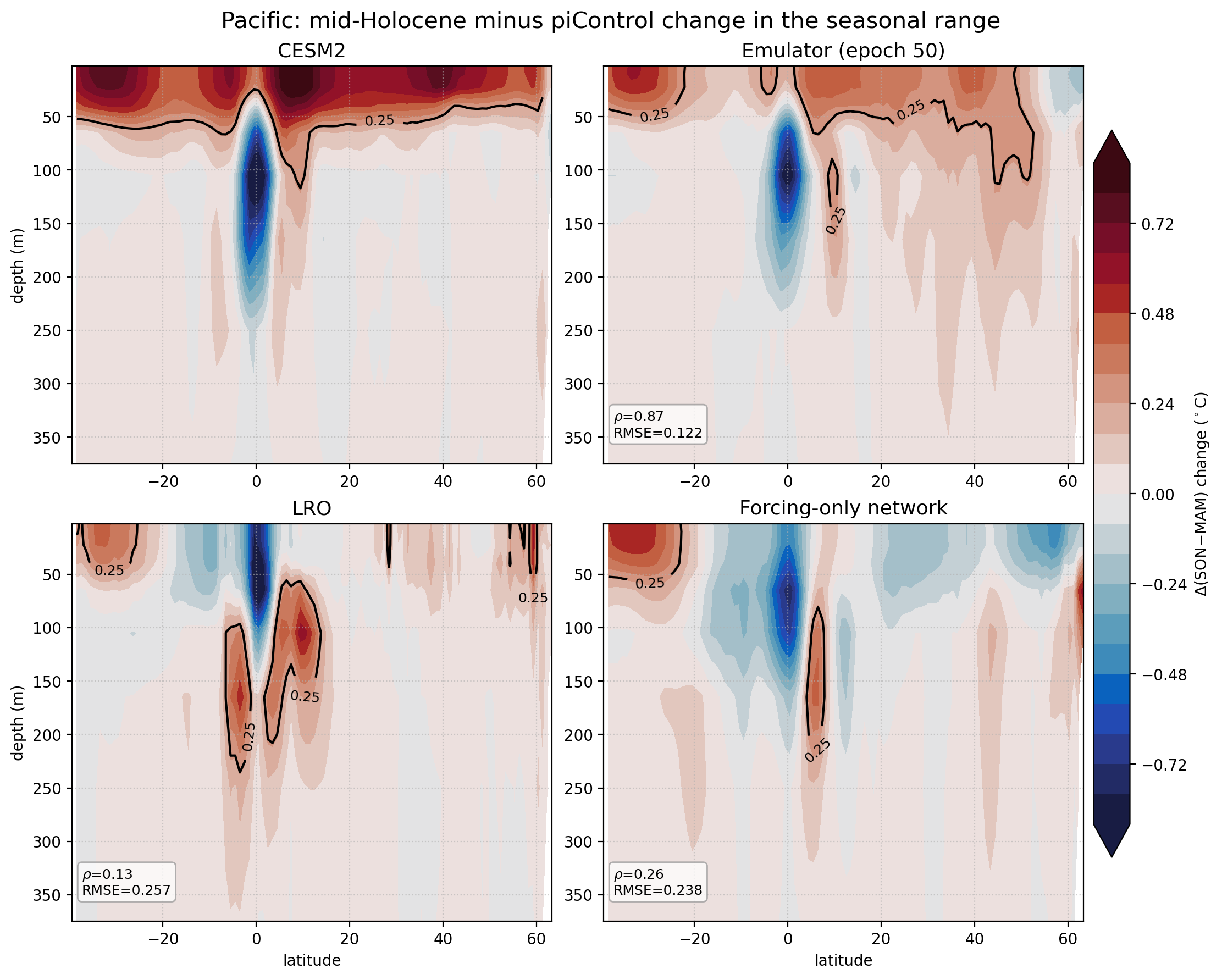}
    \caption{Depth profiles detailing the change in the seasonal range of potential temperature (SON minus MAM) between the midHolocene and piControl climates (Equation 7) in the Pacific Ocean, comparing the emulator against the two baselines. A: change in range between the two CESM2 experiments. For B-D: we use the title to indicate the model shown in each panel, and each panel gives the same difference as A, computed between that model driven by midHolocene and by piControl boundary forcings. B: $\mathcal{F}_{\mathrm{pi}}$, at the checkpoint reported throughout, using initial conditions and insolation from the same experiment as the boundary forcings. C: the local linear regression operator. D: the forcing-only network. Both baselines are fit on piControl data and applied to each set of boundary forcings (Section S2). To help with comparing panels, we include a contour where $\Delta\left(\mathrm{SON} - \mathrm{MAM} \right)=0.25^\circ \mathrm{C}$. We evaluate the climatology over the final 50 years of each 100-year rollout. For each panel, we compute the correlation, $\rho$, and RMSE with respect to the CESM2 difference in A.}
    \label{fig:seasonal_cycle_baselines}
\end{figure}

\begin{figure}
    \centering
    \includegraphics[width=\linewidth]{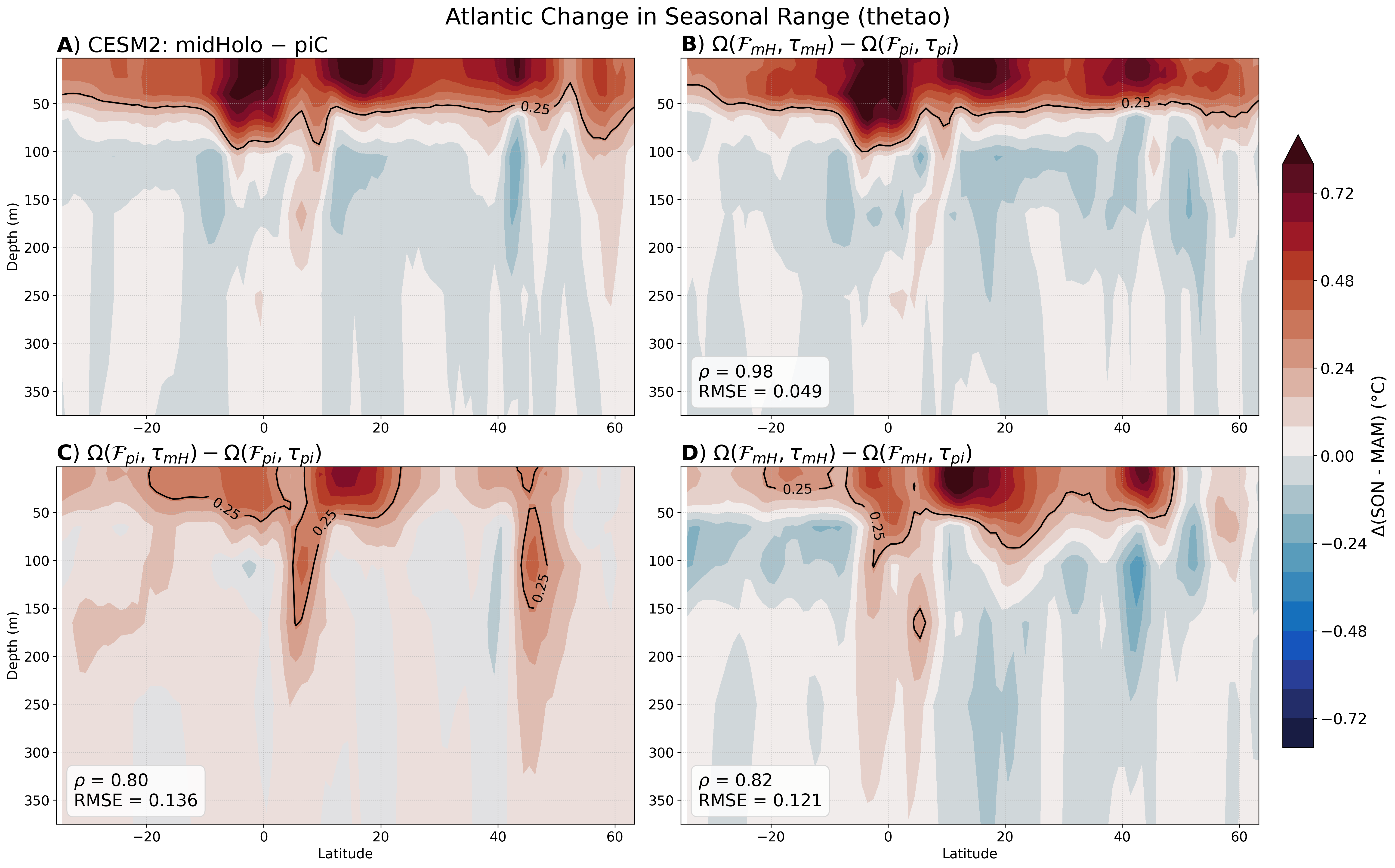}
    \caption{Depth profiles detailing the change in the seasonal range of potential temperature (SON minus MAM) between the midHolocene and piControl climates (Equation 7) in the Atlantic Ocean. A: change in range between the two CESM2 experiments. For B–D: we use the title to indicate the networks and rollout parameters using  abbreviated forms of the  notation defined in Section 2.5.3. For each rollout, we use initial conditions and insolation from the same experiment listed for the boundary forcings (e.g., where we state $\boldsymbol{\tau}_{\mathrm{pi}}$, we also use $\boldsymbol{\Phi}_{\mathrm{pi}}^{[0]}$ and $I_{\mathrm{pi}}$).   B: difference between the climatology  of $\mathcal{F}_{\mathrm{mH}}$ rolled out for the midHolocene climate and  $\mathcal{F}_{\mathrm{pi}}$ rolled out for the piControl climate. C: difference between $\mathcal{F}_{\mathrm{pi}}$ rolled out for the midHolocene and the piControl climate. D:  same difference  for $\mathcal{F}_{\mathrm{mH}}$. We evaluate the climatology over the final 50 years of each 100-year rollout. For each panel, we compute the correlation, $\rho$, and RMSE with respect to the CESM2 difference in A.}
    \label{fig:seasonal_cycle_1}
\end{figure}

\begin{figure}
    \centering
    \includegraphics[width=\linewidth]{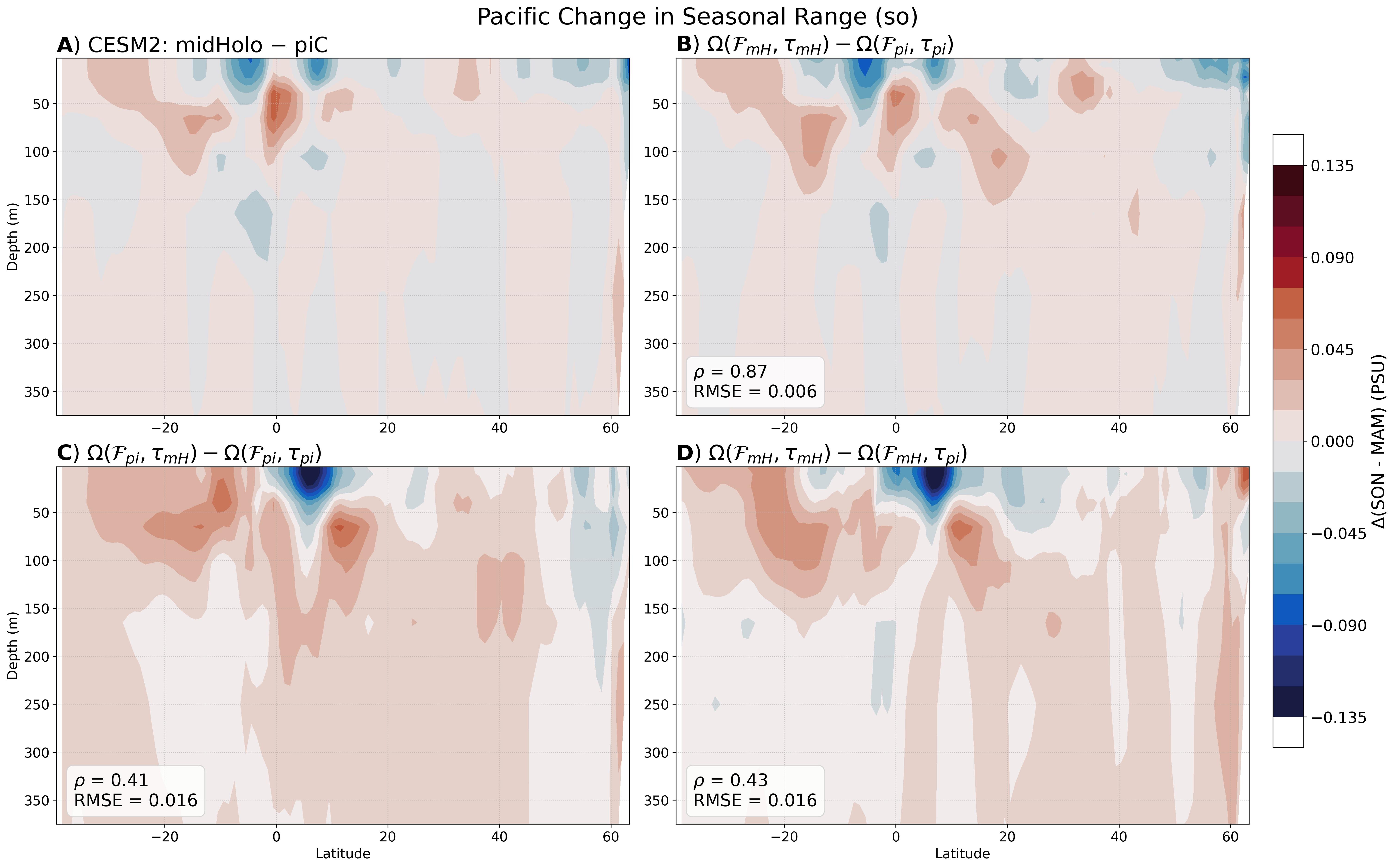}
    \caption{Depth profiles detailing the change in the seasonal range of salinity (SON minus MAM) between the midHolocene and piControl climates (Equation 7) in the Pacific Ocean. A: change in range between the two CESM2 experiments. For B–D: we use the title to indicate the networks and rollout parameters using  abbreviated forms of the  notation defined in Section 2.5.3. For each rollout, we use initial conditions and insolation from the same experiment listed for the boundary forcings (e.g., where we state $\boldsymbol{\tau}_{\mathrm{pi}}$, we also use $\boldsymbol{\Phi}_{\mathrm{pi}}^{[0]}$ and $I_{\mathrm{pi}}$).   B: difference between the climatology  of $\mathcal{F}_{\mathrm{mH}}$ rolled out for the midHolocene climate and  $\mathcal{F}_{\mathrm{pi}}$ rolled out for the piControl climate. C: difference between $\mathcal{F}_{\mathrm{pi}}$ rolled out for the midHolocene and the piControl climate. D:  same difference  for $\mathcal{F}_{\mathrm{mH}}$. We evaluate the climatology over the final 50 years of each 100-year rollout. For each panel, we compute the correlation, $\rho$, and RMSE with respect to the CESM2 difference in A.}
    \label{fig:seasonal_cycle_2}
\end{figure}

\begin{figure}
    \centering
    \includegraphics[width=\linewidth]{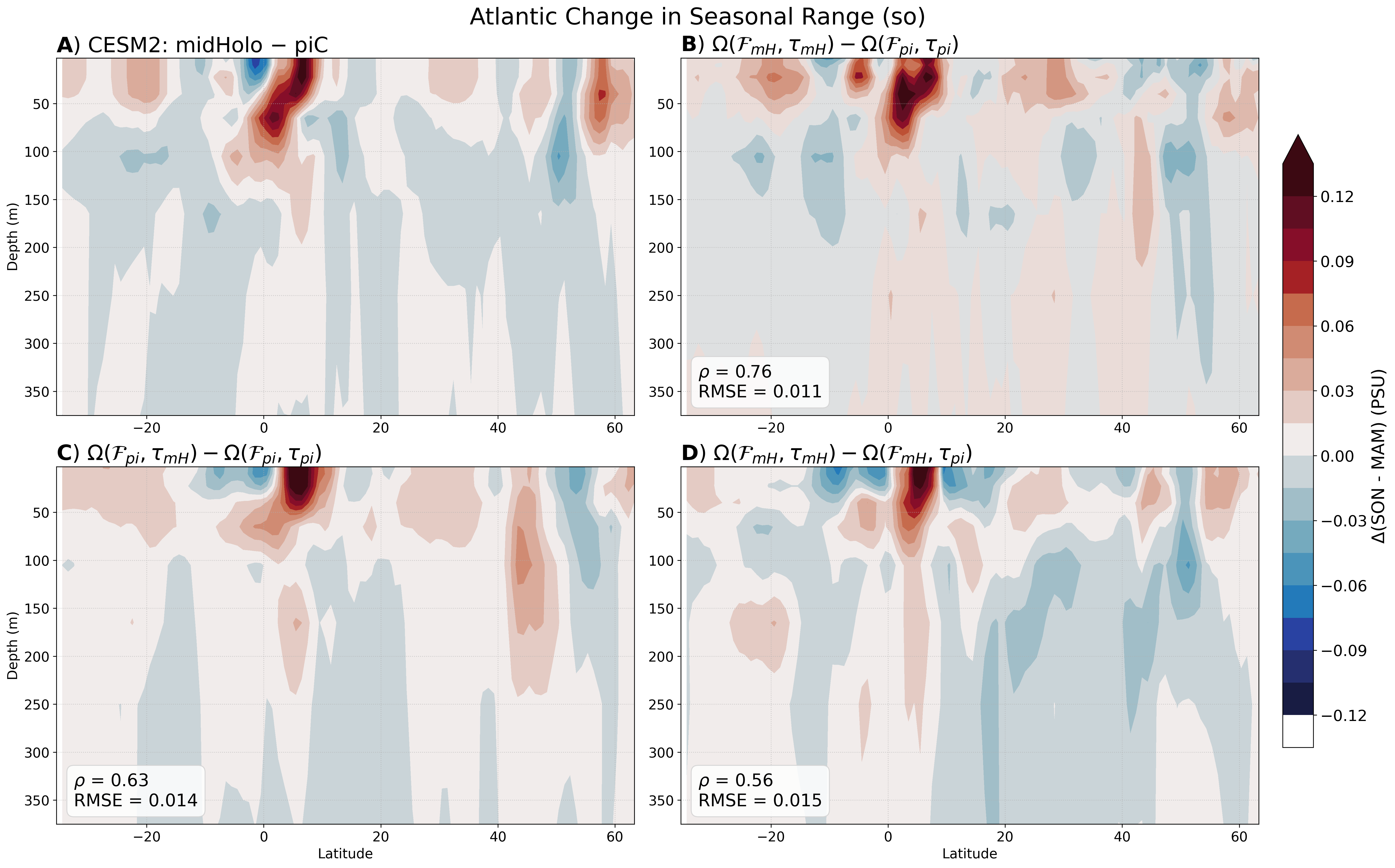}
    \caption{Depth profiles detailing the change in the seasonal range of salinity (SON minus MAM) between the midHolocene and piControl climates (Equation 7) in the Atlantic Ocean. A: change in range between the two CESM2 experiments. For B–D: we use the title to indicate the networks and rollout parameters using  abbreviated forms of the  notation defined in Section 2.5.3. For each rollout, we use initial conditions and insolation from the same experiment listed for the boundary forcings (e.g., where we state $\boldsymbol{\tau}_{\mathrm{pi}}$, we also use $\boldsymbol{\Phi}_{\mathrm{pi}}^{[0]}$ and $I_{\mathrm{pi}}$).   B: difference between the climatology  of $\mathcal{F}_{\mathrm{mH}}$ rolled out for the midHolocene climate and  $\mathcal{F}_{\mathrm{pi}}$ rolled out for the piControl climate. C: difference between $\mathcal{F}_{\mathrm{pi}}$ rolled out for the midHolocene and the piControl climate. D:  same difference  for $\mathcal{F}_{\mathrm{mH}}$. We evaluate the climatology over the final 50 years of each 100-year rollout. For each panel, we compute the correlation, $\rho$, and RMSE with respect to the CESM2 difference in A.}
    \label{fig:seasonal_cycle_3}
\end{figure}

\begin{figure}
    \centering
    \includegraphics[width=\linewidth]{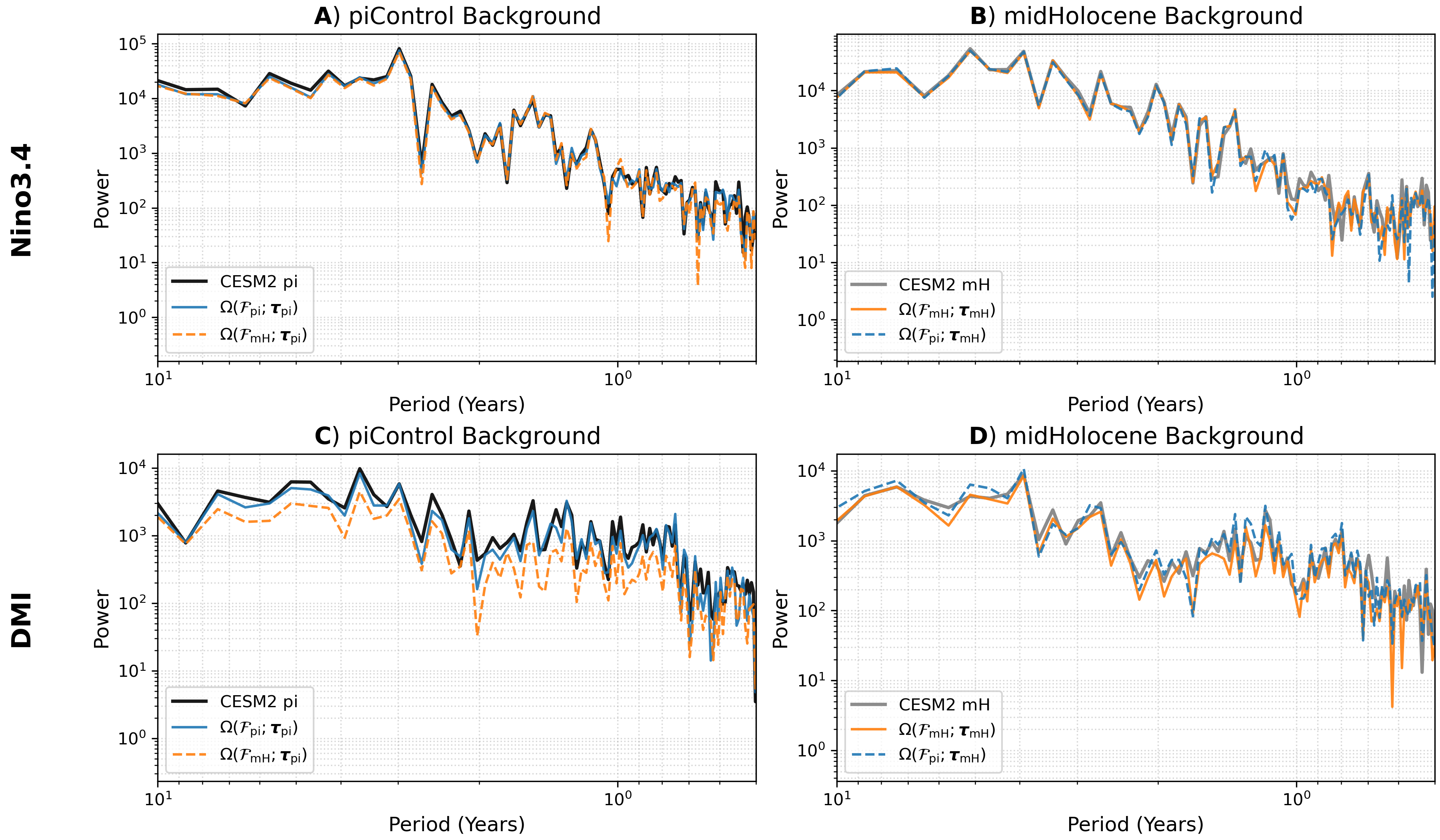}
    \caption{Comparison of the temporal power spectrum for the timeseries of Nino3.4 (A and B) and the DMI (C and D) for both the piControl (A and C) and midHolocene (B and D) experiments. Each panel shows the range of the spectrum with a return period between 15 years and 3 months. These spectra are computed from a 100-year rollout on the respective climates using unseen data for all emulators. For legend labels, we indicate the networks and rollout parameters using abbreviated forms of the notation defined in Section 2.5.3. For each rollout, we use initial conditions and insolation from the same experiment listed for the boundary forcings (e.g., where we state $\boldsymbol{\tau}_{\mathrm{pi}}$, we also use $\boldsymbol{\Phi}_{\mathrm{pi}}^{[0]}$ and $I_{\mathrm{pi}}$). }
    \label{fig:power spectra_nino}
\end{figure}

\begin{figure}
    \centering
    \includegraphics[width=\linewidth]{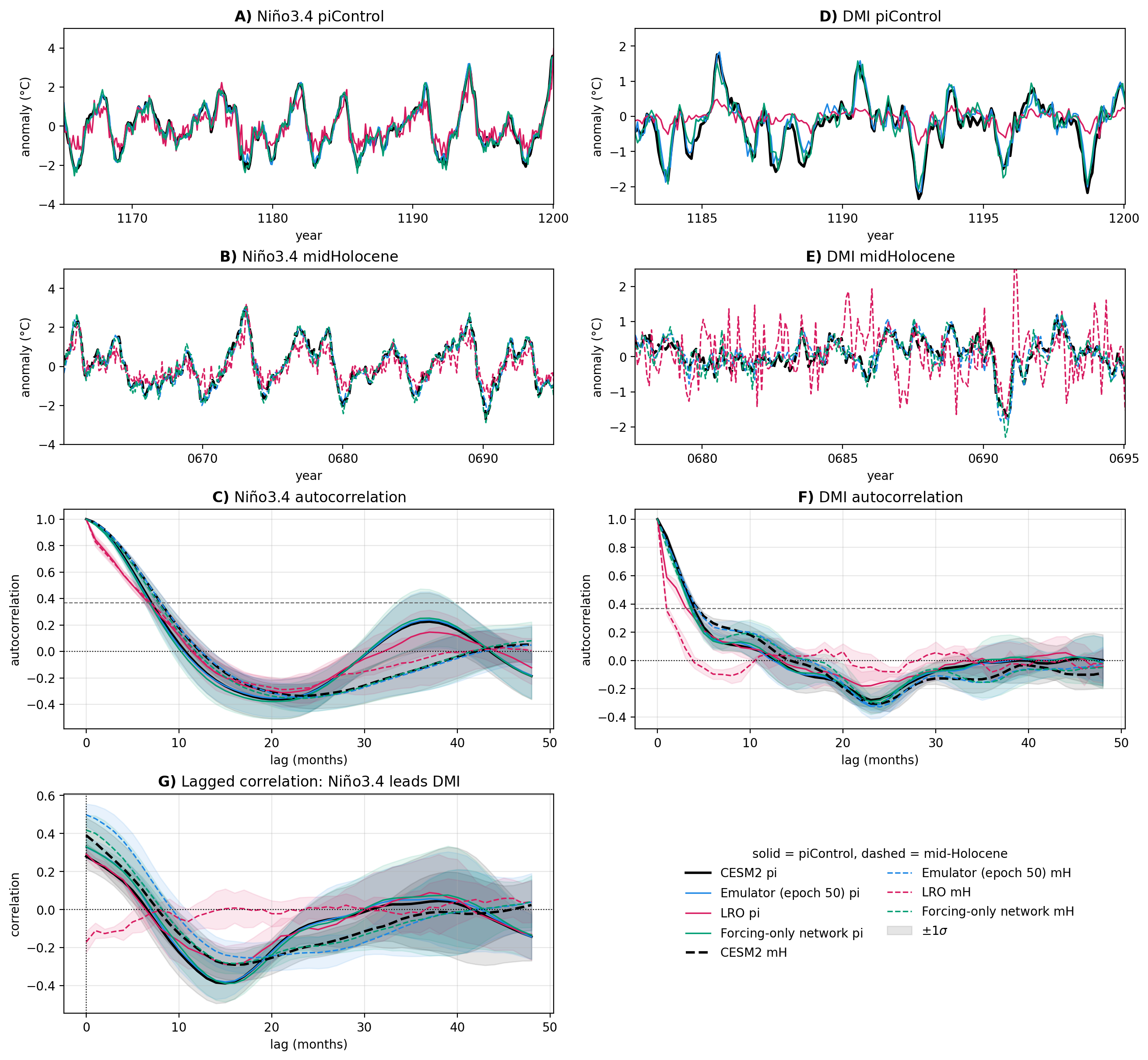}
    \caption{Comparison of dominant modes of ocean variability across the piControl and midHolocene experiments, including the two baselines. A-C: the timeseries of Ni\~{n}o3.4 in the piControl climate (A), the midHolocene climate (B), and the lagged autocorrelation (C). D-F: similar results for the DMI, a measure of the Indian Ocean Dipole. G: the lagged correlation between Ni\~{n}o3.4 and the DMI. In each panel we compare CESM2 and $\mathcal{F}_{\mathrm{pi}}$ to the local linear regression operator and the forcing-only network described in Section \ref{appendix:lro}, both applied to the boundary forcings of each climate. Shading in C, F, and G gives the $\pm 1\sigma$ spread across subsets of the data.}
    \label{fig:modes_variability_baselines}
\end{figure}

\begin{figure}
    \centering
    \includegraphics[width=\linewidth]{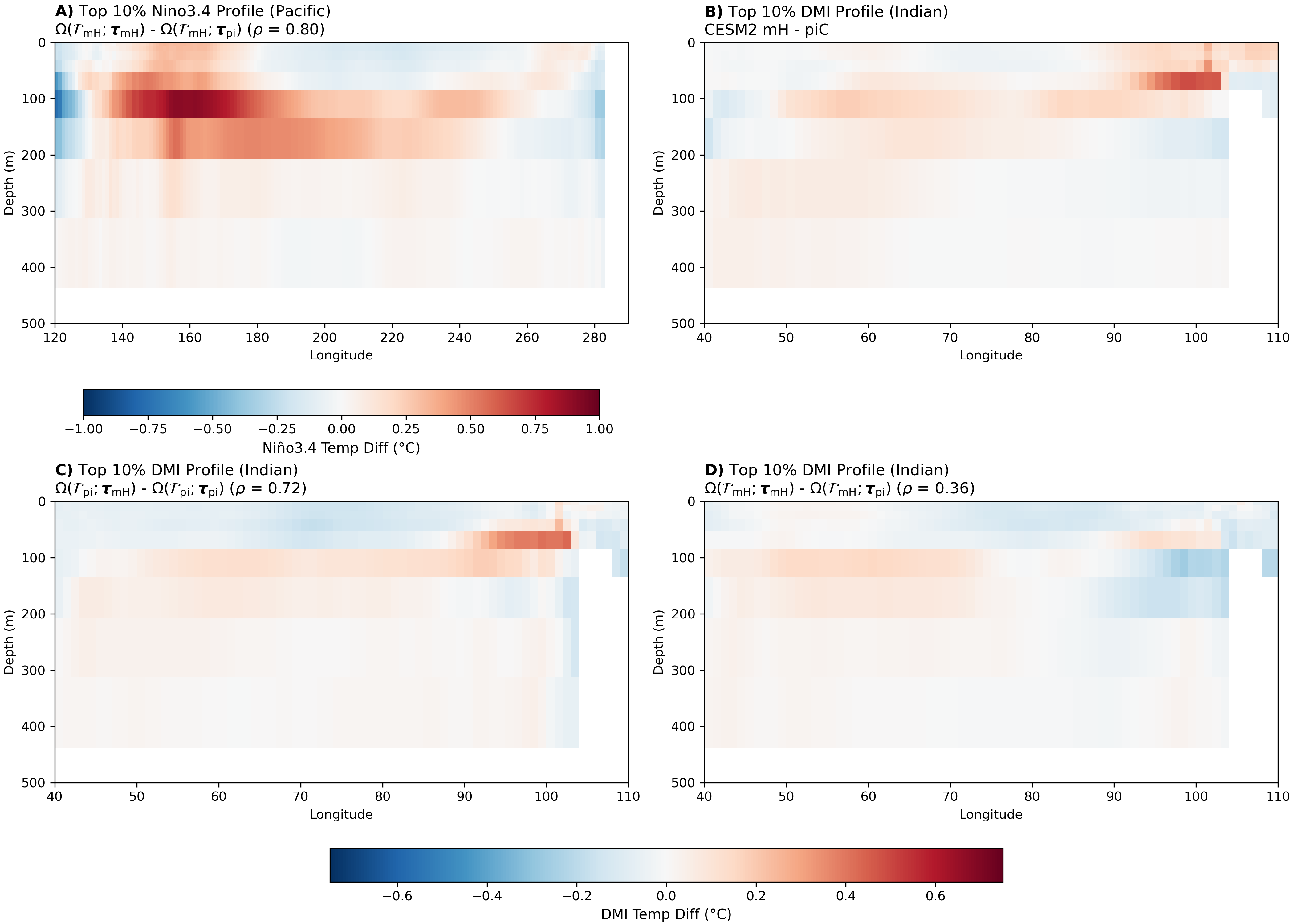}
    \caption{A:  the upper ocean structural changes between the piControl and midHolocene, highlighting the difference in anomaly patterns between the midHolocene and piControl experiments for $\mathcal{F}_{\mathrm{mH}}$. We include the spatial correlation between A and the CESM2 difference in Figure 4 panel I .  B-D: the upper ocean structural changes in the Indian Ocean for the CESM2 data, $\mathcal{F}_{\mathrm{pi}}$, and $\mathcal{F}_{\mathrm{mH}}$ in panels B, C and D respectively. In the title of panels C and D, we include the spatial correlation to the CESM2 difference in panel B.}
    \label{fig:subsurface_profiles}
\end{figure}

\begin{figure}
    \centering
    \includegraphics[width=\linewidth]{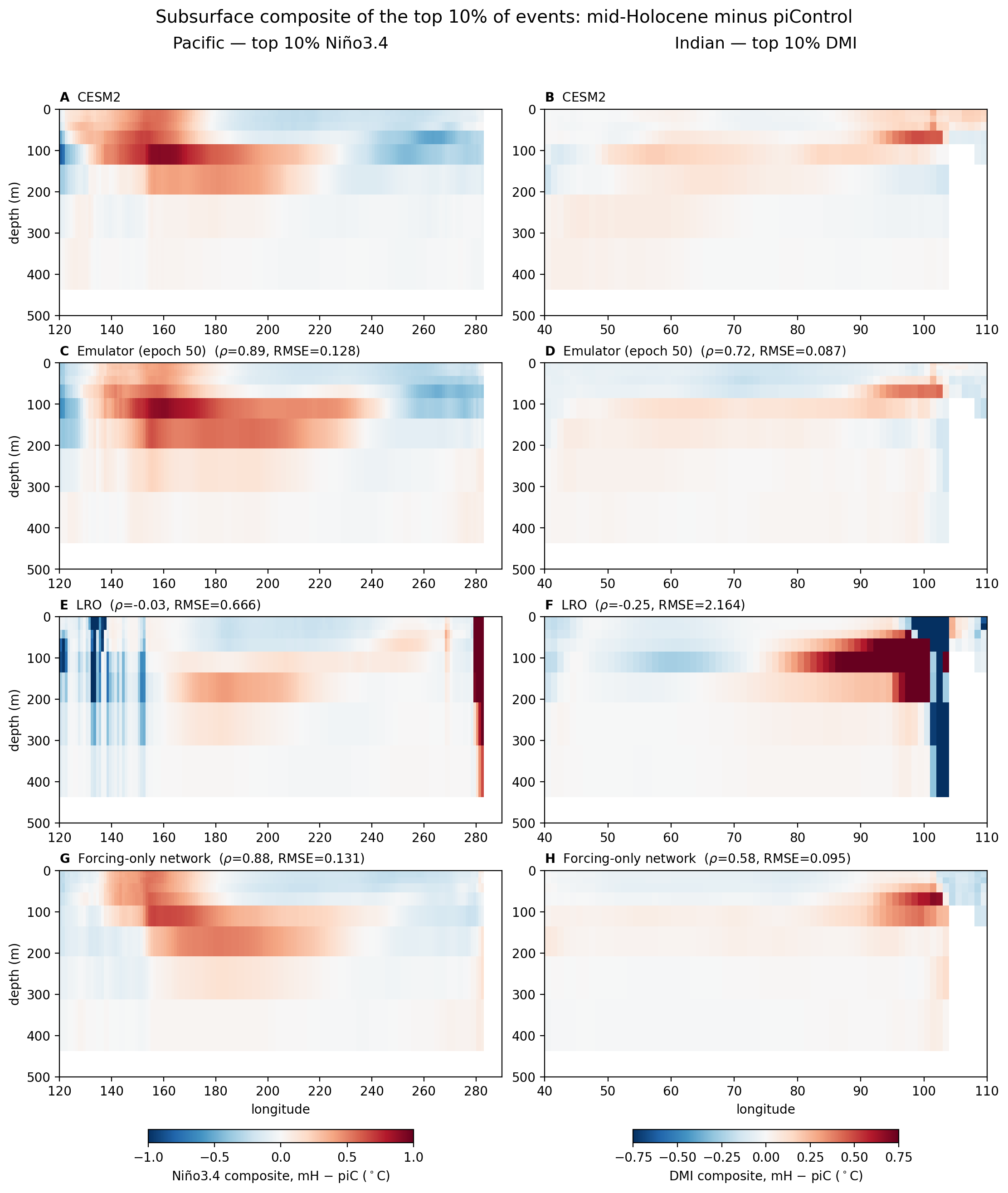}
    \caption{Comparison of the subsurface structure of the strongest events between the piControl and midHolocene experiments. Following Section 3.2 of the main text, we take the ocean anomalies for the 10\% of states with the largest index value, average them into a mean profile, and difference these profiles between the two experiments. The left column (A, C, E, and G) shows the Pacific composite for Ni\~{n}o3.4 and the right column (B, D, F, and H) the Indian Ocean composite for the DMI. A and B: the difference between the two CESM2 experiments. C and D: the same difference for $\mathcal{F}_{\mathrm{pi}}$. E and F: the local linear regression operator. G and H: the forcing-only network. Both baselines are described in Section \ref{appendix:lro} and applied to the boundary forcings of each climate. In the title of each panel we include the correlation, $\rho$, and RMSE with respect to the CESM2 difference in A and B.}
    \label{fig:modes_subsurface}
\end{figure}

\begin{figure}
    \centering
    \includegraphics[width=\linewidth]{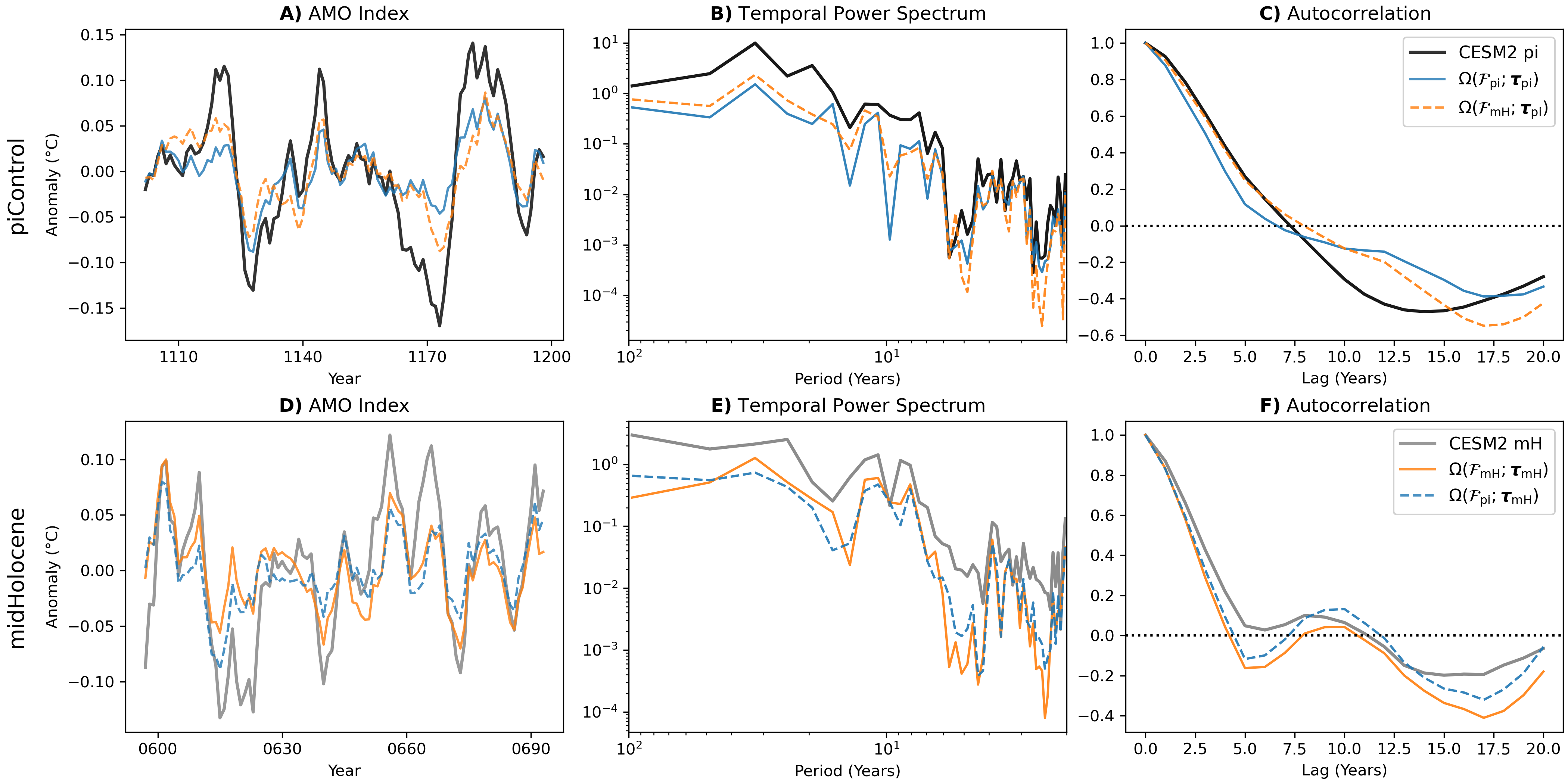}
    \caption{Comparison of skill in reproducing the Atlantic Multidecadal Oscillation. A and D: 5-year running mean of the timeseries of the AMO in the piControl climate, A, the midHolocene climate, D. B and E: temporal power spectrum of the AMO timeseries. C and F: the autocorrelation of the AMO timeseries. For legend labels, we indicate the networks and rollout parameters using abbreviated forms of the notation defined in Section 2.5.3. For each rollout, we use initial conditions and insolation from the same experiment listed for the boundary forcings (e.g., where we state $\boldsymbol{\tau}_{\mathrm{pi}}$, we also use $\boldsymbol{\Phi}_{\mathrm{pi}}^{[0]}$ and $I_{\mathrm{pi}}$).}
    \label{fig:AMO_Panels}
\end{figure}

\begin{figure}
    \centering
    \includegraphics[width=\linewidth]{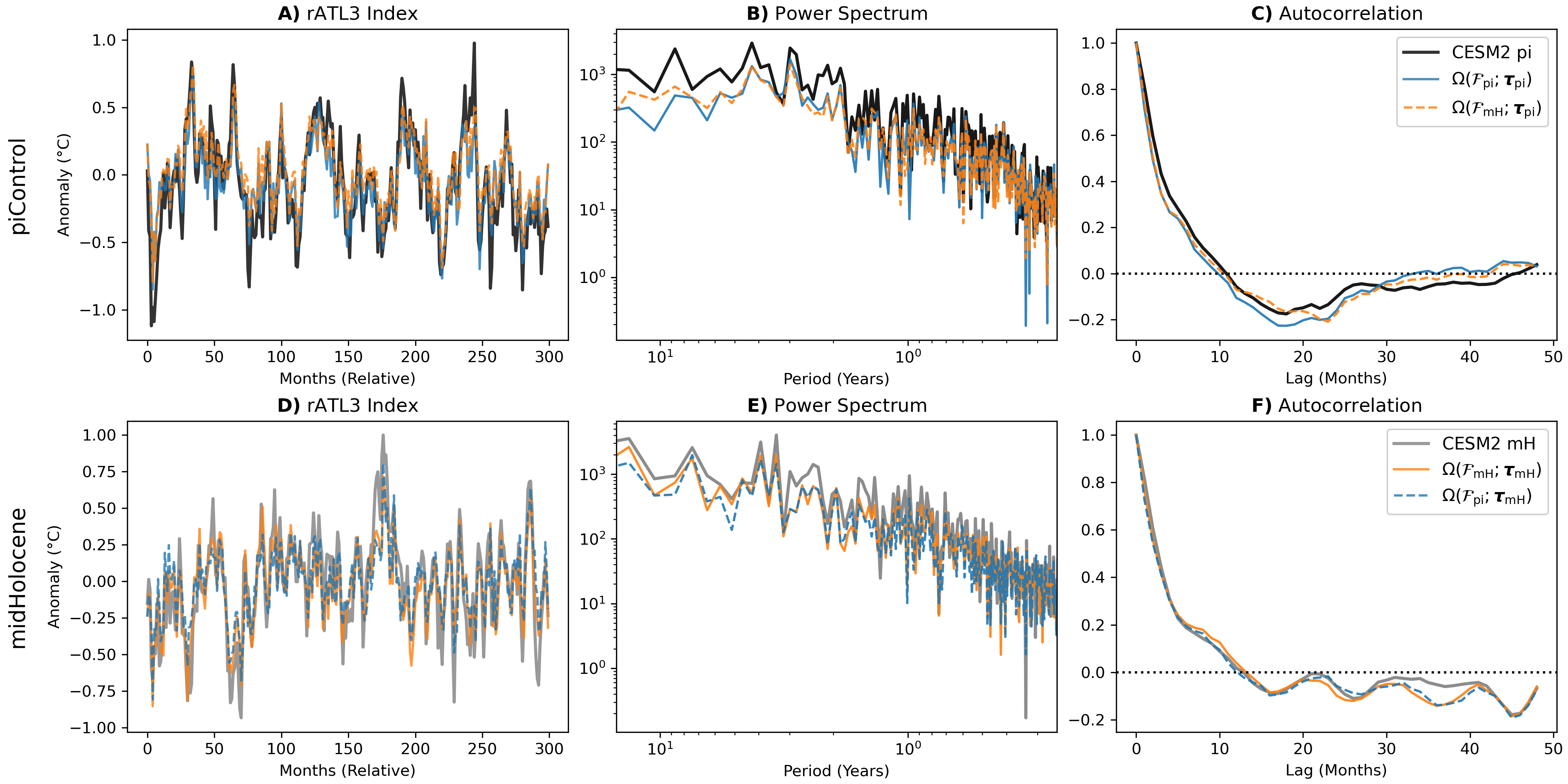}
    \caption{Comparison of skill in reproducing the relative Atlantic Nino Index (rATL3). The rATL3 is computed as the difference between the surface temperature anomaly over $20-0^\circ \mathrm{W}$ and $3^\circ \mathrm{S} -3^\circ \mathrm{N}$ and the global anomaly. A and D:  the 5-year running mean of the timeseries of the rATL3 in the piControl climate, A, the midHolocene climate, D. B and E: temporal power spectrum of the rATL3 timeseries.  C and F show the autocorrelation of the rATL3 timeseries. For legend labels, we indicate the networks and rollout parameters using abbreviated forms of the notation defined in Section 2.5.3. For each rollout, we use initial conditions and insolation from the same experiment listed for the boundary forcings (e.g., where we state $\boldsymbol{\tau}_{\mathrm{pi}}$, we also use $\boldsymbol{\Phi}_{\mathrm{pi}}^{[0]}$ and $I_{\mathrm{pi}}$).}
    \label{fig:ATL3 panels}
\end{figure}

\begin{figure}
    \centering
    \includegraphics[width=\linewidth]{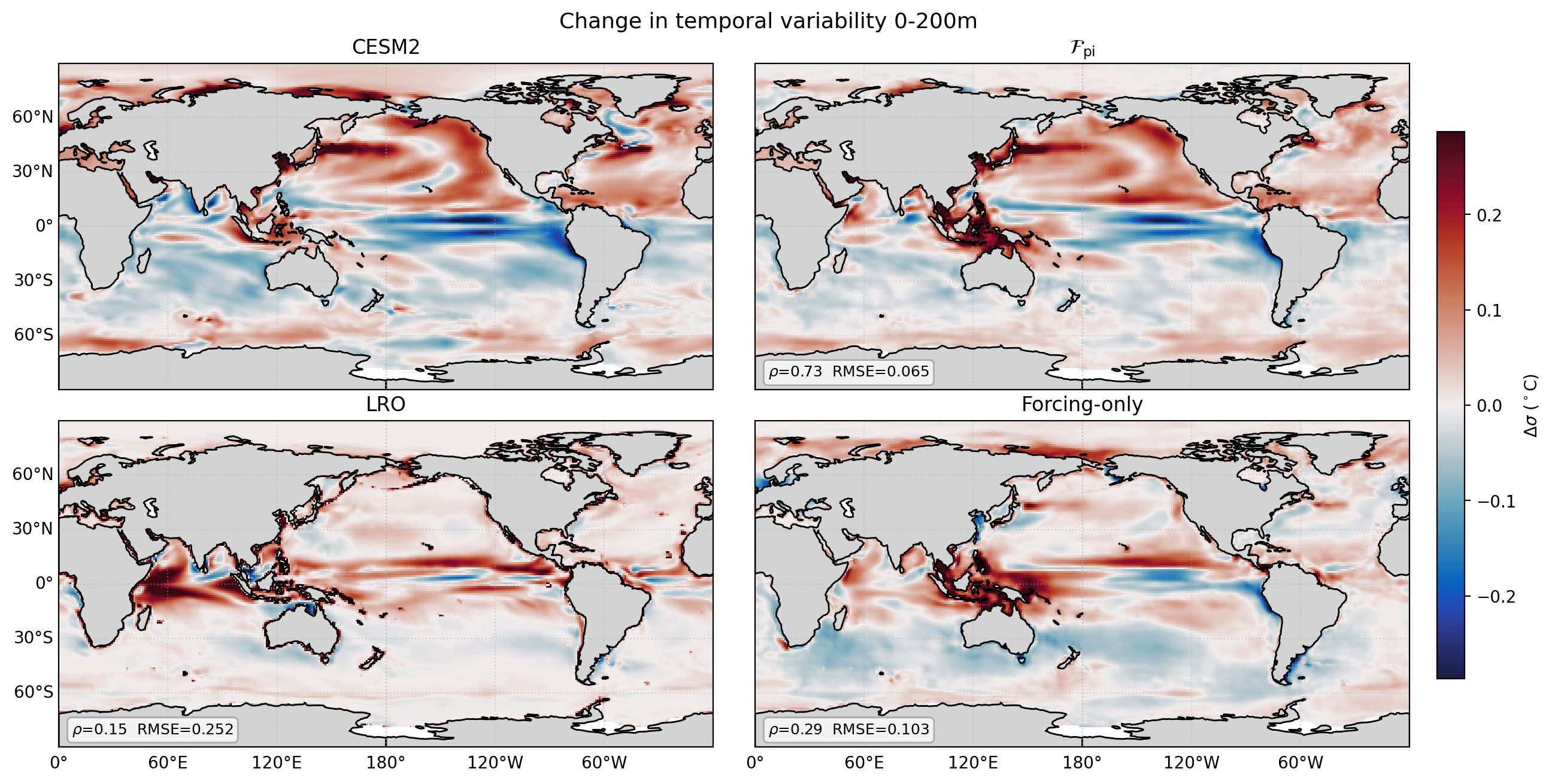}
    \caption{Comparison of the change in the potential temperature variability over the upper 200m. Here we compare the emulator from the main text, $\mathcal{F}_{\mathrm{pi}}$, to the local linear regression operator and the forcing-only network described in Section \ref{appendix:lro}. We compute the standard deviation of the monthly state over the final 50 years of each 100-year rollout. A: true difference between the two CESM2 experiments. B: the same difference for $\mathcal{F}_{\mathrm{pi}}$ rolled out with midHolocene, $\boldsymbol{\tau}_{\mathrm{mH}}$, and piControl, $\boldsymbol{\tau}_{\mathrm{pi}}$, boundary forcings. For each rollout, we use initial conditions and insolation from the same experiment listed for the boundary forcings (e.g., where we state $\boldsymbol{\tau}_{\mathrm{pi}}$, we also use $\boldsymbol{\Phi}_{\mathrm{pi}}^{[0]}$ and $\operatorname{I}_{\mathrm{pi}}$). C: the same difference for the linear regression operator. D: the same difference for the forcing-only network. Both baselines are fit on piControl data and applied to the boundary forcings of each climate. For the LRO we exclude the small number of points at which the change exceeds $2.25\,^\circ$C. For each panel, we compute the correlation, $\rho$, and RMSE with respect to the CESM2 difference in A.}
    \label{fig:variability_maps_baseline}
\end{figure}

\begin{figure}
    \centering
    \includegraphics[width=\linewidth]{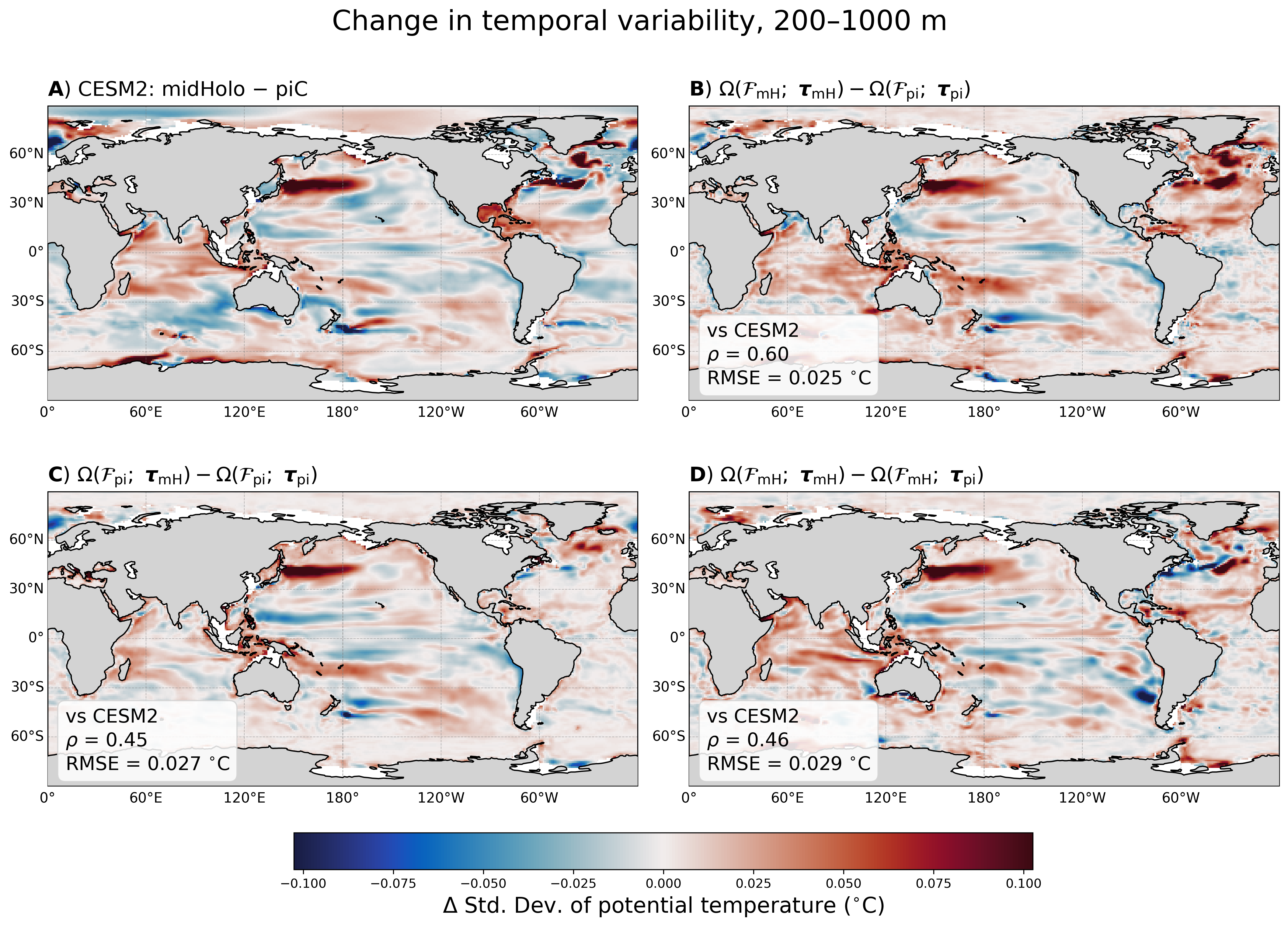}
    \caption{Comparison of the change in the potential temperature variability over 200-1000m. A: true difference between the two CESM2 experiments. For B-D, we use the title to indicate the networks and rollout parameters, using abbreviated notation defined in Section 2.5.3. For each rollout, we use initial conditions and insolation from the same experiment listed for the boundary forcings (e.g., where we state $\boldsymbol{\tau}_{\mathrm{pi}}$, we also use $\boldsymbol{\Phi}_{\mathrm{pi}}^{[0]}$ and $\operatorname{I}_{\mathrm{pi}}$).  B: difference between variability when $\mathcal{F}_{\mathrm{mH}}$ is rolled out for the midHolocene climate and $\mathcal{F}_{\mathrm{pi}}$ is rolled out for the piControl climate. C: difference for $\mathcal{F}_{\mathrm{pi}}$ rolled out for the midHolocene and piControl climate. D: the same difference  for $\mathcal{F}_{\mathrm{mH}}$. We evaluate the temporal variability over the final 50 years of each 100-year rollout. For each panel, we compute the correlation, $\rho$, and RMSE with respect to the CESM2 difference in A.}
    \label{fig:spatial_variability_200_1000}
\end{figure}

\begin{figure}
    \centering
    \includegraphics[width=\linewidth]{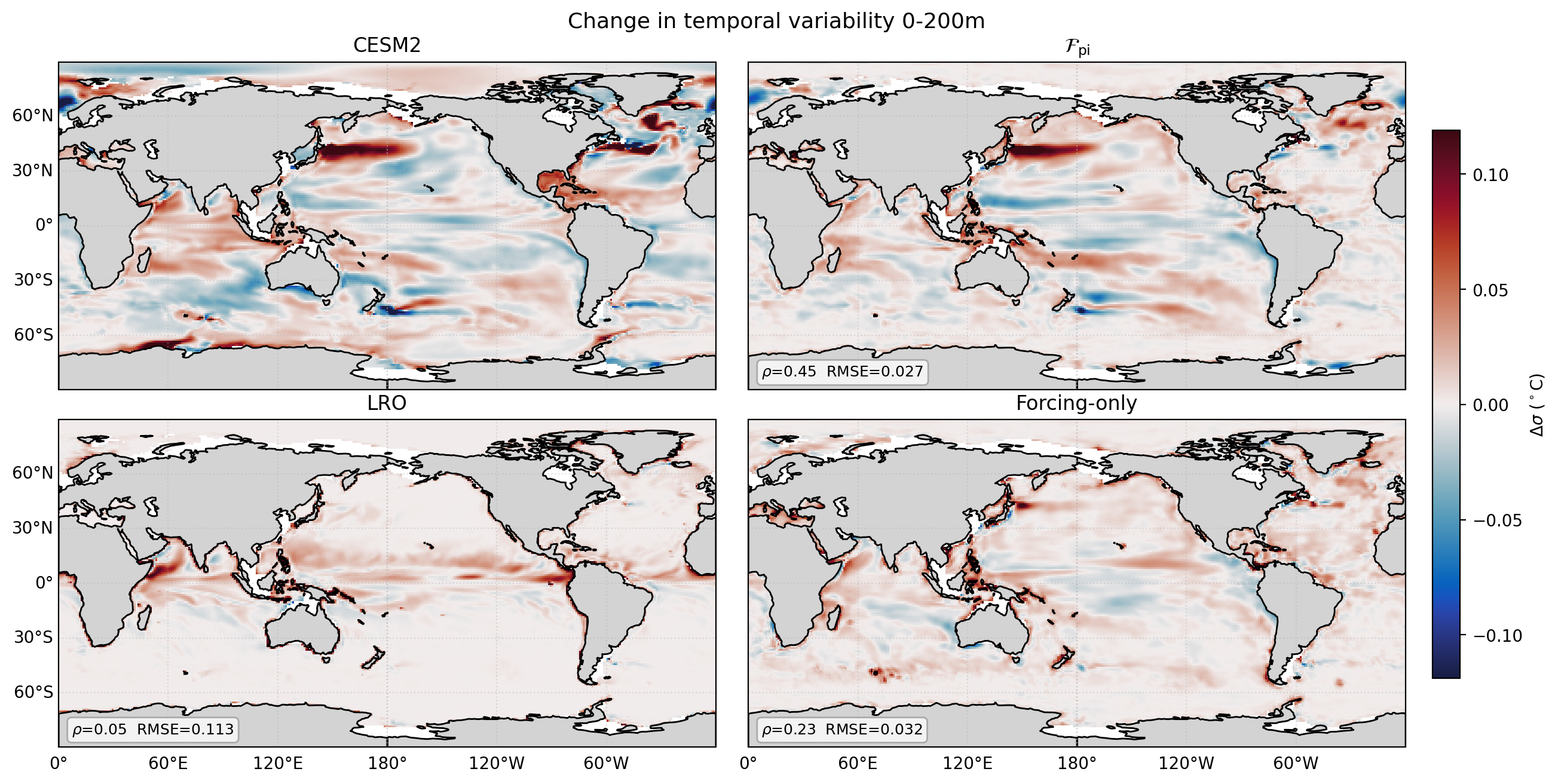}
    \caption{Comparison of the change in the potential temperature variability over 200-1000m. Here we compare the emulator from the main text, $\mathcal{F}_{\mathrm{pi}}$, to the local linear regression operator and the forcing-only network described in Section \ref{appendix:lro}. We compute the standard deviation of the monthly state over the final 50 years of each 100-year rollout. A: true difference between the two CESM2 experiments. B: the same difference for $\mathcal{F}_{\mathrm{pi}}$ rolled out with midHolocene, $\boldsymbol{\tau}_{\mathrm{mH}}$, and piControl, $\boldsymbol{\tau}_{\mathrm{pi}}$, boundary forcings. For each rollout, we use initial conditions and insolation from the same experiment listed for the boundary forcings (e.g., where we state $\boldsymbol{\tau}_{\mathrm{pi}}$, we also use $\boldsymbol{\Phi}_{\mathrm{pi}}^{[0]}$ and $\operatorname{I}_{\mathrm{pi}}$). C: the same difference for the linear regression operator. D: the same difference for the forcing-only network. Both baselines are fit on piControl data and applied to the boundary forcings of each climate. For the LRO we exclude the small number of points at which the change exceeds $2.25\,^\circ$C. For each panel, we compute the correlation, $\rho$, and RMSE with respect to the CESM2 difference in A.}
    \label{fig:variability_maps_baseline_200_1000}
\end{figure}

\begin{figure}
    \centering
    \includegraphics[width=\linewidth]{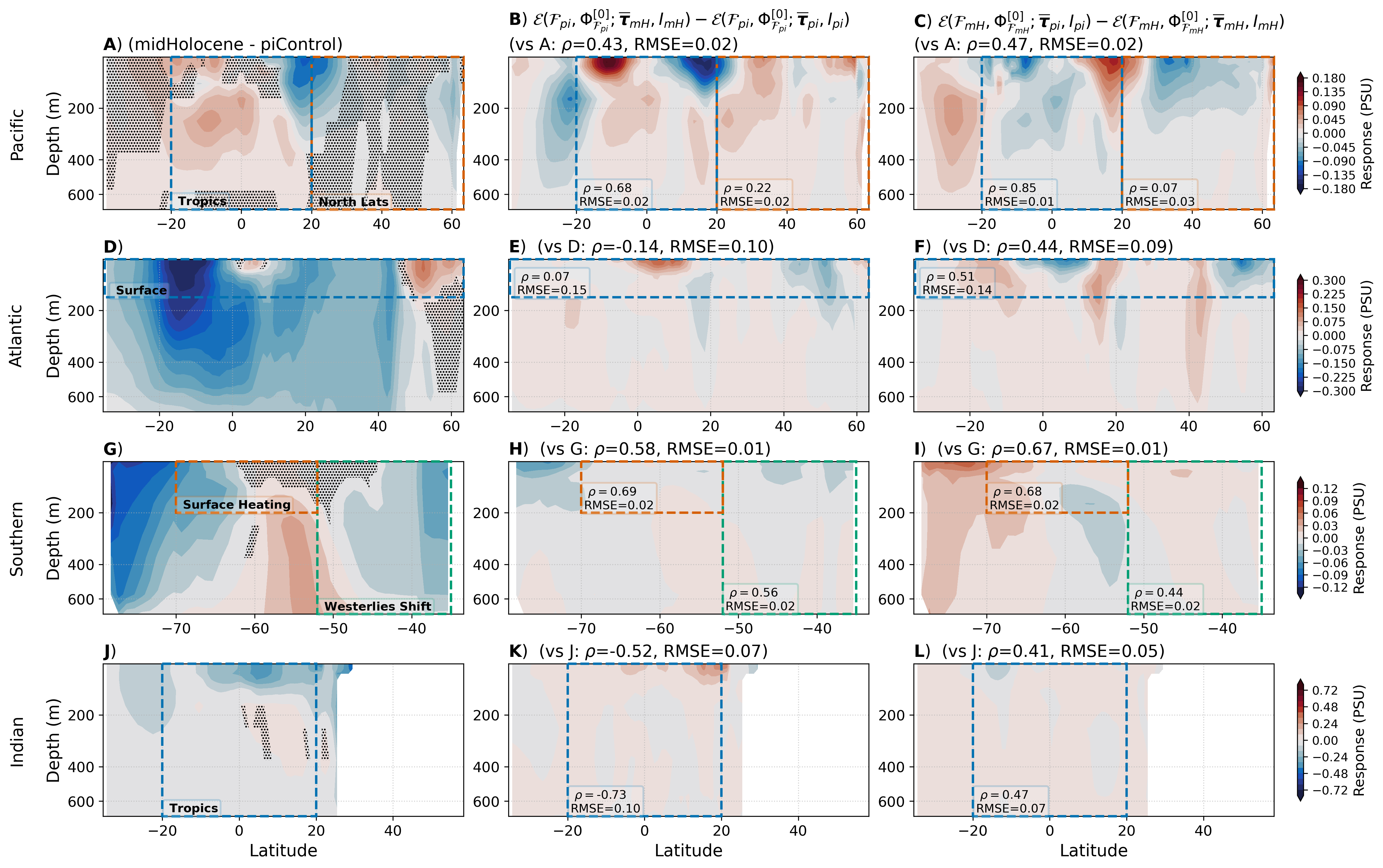}
    \caption{Comparison of the zonally averaged salinity response in each ocean basin when perturbing an emulator using climatological forcings taken from midHolocene/piControl CESM2 data. A, D, G, and J: difference between the CESM2 midHolocene and piControl experiments as an estimate of the true response. B, E, H, and K: response of the emulator trained on piControl data when perturbed with midHolocene boundary forcings, $R(\mathcal{F}_{\mathrm{pi}})$. C, F, I, and L: response of the emulator trained on midHolocene data when perturbed with piControl boundary forcings, $R(\mathcal{F}_{\mathrm{mH}})$. We show responses in the Pacific (A-C), Atlantic (D-F), Southern (G-I), and Indian Ocean (J-L). For each emulator response we include the correlation, $\rho$,  and RMSE with respect to the true change in the respective basin, A,D,G, and J. Note that for $R(\mathcal{F}_{\mathrm{mH}})$, we flip the sign of the true response before computing the correlation and RMSE. We additionally highlight specific regions in each basin, and again include the correlation and RMSE over each region with respect to the change between CESM2 experiments.  We stipple the CESM2 panels to mark regions where the reference response is not significant relative to internal variability (two-tailed Student's $t$-test at the 95\% level, with internal variability estimated from ten 25-year chunks of each experiment).}
    \label{fig:Full_Responses_All_Basins_salinity}
\end{figure}

\begin{figure}
    \centering
    \includegraphics[width=\linewidth]{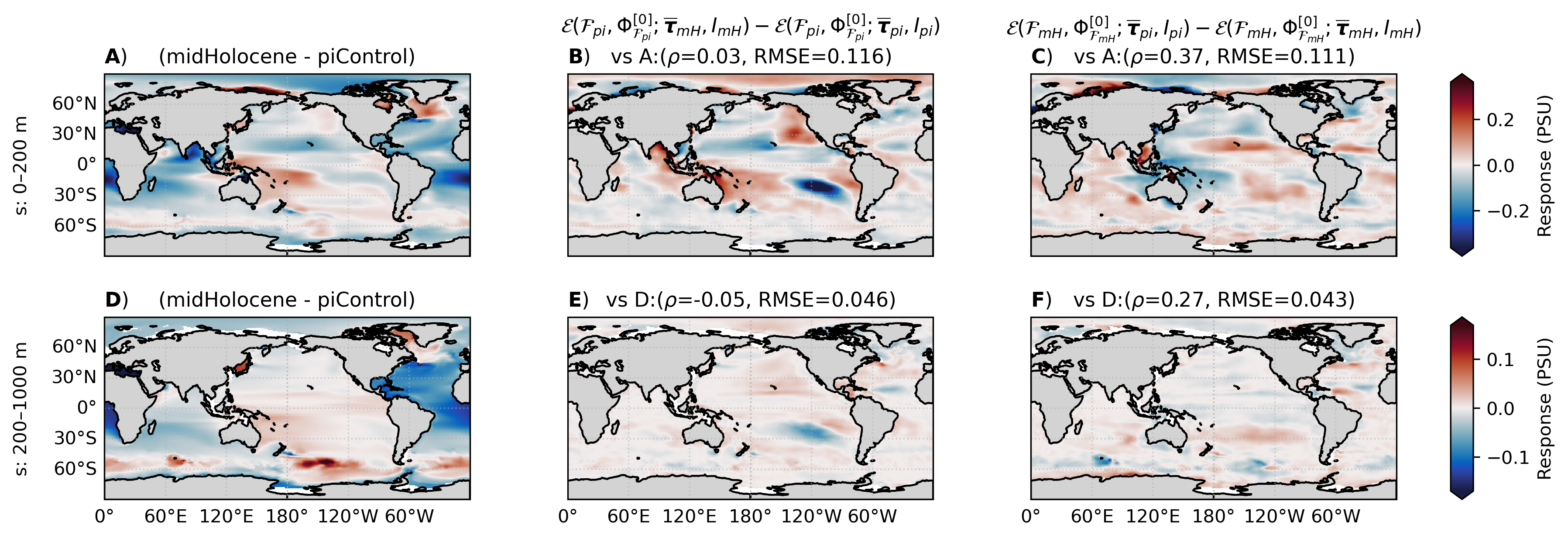}
    \caption{Comparison of the depth averaged salinity response over the upper 200m (A-C) and 200-1000m (D-F) when perturbing an emulator with climatological forcings from the midHolocene/piControl CESM2 experiments. A and D: difference between the CESM2 midHolocene and piControl experiments as an estimate of the true response. B and E: responses of the emulator trained on piControl data when perturbed with midHolocene boundary forcings, $R(\mathcal{F}_{\mathrm{pi}})$. C and F: responses of the emulator trained on midHolocene data when perturbed with piControl boundary forcings, $R(\mathcal{F}_{\mathrm{mH}})$.  For each emulator response we include the correlation, $\rho$, and RMSE with respect to the difference between CESM2 experiments in A and D. Note that for $R(\mathcal{F}_{\mathrm{mH}})$, we flip the sign of the true response before computing the correlation and RMSE.}
    \label{fig:Full_Responses_Map_salinity}
\end{figure}

\begin{figure}[htbp]
    \makebox[\textwidth][c]{\includegraphics[width=7.5in]{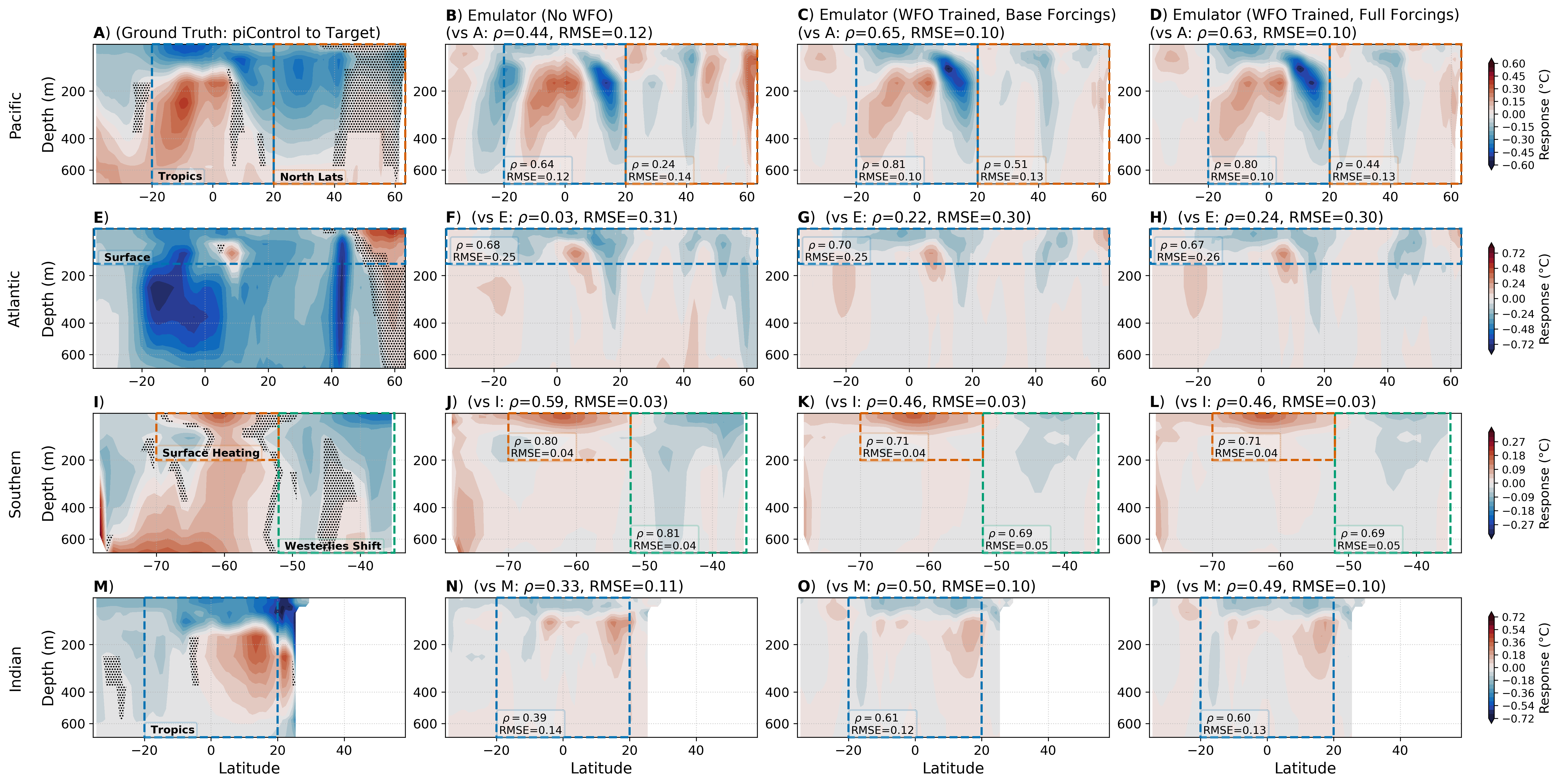}}
    \caption{Comparison of the zonally averaged potential temperature response in each ocean basin when perturbing a piControl baseline climate using climatological forcings taken from midHolocene data. Here we compare the response of the emulator from the main text $\mathcal{F}_{\mathrm{pi}}$ to that of a similar emulator trained with freshwater flux as an additional boundary condition, $\mathcal{F}_{\mathrm{pi}}^{\mathrm{FW}}$. A, E, I, and M: difference between the CESM2 midHolocene and piControl experiments as an estimate of the true response. B, F, J, and N: response of the original emulator trained without freshwater flux, $R(\mathcal{F}_{\mathrm{pi}})$. C, G, K, and O: response of $\mathcal{F}_{\mathrm{pi}}^{\mathrm{FW}}$ perturbed with base midHolocene forcings, keeping the freshwater flux at the piControl climatology. D, H, L, and P: response $\mathcal{F}_{\mathrm{pi}}^{\mathrm{FW}}$ when perturbed with the full set of midHolocene forcings, including the freshwater flux boundary condition. We show responses in the Pacific (A-D), Atlantic (E-H), Southern (I-L), and Indian Ocean (M-P). For each emulator response, we include the correlation, $\rho$, and RMSE with respect to the true change in the respective basin. We additionally highlight specific regions in each basin and include the correlation and RMSE over each region with respect to the change between CESM2 experiments. Stippling denotes regions where the response is not statistically significant at the 95\% confidence level, evaluated using a two-tailed Student's t-test and the standard error of the mean.}
    \label{fig:Full_Responses_All_Basins_wfo}
\end{figure}

\begin{figure}[htbp]
    \makebox[\textwidth][c]{\includegraphics[width=7.5in]{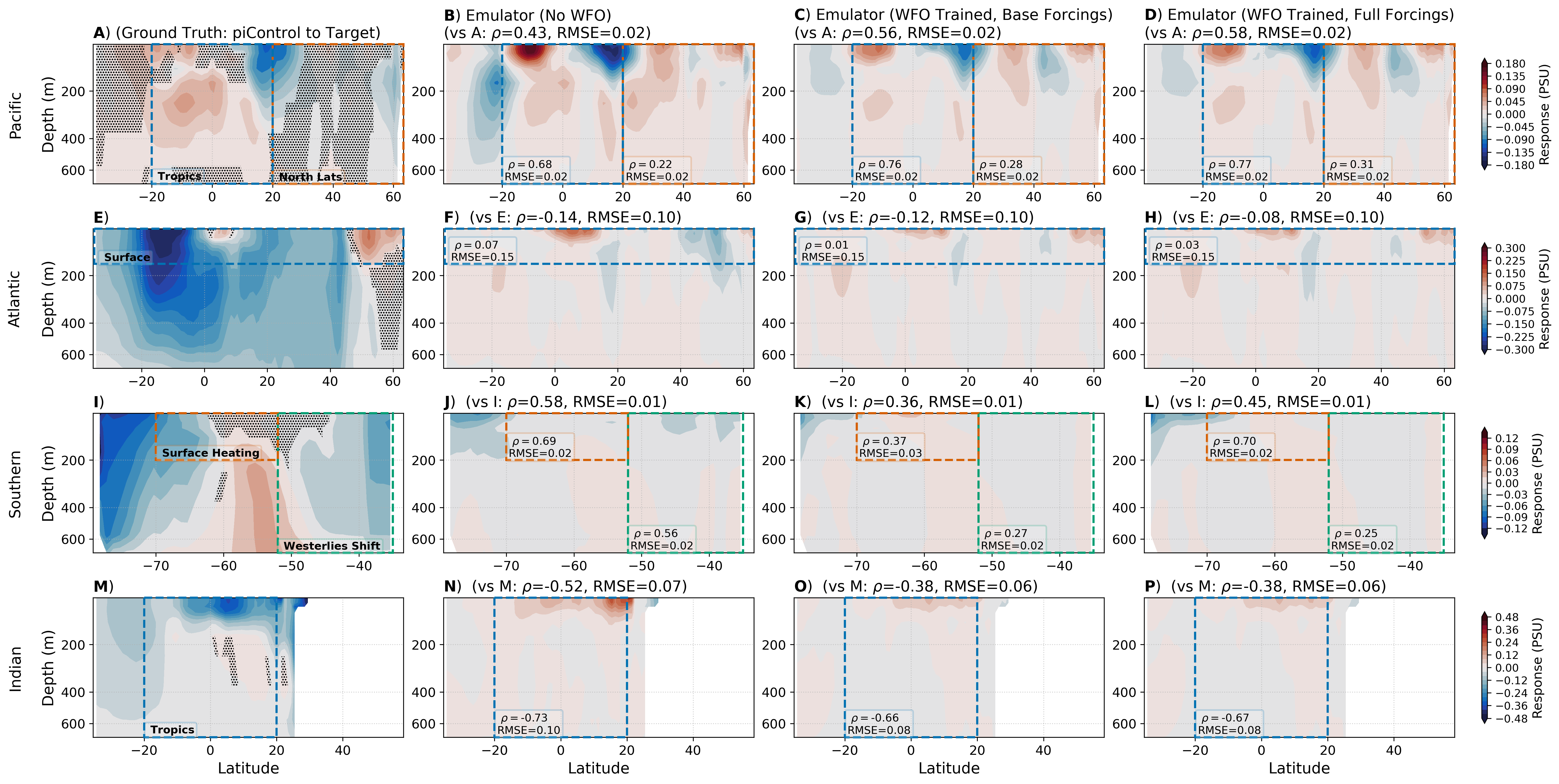}}
    \caption{Comparison of the zonally averaged salinity response in each ocean basin when perturbing a piControl baseline climate using climatological forcings taken from midHolocene data. Here we compare the response of the emulator from the main text $\mathcal{F}_{\mathrm{pi}}$ to that of a similar emulator trained with freshwater flux as an additional boundary condition, $\mathcal{F}_{\mathrm{pi}}^{\mathrm{FW}}$. A, E, I, and M: difference between the CESM2 midHolocene and piControl experiments as an estimate of the true response. B, F, J, and N: response of the original emulator trained without freshwater flux, $R(\mathcal{F}_{\mathrm{pi}})$. C, G, K, and O: response of $\mathcal{F}_{\mathrm{pi}}^{\mathrm{FW}}$ perturbed with base midHolocene forcings, keeping the freshwater flux at the piControl climatology. D, H, L, and P: response $\mathcal{F}_{\mathrm{pi}}^{\mathrm{FW}}$ when perturbed with the full set of midHolocene forcings, including the freshwater flux boundary condition. We show responses in the Pacific (A-D), Atlantic (E-H), Southern (I-L), and Indian Ocean (M-P). For each emulator response, we include the correlation, $\rho$, and RMSE with respect to the true change in the respective basin. We additionally highlight specific regions in each basin and include the correlation and RMSE over each region with respect to the change between CESM2 experiments. Stippling denotes regions where the response is not statistically significant at the 95\% confidence level, evaluated using a two-tailed Student's t-test and the standard error of the mean.}
    \label{fig:Full_Responses_All_Basins_wfo_so}
\end{figure}

\begin{figure}
    \centering
    \includegraphics[width=\linewidth]{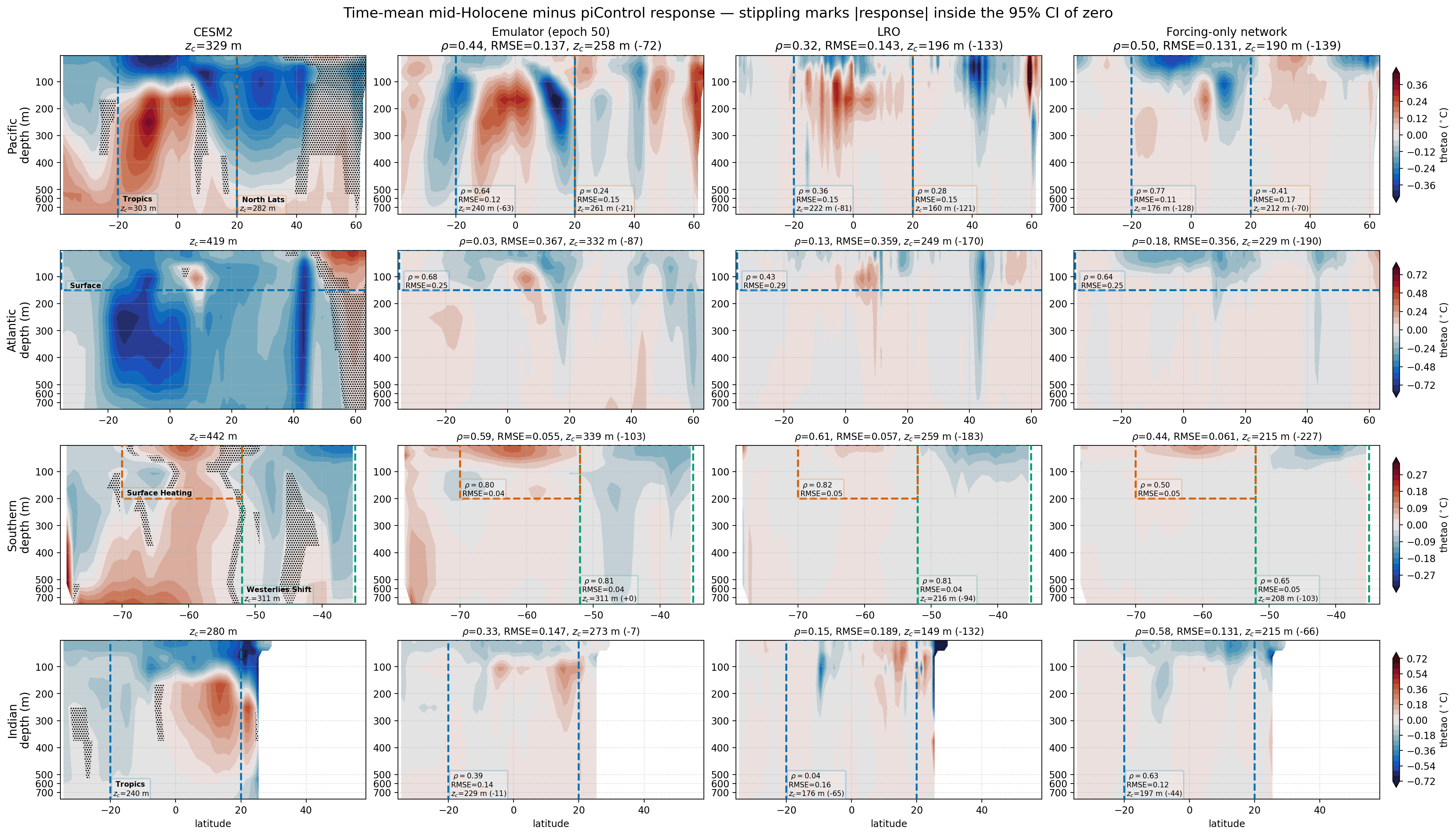}
    \caption{Comparison of the zonally averaged potential temperature response in each ocean basin when perturbing an emulator using climatological forcings taken from midHolocene/piControl CESM2 data, alongside the two baselines. A, E, I, and M: difference between the CESM2 midHolocene and piControl experiments as an estimate of the true response. B, F, J, and N: response of the emulator trained on piControl data when perturbed with midHolocene boundary forcings, $R(\mathcal{F}_{\mathrm{pi}})$, at the checkpoint reported throughout. C, G, K, and O: the same response from the local linear regression operator. D, H, L, and P: the same response from the forcing-only network. Both baselines are fit on piControl data and applied to each set of boundary forcings (Section S2). We show responses in the Pacific (A-D), Atlantic (E-H), Southern (I-L), and Indian Ocean (M-P) over the upper 1000m, and we set the color scale in each row from the corresponding CESM2 panel. For each model response, we include the correlation, $\rho$, and RMSE with respect to the true change in the respective basin, A, E, I, and M. We additionally highlight specific regions in each basin, and again include the correlation and RMSE over each region with respect to the change between CESM2 experiments. Alongside these we report the centroid depth, $z_c$, the first moment with depth of the response amplitude, where the amplitude at each level is the latitude-weighted root mean square of the zonal-mean response. Because the amplitude appears in both the numerator and the denominator, $z_c$ is unchanged by a uniform rescaling of the response and so measures where the response sits rather than how large it is; a value shallower than the CESM2 reference indicates a response confined closer to the surface. We give $z_c$ in meters for every panel and, for the model panels, its difference from the CESM2 value in parentheses, omitting it for the two regions that do not extend below 200m. We stipple panels to mark where the response is not significant relative to internal variability (two-tailed Student's $t$-test at the 95\% level, with internal variability estimated from ten 10-year means of the overlapping period between the two CESM2 experiments and from the ensemble members of each emulator rollout). The local linear operator and the forcing-only network produce a single realization, so no significance test is available for them.}
    \label{fig:Responses_basins_baselines}
\end{figure}

\begin{figure}[htbp]
    \makebox[\textwidth][c]{\includegraphics[width=7.5in]{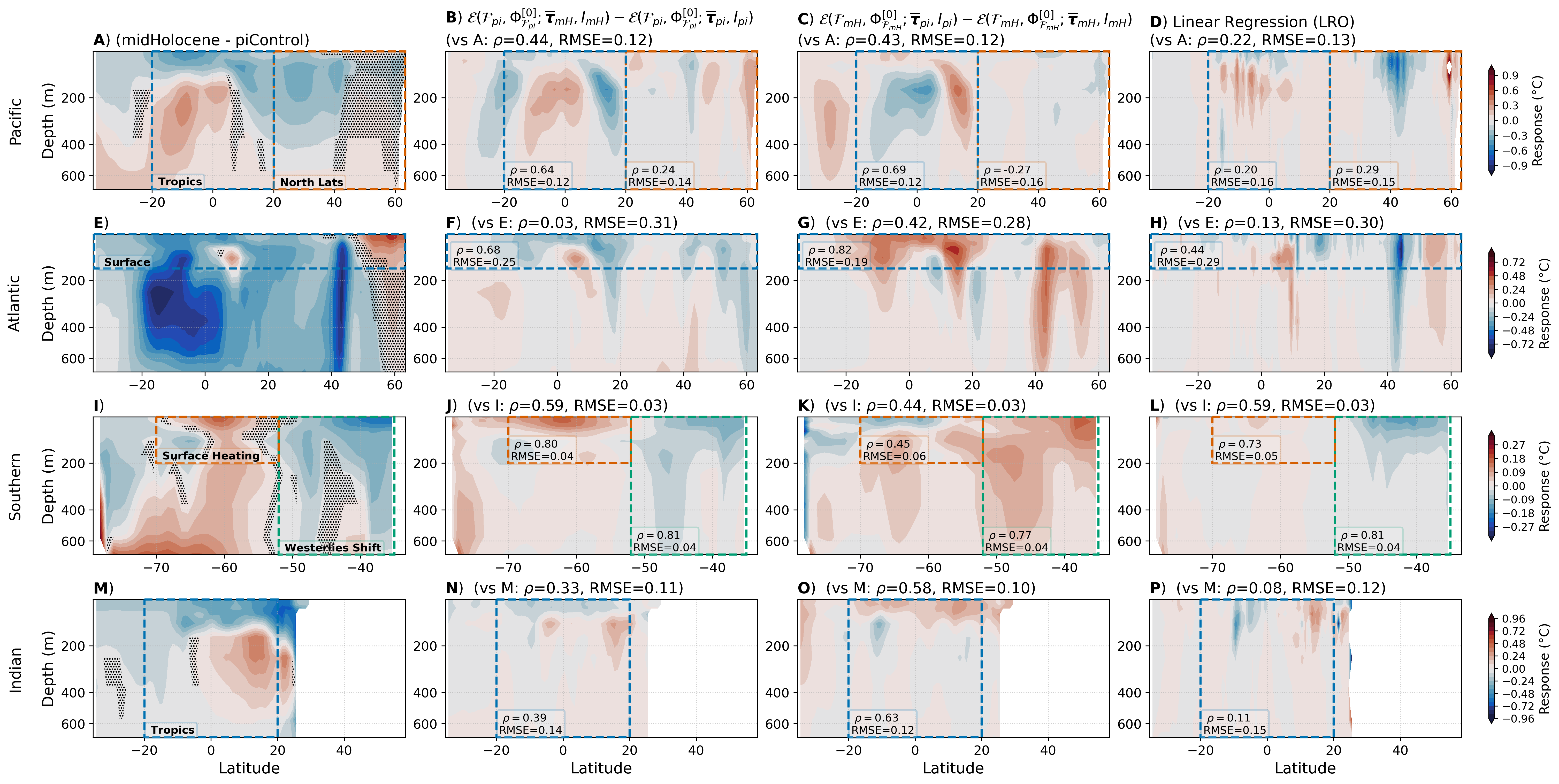}}
    \caption{Comparison of the zonally averaged potential temperature response in each ocean basin when perturbing a piControl baseline climate using climatological forcings taken from midHolocene data. Here we compare the response of the emulators from the main text $\mathcal{F}_{\mathrm{pi}}$ and $\mathcal{F}_{\mathrm{mH}}$ to the LRO described in Section \ref{appendix:lro}. A, E, I, and M: difference between the CESM2 midHolocene and piControl experiments as an estimate of the true response. B, F, J, and N: response of the original emulator trained without freshwater flux, $R(\mathcal{F}_{\mathrm{pi}})$. C, G, K, and O: response of the emulator trained on midHolocene data when perturbed with piControl boundary forcings, $R(\mathcal{F}_{\mathrm{mH}})$. D, H, L, and P: response of the linear model when provided midHolocene boundary conditions. We show responses in the Pacific (A-D), Atlantic (E-H), Southern (I-L), and Indian Ocean (M-P). For each emulator response, we include the correlation, $\rho$, and RMSE with respect to the true change in the respective basin. We additionally highlight specific regions in each basin and include the correlation and RMSE over each region with respect to the change between CESM2 experiments. Stippling denotes regions where the response is not statistically significant at the 95\% confidence level, evaluated using a two-tailed Student's t-test and the standard error of the mean.}
    \label{fig:Full_Responses_All_Basins_linear}
\end{figure}

\begin{figure}[htbp]
    \makebox[\textwidth][c]{\includegraphics[width=7.5in]{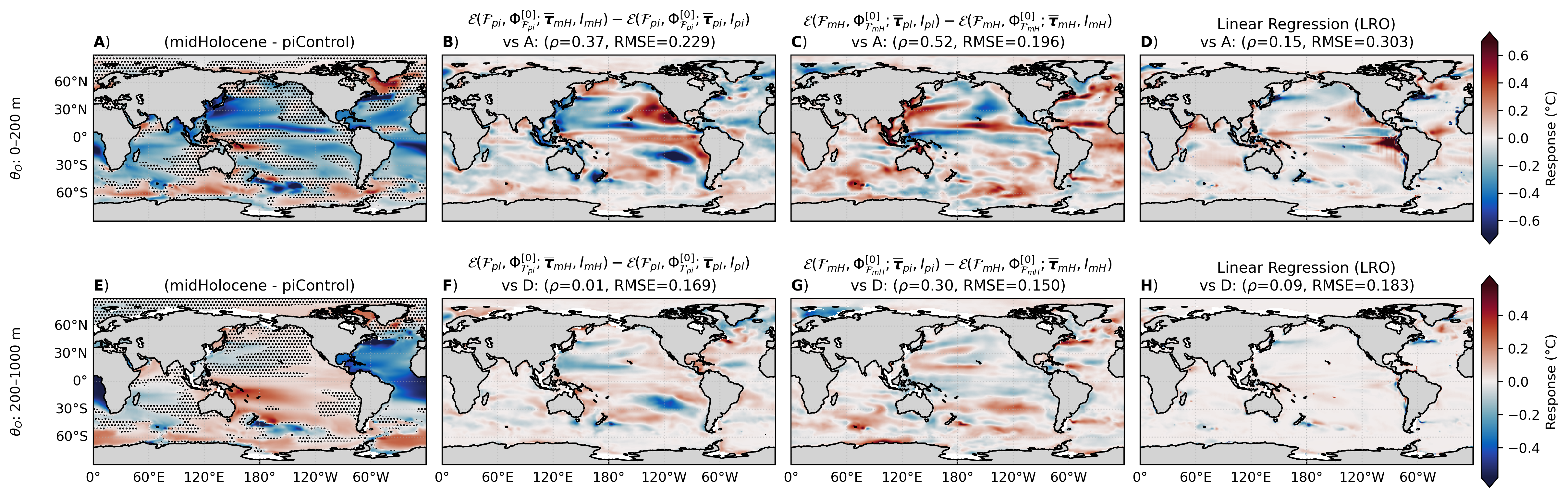}}
    \caption{Comparison of the depth-averaged potential temperature response over the upper 200m (A-D) and 200-1000m (E-H) when perturbing an emulator with climatological forcings from the midHolocene/piControl CESM2 experiments. Here we compare the response of the emulators from the main text $\mathcal{F}_{\mathrm{pi}}$ and $\mathcal{F}_{\mathrm{mH}}$ to the LRO described in Section \ref{appendix:lro}. A and E: difference between the CESM2 midHolocene and piControl experiments as an estimate of the true response. B and F: responses of the emulator trained on piControl data when perturbed with midHolocene boundary forcings, $R(\mathcal{F}_{\mathrm{pi}})$. C and G: responses of the emulator trained on midHolocene data when perturbed with piControl boundary forcings, $R(\mathcal{F}_{\mathrm{mH}})$. D and H: response of the linear model when provided midHolocene boundary conditions. For each emulator response and for the linear model, we include the correlation, $\rho$, and RMSE with respect to the difference between CESM2 experiments in A and E. Note that for $R(\mathcal{F}_{\mathrm{mH}})$, we flip the sign of the true response before computing the correlation and RMSE. For the CESM2 panels, significance is assessed against internal variability estimated from ten 25-year chunks of each experiment; for the emulator responses, against the spread across the five ensemble members (standard error of the mean).}
    \label{fig:Full_Responses_Map_LRO}
\end{figure}

\begin{figure}
    \centering
    \includegraphics[width=.9\linewidth]{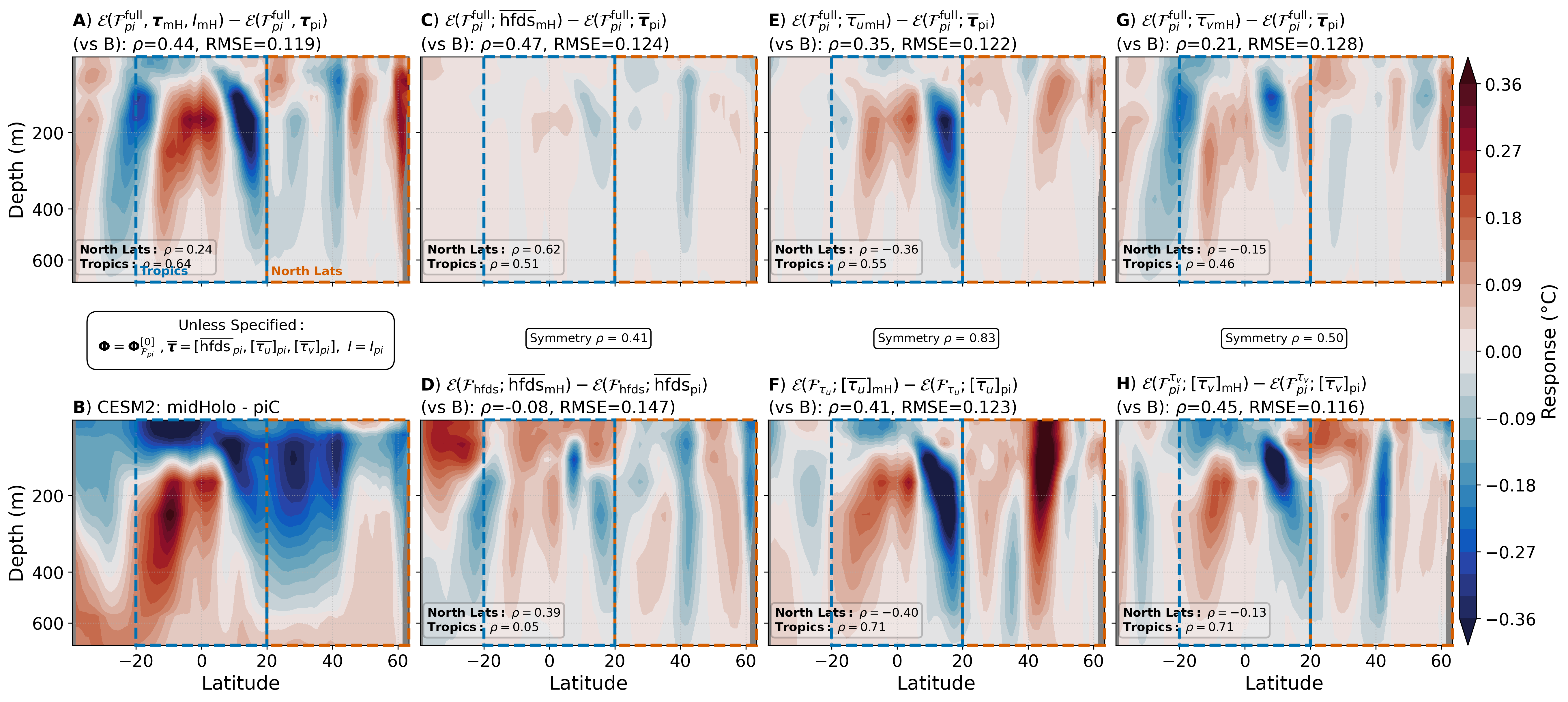}
    \caption{Comparison of the zonally averaged potential temperature response in the  Pacific Ocean when perturbing piControl trained emulators with boundary forcing components of  climatological values from the midHolocene CESM2 data. The title of each panel follows the notation in Section 2.5.3; however, for brevity, unless specified, the rollouts are initialized from a piControl rollout, $\boldsymbol{\Phi}_{\mathrm{pi}}^{[0]}$, each forcing component comes from piControl climatology, $\overline{\boldsymbol{\tau}}=[\overline{\operatorname{hfds}}_{\mathrm{pi}},[\overline{\tau_u}]_{\mathrm{pi}},[\overline{\tau_v}]_{\mathrm{pi}}]$, and the insolation is computed with piControl orbital parameters, $I_{\mathrm{pi}}$. A: total response of $\mathcal{F}_{\mathrm{pi}}^{All}$ when all boundary forcing components are taken from the midHolocene climatology: $R(\mathcal{F}_{\mathrm{pi}}^{All};\overline{\boldsymbol{\tau}}_{\mathrm{mH}},I_{\mathrm{mH}})$. B: difference between the CESM2 midHolocene and piControl runs to provide a reference. C, E, and G: response of the $\mathcal{F}_{\mathrm{pi}}^{All}$ emulator when perturbing individual components of the boundary forcing, (e.g. $R(\mathcal{F}_{\mathrm{pi}}^{All};\overline{\operatorname{hfds}}_{\mathrm{mH}})$.  D, F, and H: response of an emulator trained on a single forcing component to the corresponding boundary component (e.g. $R(\mathcal{F}_{\mathrm{pi}}^{\operatorname{hfds}};\overline{\operatorname{hfds}}_{\mathrm{mH}})$). We show the response for $\overline{\operatorname{hfds}}$ (C and D), $\overline{\tau_u}$(E and F), and $\overline{\tau_v}$ (G and H). For each response we show the correlation, $\rho$,  and RMSE relative to the ground truth response, B. In the middle of the plot, we include the correlation between the component emulator and $\mathcal{F}_{\mathrm{pi}}^{All}$ forced by the respective boundary component (e.g. $\rho(R(\mathcal{F}_{\mathrm{pi}}^{All};\overline{\operatorname{hfds}}_{\mathrm{mH}}),R(\mathcal{F}_{\mathrm{pi}}^{\operatorname{hfds}};\overline{\operatorname{hfds}}_{\mathrm{mH}}))$. Within each emulator response panel, we additionally include the correlation for the two highlighted regions, again with respect to B.}
    \label{fig:Response_Pacific_Thetao}
\end{figure}

\begin{figure}
    \centering
    \includegraphics[width=.9\linewidth]{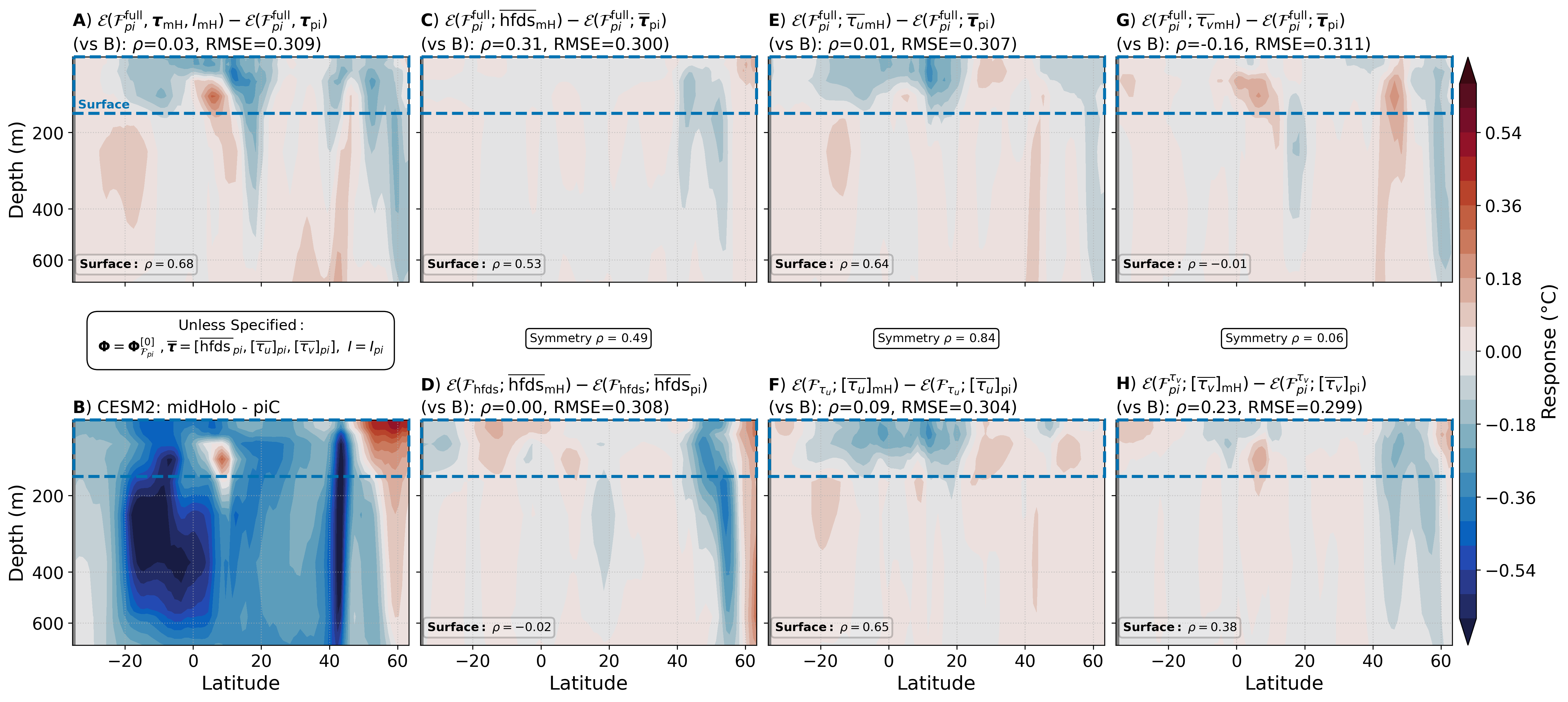}
   \caption{Comparison of the zonally averaged potential temperature response in the  Atlantic Ocean when perturbing piControl trained emulators with boundary forcing components of  climatological values from the midHolocene CESM2 data. The title of each panel follows the notation in Section 2.5.3; however, for brevity, unless specified, the rollouts are initialized from a piControl rollout, $\boldsymbol{\Phi}_{\mathrm{pi}}^{[0]}$, each forcing component comes from piControl climatology, $\overline{\boldsymbol{\tau}}=[\overline{\operatorname{hfds}}_{\mathrm{pi}},[\overline{\tau_u}]_{\mathrm{pi}},[\overline{\tau_v}]_{\mathrm{pi}}]$, and the insolation is computed with piControl orbital parameters, $I_{\mathrm{pi}}$. A: total response of $\mathcal{F}_{\mathrm{pi}}^{All}$ when all boundary forcing components are taken from the midHolocene climatology: $R(\mathcal{F}_{\mathrm{pi}}^{All};\overline{\boldsymbol{\tau}}_{\mathrm{mH}},I_{\mathrm{mH}})$. B: difference between the CESM2 midHolocene and piControl runs to provide a reference. C, E, and G: response of the $\mathcal{F}_{\mathrm{pi}}^{All}$ emulator when perturbing individual components of the boundary forcing, (e.g. $R(\mathcal{F}_{\mathrm{pi}}^{All};\overline{\operatorname{hfds}}_{\mathrm{mH}})$.  D, F, and H: response of an emulator trained on a single forcing component to the corresponding boundary component (e.g. $R(\mathcal{F}_{\mathrm{pi}}^{\operatorname{hfds}};\overline{\operatorname{hfds}}_{\mathrm{mH}})$). We show the response for $\overline{\operatorname{hfds}}$ (C and D), $\overline{\tau_u}$(E and F), and $\overline{\tau_v}$ (G and H). For each response we show the correlation, $\rho$,  and RMSE relative to the ground truth response, B. In the middle of the plot, we include the correlation between the component emulator and $\mathcal{F}_{\mathrm{pi}}^{All}$ forced by the respective boundary component (e.g. $\rho(R(\mathcal{F}_{\mathrm{pi}}^{All};\overline{\operatorname{hfds}}_{\mathrm{mH}}),R(\mathcal{F}_{\mathrm{pi}}^{\operatorname{hfds}};\overline{\operatorname{hfds}}_{\mathrm{mH}}))$. Within each emulator response panel, we additionally include the correlation for the two highlighted regions, again with respect to B.}
    \label{fig:Response_Atlantic_Thetao}
\end{figure}

\begin{figure}
    \centering
    \includegraphics[width=.9\linewidth]{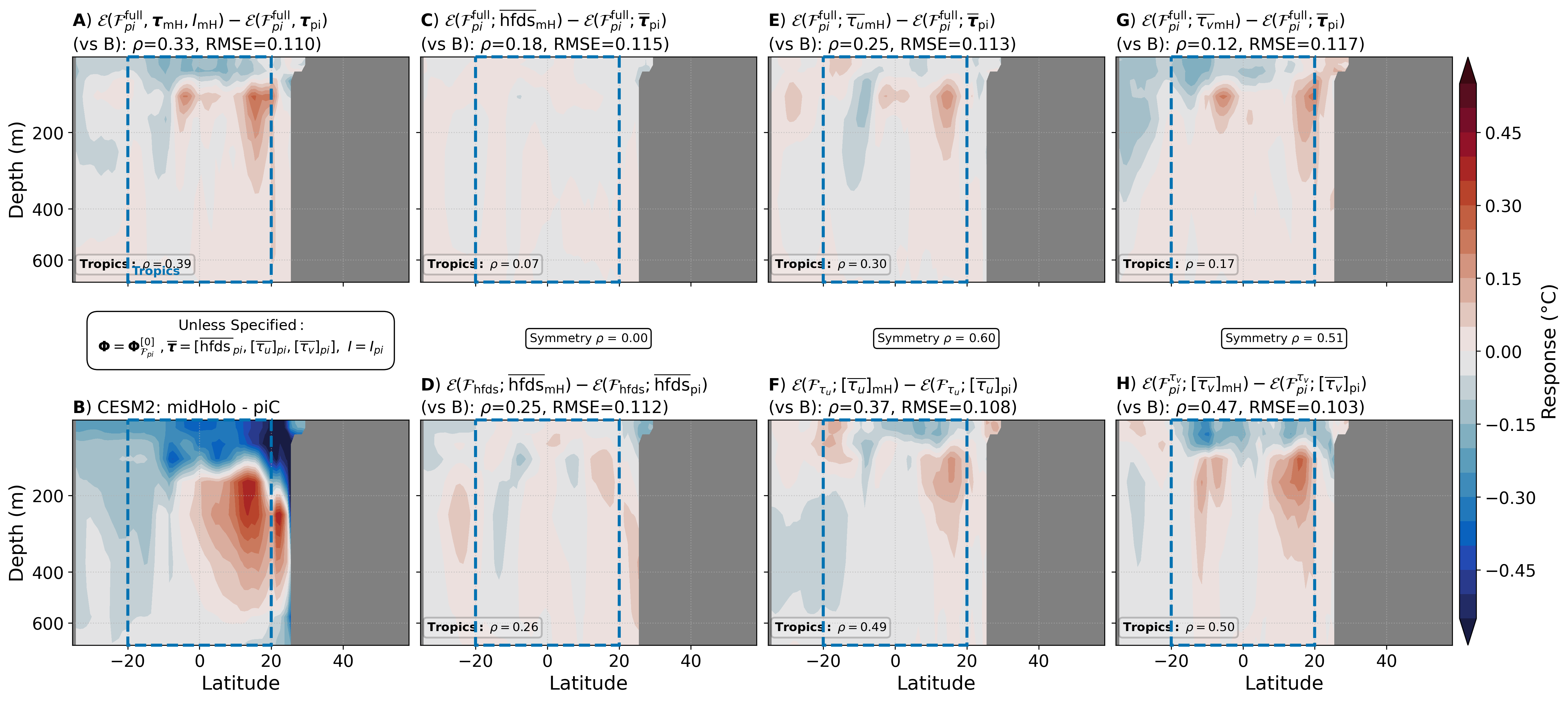}
    \caption{Comparison of the zonally averaged potential temperature response in the  Indian Ocean when perturbing piControl trained emulators with boundary forcing components of  climatological values from the midHolocene CESM2 data. The title of each panel follows the notation in Section 2.5.3; however, for brevity, unless specified, the rollouts are initialized from a piControl rollout, $\boldsymbol{\Phi}_{\mathrm{pi}}^{[0]}$, each forcing component comes from piControl climatology, $\overline{\boldsymbol{\tau}}=[\overline{\operatorname{hfds}}_{\mathrm{pi}},[\overline{\tau_u}]_{\mathrm{pi}},[\overline{\tau_v}]_{\mathrm{pi}}]$, and the insolation is computed with piControl orbital parameters, $I_{\mathrm{pi}}$. A: total response of $\mathcal{F}_{\mathrm{pi}}^{All}$ when all boundary forcing components are taken from the midHolocene climatology: $R(\mathcal{F}_{\mathrm{pi}}^{All};\overline{\boldsymbol{\tau}}_{\mathrm{mH}},I_{\mathrm{mH}})$. B: difference between the CESM2 midHolocene and piControl runs to provide a reference. C, E, and G: response of the $\mathcal{F}_{\mathrm{pi}}^{All}$ emulator when perturbing individual components of the boundary forcing, (e.g. $R(\mathcal{F}_{\mathrm{pi}}^{All};\overline{\operatorname{hfds}}_{\mathrm{mH}})$.  D, F, and H: response of an emulator trained on a single forcing component to the corresponding boundary component (e.g. $R(\mathcal{F}_{\mathrm{pi}}^{\operatorname{hfds}};\overline{\operatorname{hfds}}_{\mathrm{mH}})$). We show the response for $\overline{\operatorname{hfds}}$ (C and D), $\overline{\tau_u}$(E and F), and $\overline{\tau_v}$ (G and H). For each response we show the correlation, $\rho$,  and RMSE relative to the ground truth response, B. In the middle of the plot, we include the correlation between the component emulator and $\mathcal{F}_{\mathrm{pi}}^{All}$ forced by the respective boundary component (e.g. $\rho(R(\mathcal{F}_{\mathrm{pi}}^{All};\overline{\operatorname{hfds}}_{\mathrm{mH}}),R(\mathcal{F}_{\mathrm{pi}}^{\operatorname{hfds}};\overline{\operatorname{hfds}}_{\mathrm{mH}}))$. Within each emulator response panel, we additionally include the correlation for the two highlighted regions, again with respect to B.}
    \label{fig:Response_Indian_Thetao}
\end{figure}

\begin{figure}
    \centering
    \includegraphics[width=.6\linewidth]{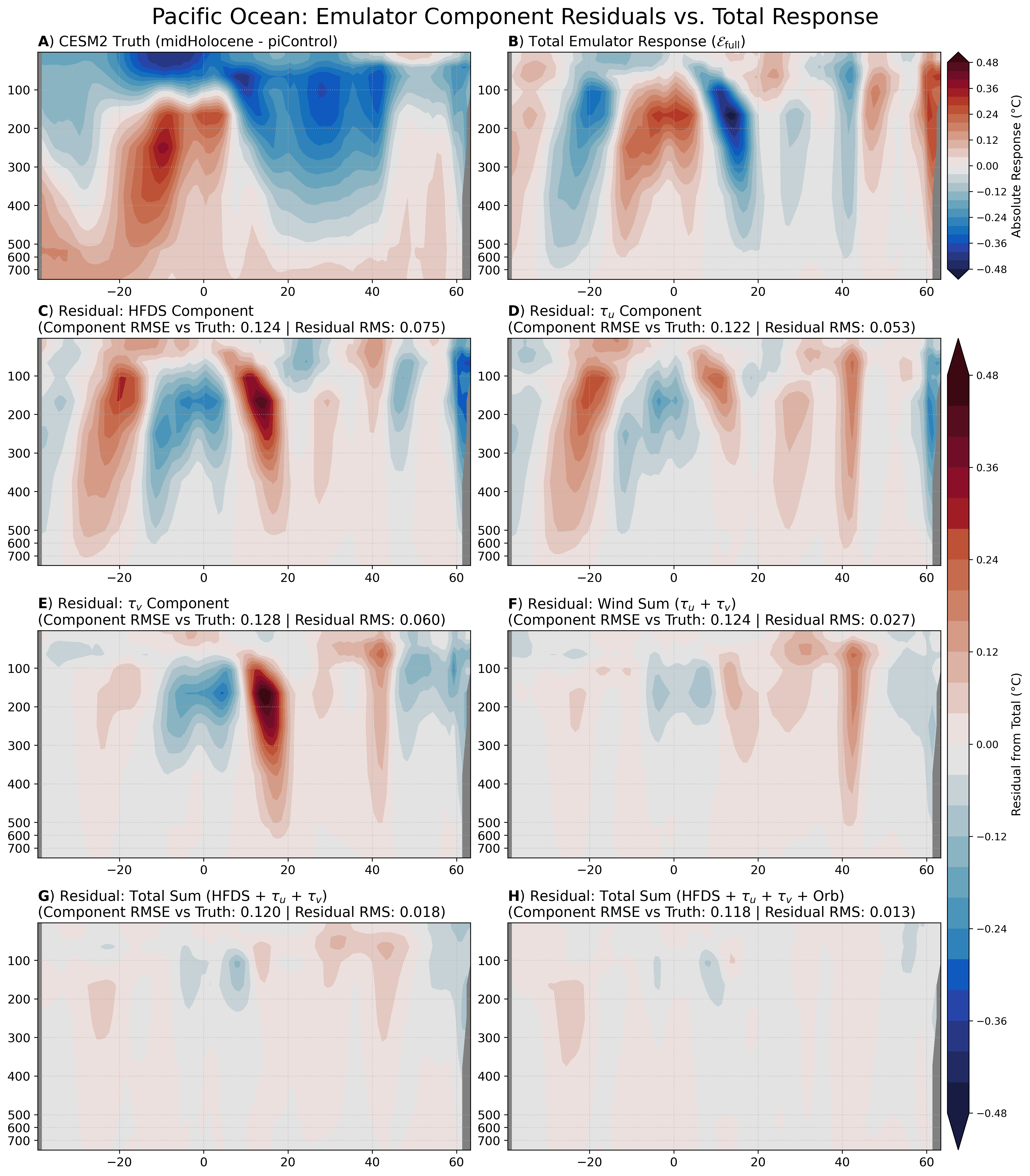}
    \caption{Comparison of the total Pacific Ocean potential temperature response to midHolocene boundary forcings against the linear superposition of individual component responses. A: Difference between the CESM2 midHolocene and piControl numerical experiments for reference. B: Total potential temperature response of the full emulator ($\mathcal{F}_{pi}^{All}$) to all midHolocene boundary forcings, denoted as $R(\mathcal{F}_{pi}^{All};\overline{\tau}_{mH},I_{mH})$. C-E: Residuals computed between a component response of the emulator to the total response. The component responses are heat flux (C: $R(\mathcal{F}_{pi}^{All};\overline{hfds}_{mH})$), zonal surface stress (D: $R(\mathcal{F}_{pi}^{All};[\overline{\tau_{u}}]_{mH})$), and meridional surface stress (E: $R(\mathcal{F}_{pi}^{All};[\overline{\tau_{v}}]_{mH})$). F-H: Residuals computed between linear superpositions of component responses of the emulator to the total response. The superpositions are both surface stress components (F), all dynamic boundary forcings consisting of zonal surface stress, meridional surface stress, and heat flux (G), and all boundary forcings including solar insolation (H).}
    \label{fig:residuals_Pacific}
\end{figure}

\begin{figure}
    \centering
    \includegraphics[width=.6\linewidth]{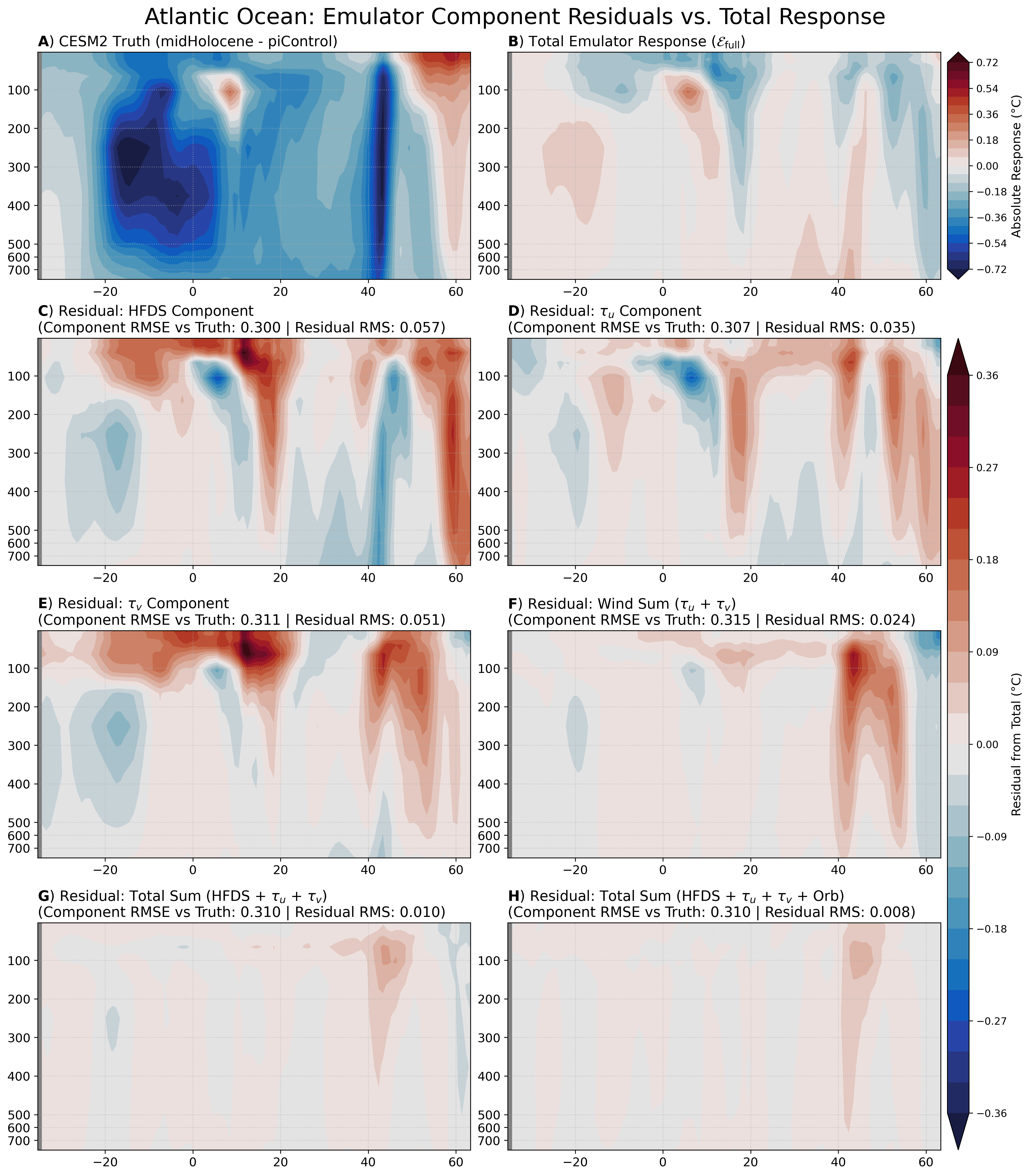}
    \caption{Comparison of the total Atlantic Ocean potential temperature response to midHolocene boundary forcings against the linear superposition of individual component responses. A: Difference between the CESM2 midHolocene and piControl numerical experiments for reference. B: Total potential temperature response of the full emulator ($\mathcal{F}_{pi}^{All}$) to all midHolocene boundary forcings, denoted as $R(\mathcal{F}_{pi}^{All};\overline{\tau}_{mH},I_{mH})$. C-E: Residuals computed between a component response of the emulator to the total response. The component responses are heat flux (C: $R(\mathcal{F}_{pi}^{All};\overline{hfds}_{mH})$), zonal surface stress (D: $R(\mathcal{F}_{pi}^{All};[\overline{\tau_{u}}]_{mH})$), and meridional surface stress (E: $R(\mathcal{F}_{pi}^{All};[\overline{\tau_{v}}]_{mH})$). F-H: Residuals computed between linear superpositions of component responses of the emulator to the total response. The superpositions are both surface stress components (F), all dynamic boundary forcings consisting of zonal surface stress, meridional surface stress, and heat flux (G), and all boundary forcings including solar insolation (H).}
    \label{fig:residuals_Atlantic}
\end{figure}

\end{document}